# Design Limits on Large Space Stations

David W. Jensen, Ph.D.
*Technical Fellow, Retired*
*Rockwell Collins / Collins Aerospace*
*Cedar Rapids, IA 52302*
*david.jensen@alumni.iastate.edu*

## Abstract

*As the space industry matures, large space stations will be built. This paper organizes and documents constraints on the size of these space stations. Human frailty, station design, and construction impose these constraints. Human limitations include gravity, radiation, air pressure, rotational stability, population, and psychology. Station design limitations include gravity, population, material, geometry, mass, air pressure, and rotational stability. Limits on space station construction include construction approaches, very large stations, and historic station examples. This paper documents all these constraints for thoroughness and review; however, only a few constraints significantly limit the station size. This paper considers rotating stations with radii greater than 10 kilometers. Such stations may seem absurd today; however, with robotic automation and artificial intelligence, such sizes may become feasible in the future.*

## 1 Introduction

This introduction first overviews the organization of this paper. It then offers an illustration and description of a large space station. It provides a historical review of space stations and the four main types of station geometries. It covers station sizes and example metrics on the station size. This introduction concludes with this work's background foundation.

Section 2 covers human frailty limitations and their effect on the station design. These limitations include tolerance to rotation, gravity, and air pressure. The rotation and centripetal gravity directly limit the station size and design. The centripetal gravity also defines the livable air pressure, which imposes a human limit. Radiation and psychology affect humans and impose station limits.

Section 3 covers station design constraints. Such constraints include centripetal gravity and air pressure, which both significantly impact the station design. Population limits are defined by the target purpose, floor space allocation, and multiple floors. Building material imposes limits on the station design. These limits produce a set of refined station geometries. For this analysis, these geometries are decomposed into their major components. Their masses are analyzed and used to determine the station's rotational stability.

Section 4 considers space station construction. This includes a review of common construction approaches planned for space stations. This section considers very large space stations and constraints such as material, atmosphere, and population. It also includes examples of large stations and large historic space stations. This section provides summaries of the large station limitations.

Section 5 summarizes this paper. It contains summaries on a broad set of constraints, including gravity, radiation, material, air pressure, stability, mass, population, and psychology. This section includes graphical charts of these large station limitations and our conclusions from those limitations.

### 1.1 Example Space Station

The cutaway rendering in Figure 1-1 helps to understand our vision of a large rotating space station. This rendering shows or implies many of the topics to be considered in this paper. The rendering shows a part of a large elliptical torus space station. It shows a central floor curving upward. This station provides open space and an excellent vista for the residents. Below that central floor are multiple floors. These floors provide space for agriculture, industry, commercial, residential,

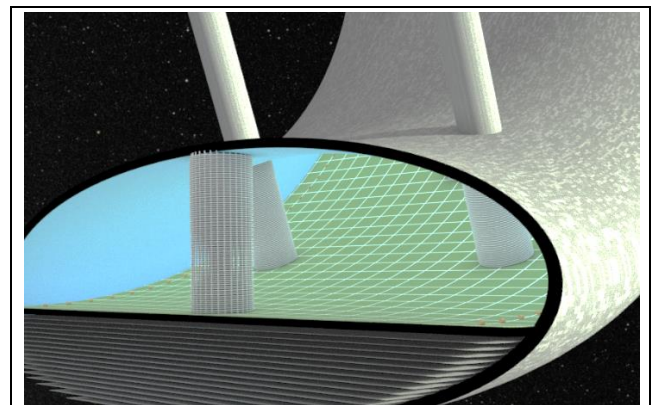

*Credit: Self-produced using Blender. Background Milky Way from NASA SVS Deep Star Maps [NASA Image Public Domain]*

*Large Torus Space Station: Cutaway Diagram showing 17 subfloors and 82 floors in spoke towers*

**Figure 1-1 – Multiple Floors in Large Space Station**

and services. The station is designed to provide an adequate earth-like gravity range from the main floor to the bottom outer rim. The station design provides passive rotational stability. The station uses high-tensile strength structures for its construction. Such a station could be built using construction material from asteroids; however, this station could also use material from moons or dwarf planets.

## 1.2 Historic Limits

Space stations have been researched for over 150 years. Table 1-1 provides a historical summary of space station geometries. The table offers significant events and years for four common space station geometries. Many of the earliest researchers were inspired by science fiction works of the late 1800s, such as those by Jules Verne [Verne 1865] and H.G. Wells [Wells 1898]. The early 1900s introduced the concepts of chemically fueled and multiple-stage rockets to lift people and equipment to Earth's orbit and beyond. Early researchers first described living in a weightless environment and then used rotating stations and centripetal force to replace gravity.

## 1.3 Geometries

There are four common space station geometries: Sphere, Dumbbell, Torus, and Cylinder [Johnson and Holbrow 1977]. Table 1-1 shows over 100 years of historical examples for each of the four geometries. Later sections describe and provide diagrams of these geometries. All four of these geometries have symmetry to support spinning and the production of centripetal gravity. They also have hollow enclosed regions to hold the atmosphere. From the 1900s to today, the preferred space station geometry has varied from the torus, to spheres, and to cylinders.

## 1.4 Station Sizes

Space station designs have ranged from tiny enclosures to huge habitats.

**Small:** The Kalpana One space structure is an example of a small space station. It is a rotating cylinder with a radius of 250 meters and a length of 325 meters [Globus et al. 2007]. It would rotate two times every minute to create an Earth-like gravity on the surface. This design would have at least 0.5 square kilometers of surface area and support 3,000 people.

**Medium:** A 1974 Stanford study detailed a rotating torus design with a major radius of 830 meters and a minor radius of 65 meters [Johnson and Holbrow 1977]. This torus would rotate at 1 revolution per minute (rpm) to produce the effect of one Earth gravity on the outer surface. Their study suggested 67 square meters of projected area and 1738 cubic meters of habitat volume were needed for each inhabitant. The Stanford Torus would have 678,000 square meters of surface area and support a population of 10,000.

**Large:** The space station in the 2013 movie Elysium was a large rotating torus with a major radius of 30 kilometers and a minor radius of 1.5 kilometers. It would rotate about once every 6 minutes to create an Earth-like gravity on the surface [Brody 2013]. With one floor, this station would have at least 565 square kilometers of surface area. Elysium supports 500,000 people; as such, it has a population density of 1131 square meters per resident.

## 1.5 Metrics on Station Limits

Our studies have found multiple metrics used to evaluate and select a station geometry. One metric was to maximize the station volume, and O'Neill found that the sphere geometry was superior [O'Neill 1976]. Another was to minimize the mass for a given population and the cylinder geometry was superior [O'Neill 1976] [Globus et al. 2007]. A third example metric was used to minimize the station mass for a given rotation radius, and the torus geometry was superior [Johnson and Holbrow 1977] [Misra 2010]. Providing passive rotational balance imposes a geometry constraint on space station designs [Brown 2002]. This constraint dramatically changed the shape of cylinder space stations [Globus et al. 2007].

These assessments and selections were typically based on thin space station shells. These historic assessments were usually based on a single projected floor. Because this study uses thick shells and multiple floors, it uses slightly different geometries than the historical studies. Our study uses ellipsoids instead of spheres, short cylinders instead of long cylinders, elliptical cross-section toruses instead of circular

| Table 1-1 – Space Station Geometries and History ||||
|---|---|---|---|
| *Sphere* | *Dumbbell* | *Torus* | *Cylinder* |
| 1869 - "The Brick Moon" by Edward Everett Hale [Hale 1869] | 1923 - Two Supply Ships with cable several kilometers long [Oberth 1923] | 1929 - "Wohnrad" (Living Wheel); Herman Potočnik (1929) | 1920 - Rotating (Traverse Axis) [Tsiolkovsky 1920] |
| 1883 - Spherical spacecraft - Konstantin Tsiolkovsky [Tsiolkovsky 1883] | 1960s - NASA research and designs using spent Apollo stages | 1952 - von Braun's Space Station; [von Braun 1952] | 1933 - Rotating (Center Axis) [Tsiolkovsky 1933] |
| 1929 - Bernal Sphere - not rotating [Bernal 1929] | 2007 - Economic Homesteading Bolos [Curreri 2007] | 1959 - Project Horizon; [Koelle and Williams 1959] | 1954 - Gigantic habitable cylinders – [Oberth 1954] |
| 1937 - Gauze of Light Traps - Olaf Stapledon [Stapledon 1937] | 2024 – Double dumbbell - rotationally stable; [Jensen 2024s] | 1968 - Space Station V in 2001: A Space Odyssey; [Clarke 1968] | 1956 - Large Cylinder and Disc - [Romick 1956] |
| 1960 - Dyson Shell – [Dyson 1960] | | 1974 – Stanford Torus; [Johnson and Holbrow 1977] | 1974 - O'Neill - Island Three [O'Neill 1974] |
| 1974 - O'Neill – Island One and Two [O'Neill 1974] | | 2013 - Elysium Torus in the movie Elysium; [Brody 2013] | 1990 - Hatbox - Kalpana - [Globus et al. 2007] |
| 2023 - Ellipsoid [Jensen 2023] | | | |

cross-section toruses, and avoids single dumbbell structures. We humbly include our ellipsoid station design and double dumbbell at the bottom of Table 1-1. Historical examples of the double dumbbell exist, but we have not found a dumbbell stability analysis in the literature.

## 1.6 Background Foundation

This paper reviews and extends results from our Asteroid Restructuring paper [Jensen 2023]. That paper provides a wealth of citations and analysis. We liberally reuse that publication's charts, images, and text to provide context and continuity. That paper showed that asteroids can provide nearly unlimited amounts of building material. It also showed that automation can provide almost unlimited amounts of labor to build these rotating space stations. This paper continues the investigation of the various constraints limiting the potential size of these stations. These constraints are organized by considering human frailty, station design, and construction approaches.

## 2 Human Frailty Limitations

Space stations are designed to support personnel. As such, this first set of constraints are those affecting the welfare of the occupants. These limits include gravity, air pressure, radiation, rotational stability, population, and psychological.

## 2.1 Gravity Constraints

The following subsections cover multiple aspects of gravity. They explain centripetal gravity, review Earth gravity, present gravity equations, and cover health impacts.

### 2.1.1 Centripetal Gravity

Researchers have envisioned rotating space stations to provide centripetal gravity since the 1920s [Tsiolkovsky 1920] [Oberth 1923] [Noordung 1929]. Three forces are felt in the rotating station: centripetal (inward), centrifugal (outward), and Coriolis (movement) forces. The centripetal force is a "real" force directed radially towards the center of rotation. The Coriolis and centrifugal forces are called fictitious or inertial forces and act as additional forces in the rotating station. All three forces contribute to the object's acceleration. The Coriolis force is a function of the object's velocity. The centrifugal force is a function of the object's distance from the center of rotation. Multiple sources are available to provide details on these forces [Hall 1991] [Hand and Finch 1998] [Lucas 2019].

### 2.1.2 Gravity Equations

On the Earth, the force of gravity at the Earth's surface is $GMR^2$, where G is the gravitational constant, M is the mass of the Earth, and R is the radius of the Earth. The variable $g_0$ is the gravity acceleration at Earth's surface (9.80665 meters per second squared). Earth's gravity is $g_h = g_0 (R/(R+h))^2$ where R is the Earth's radius, and h is the height above Earth's surface. The Earth's radius is 6378 kilometers, and gravity's force decreases slowly with height above the surface. Figure 2-1 shows this slow decrease as a solid black line. In a rotating station, the centripetal force is always directed radially towards the center of rotation. In the rotating station, this force is $\omega^2 R$, where $\omega$ is the station rotation speed, and R is the radius to the center. The variable $g_0$ can also represent the centripetal gravity acceleration at the station's outer rim. The station's centripetal gravity is $g_h = g_0 (R-h)/R$, where R is the rotating station's radius, and h is the height above the outer rim. Figure 2-1 shows this centripetal gravity for several station sizes. These stations rotate at speeds to produce centripetal gravity $g_0 = 1.05g$ at the outer rim (h=0). The centripetal gravity in small stations becomes smaller quickly with height. In large stations, the centripetal gravity becomes more similar to Earth's gravity.

### 2.1.3 Gravity Health Impact

Human bodies have adapted to a very narrow range of gravity on Earth's surface. This gravity only ranges from 0.996g to 1.003g over the surface of the Earth [Hirt et al. 2013]. This paper considers what range of space station gravity will be acceptable in the space station for the colonists' health.

**Microgravity:** Today, it is known that microgravity can negatively affect human health. Astronauts in the International Space Station (ISS) experience microgravity. Common initial effects include nausea, vomiting, and headaches [Hall 1997]. Longer-term microgravity negatively affects human health by weakening bones and muscles. This increases the risk of osteoporosis and cardiovascular problems.

**Low Gravity:** The human body has no problem with Earth's gravity; it has many problems with zero gravity. It is unknown if the human body can tolerate long-term gravity between 0g and 1g [NASA 2004]. Lower gravities are of interest because of planned long missions and colonies on the Moon and Mars. The gravity of Mars is 0.38g, and the gravity of the moon is 0.17g. There is debate on whether those gravities could prevent the microgravity physical problems. Boyle [Boyle 2020] writes, despite more gravity than microgravity, that *"long-duration visitors will still experience some of low gravity's deleterious effects."*

**High Gravity:** There could also be issues with high gravity. Researchers recently determined a muscle strength upper limit of 1.1g, but they also found that gravity maximum tolerance could be increased up to 3g or 4g [Poljak, Klindzic, and Kruljac 2018]. That upper range required the strength of an elite athlete.

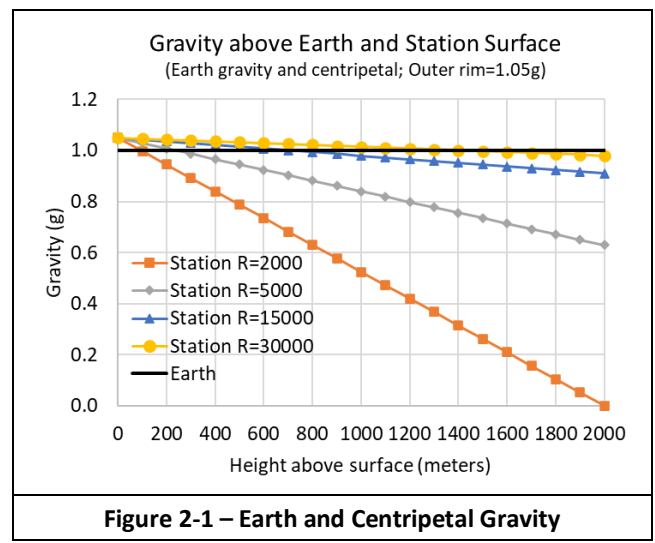

**Figure 2-1 – Earth and Centripetal Gravity**

Maintaining a near-Earth gravity range in the large rotating space stations' habitable areas should help avoid deleterious health effects. The station's artificial gravity range is optimistically limited between 0.8g to 1.2g. Long-term residential areas and most of our analysis and designs limit the gravity range from 0.95g to 1.05g.

2.1.3.1 Centripetal Gravity Benefits

Long-term microgravity exposure produces many adverse side effects and health issues. The goal of rotating a station is to produce centripetal gravity and ameliorate those issues. Some experiments have given credence to that goal. Experiments on small animals and cell cultures have yielded encouraging results. These experiments were performed in centrifuge systems in a weightless environment. Centrifuged animals in those experiments saw improvements in lifespans, hemolysis (red blood cell problem), and bone systems [Connors et al. 1985]. Microgravity impacts the immune system and causes "space anemia." Centrifuging prevents the negative impact on the T-cell lymphocytes [Diamandis 1987].

2.1.3.2 Centripetal Gravity Detriments

Motion sickness, movement errors, and throwing errors are some of the detriments from rotation and centripetal gravity. Motion sickness is the most serious problem in space settlement [Globus and Hall 2017]. Rotation creates inner ear disturbances and causes motion sickness and disorientation. Adapting to the point where no motion sickness is perceived can take hours, days, or weeks. In a rotating station, Coriolis forces can cause limb motion to a target to be inaccurate or to take an unusual path. The rotation also makes accurately throwing objects difficult. One must adapt by aiming and throwing the object to the future position of a target. The detriments are not typically debilitating. With large stations, the rotation speeds will be slow with fast adaptation.

*2.1.4 Rotation Constraints*

Theodore Hall thoroughly reviews and analyzes five historic guidelines [Hall 1997]. He combined the five studies into one "comfort chart." Al Globus and Theodore Hall recently used this comfort chart in their paper on rotation tolerance [Globus and Hall 2017]. We previously adapted that comfort chart in [Jensen 2023]. This paper includes that chart in Figure 2-2 and re-uses the summary and text from that paper. The chart shows the station radius on the y-axis from 100 to 25,600 meters on a logarithmic scale. The x-axis shows the rotation in revolutions per minute.

Figure 2-2 shows Mars's gravity of 0.38g and the moon's gravity of 0.17g for reference. There is debate on whether those gravities could prevent the low-gravity physical problems [Boyle 2020]. The Mars and Moon gravities are placed outside the tolerance zone. In the chart, we optimistically set a minimum long-term gravity of 0.8g and a maximum gravity tolerance of 1.2g.

The station rotation speed is limited to 2 rpm to prevent motion sickness. The chart includes a limit using a large ratio of Coriolis to centripetal force (Acor/Acent =25% at velocity=5 meters per second) as included by [Globus and Hall 2017]. In large stations, the Coriolis distortion and the motion sickness have minimal effect on the tolerance zone. Considering material strengths, the top boundary points are limited to a maximum hoop stress radius with a rotation radius of 10 kilometers [Jensen 2023]. The pentagon boundary points of the chart are defined in a similar style to [Globus and Hall 2017]; more details are in [Jensen 2023]. Figure 2-2 includes a 1 rpm limit. An earlier NASA study defined this as the maximum rotation velocity to support using an attached shield [Johnson and Holbrow 1977]. This implies a minimum rotation radius of almost 900 meters. Larger radius stations will keep the rotation speed lower than 1 rpm.

*2.1.5 Artificial Gravity Summary*

Multiple papers provide our foundation for artificial gravity [Johnson and Holbrow 1977] [Hall 1997] [Hall 2006] and [Globus and Hall 2017]. Centripetal gravity addresses microgravity health issues but introduces anomalous effects on moving objects and people. We agree with Hall who ultimately finds "*that it is impossible to design away the gravitational distortions inherent in rotating environments. They can be kept arbitrarily small only by keeping the radius sufficiently large.*" [Hall 1993]. This paper envisions large stations that will minimize these anomalous effects.

## 2.2 Air Pressure Constraints

Air is obviously required to support a station's population. Early space missions used low-pressure, pure oxygen to

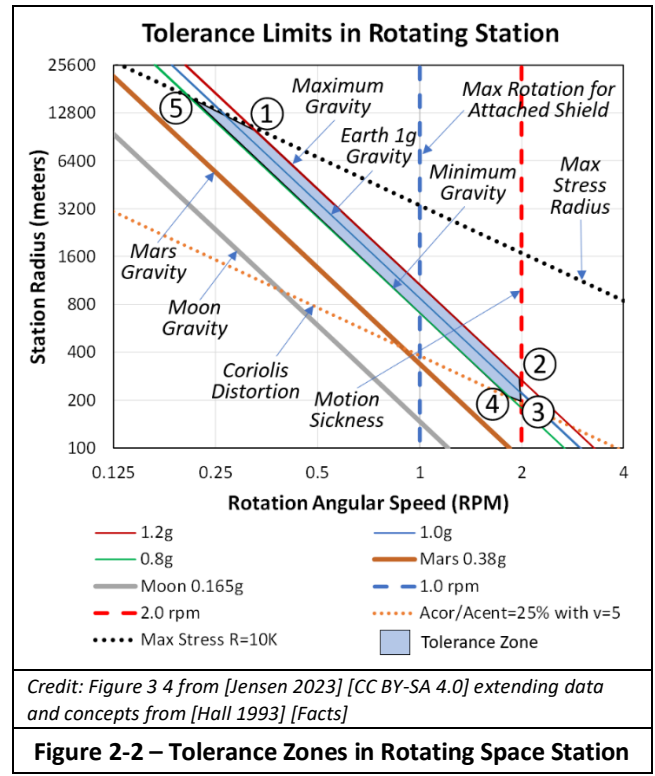

*Credit: Figure 3 4 from [Jensen 2023] [CC BY-SA 4.0] extending data and concepts from [Hall 1993] [Facts]*

**Figure 2-2 – Tolerance Zones in Rotating Space Station**

reduce weight and still provide a breathable atmosphere. This study assumes an Earth-like mixture of nitrogen, oxygen, and carbon dioxide in the station.

*2.2.1 Introduction*

This subsection introduces air pressure limits on people. It introduces the effects of low and high air pressure on people. Our human bodies have also adapted to a range of air pressures on Earth's surface. The highest air pressures are at low elevations. The Dead Sea is 430 meters below sea level and has an air pressure of 106,624 Pascals. Sea level has an air pressure of 101,325 Pascals. The highest permanent settlement, La Rinconada, at 5100 meters above sea level, has a low air pressure of 53,500 Pascals. This is where severe altitude sickness can occur (above 3,500 meters). Mountain climbers call the region above 8,000 meters the death zone with less than 35,600 Pascals air pressure. Mount Everest has an altitude of 8,850 meters and air pressure of 33,700 Pascals.

This section considers what range of air pressures will be acceptable for the space station colonists' health. Higher air pressures are denser and would use more air to fill the station. Sea level is our highest air pressure. Our lowest air pressure limit is 83,728 Pascals. This is the air pressure of Denver, which is 1609 meters above sea level. These limits are conservative. Nearly all individuals would breathe comfortably at these air pressures. The designers of the Stanford Torus planned on using 51,700 Pascals of air pressure [Johnson and Holbrow 1977]. Lower air pressure and density reduce the mass, costs, and stresses. One-half sea level pressure was deemed acceptable for long-term habitation. This is less than our Denver air pressure limit.

*2.2.2 Air Pressure Equations*

To evaluate the effect of air pressure on the station design, this subsection reviews equations to model the air pressure. This paper uses the barometric formula for the Earth from [Lente and Ősz 2020]. Their approach derives the barometric formula at a constant temperature using differential equations. This results in the following equation to calculate the Earth's air pressure at altitudes above its surface:

$$P_h = P_0 \exp\left(-\frac{Mg_0}{kT}h\right)$$

Where $P_0$ is the air pressure at sea level, and h is the height above Earth's surface. For completeness, Table 2-1 includes definitions and values of the air pressure and gravity equations variables.

This paper adapts the [Lente and Ősz 2020] approach but uses the station centripetal gravity instead of Earth's gravity. O'Neill used the same approach and produced the same equation [O'Neill 1974]. This equation calculates the rotating space station's air pressure at altitudes above its outer rotating surface:

$$P_h = P_0 \exp\left(-\frac{g_0 \rho_0}{2RP_0}(R^2 - r^2)\right)$$

| Table 2-1 – Air Pressure Equation Variables ||||
|---|---|---|---|
| **Variable** | **Value** | **Units** | **Description** |
| p or ρ | ρ₀=1.225 | kg/m3 | Air Density |
| P | P₀=101325 | Pascals | Air Pressure |
| g | g₀=9.80665 | m/s2 | Gravity Constant |
| k | 8.3144598 | J/(K·mol) | Gas Constant |
| T | T₀=288.15 | Kelvin | Temperature |
| M | 0.0289644 | kg/mol | Air molar mass |
| M/(kT) | 1.20896E-05 | kg/J | Equation Value |
| h | h₀=sea level | Meters | Height above surface |

Where $P_0$ is the air pressure and $\rho_0$ is the air density at the outer radius. R is the outer radius of the rotating station, and r is the radius of interest in the station.

The height above the outer station surface is h = R-r. Our analysis assumes an ideal gas, uses the barometric formula $\rho_0/P_0 = M/(kT)$, and assumes a constant air temperature. Those assumptions allow one to freely exchange the station equation's density and air pressure variables. Algebraic manipulation with these assumptions finds the equation:

$$P_h = P_0 \exp\left(-\frac{Mg_0}{kT}\left(h - \frac{h^2}{2R}\right)\right)$$

Where R is the outer radius of a cylinder (or ellipsoid), and h is the height above the outer rim of a cylinder. The equation can be adapted for the torus (and dumbbells):

$$P_h = P_0 \exp\left(-\frac{Mg_0}{kT}\left(h - \frac{h^2}{2(R+r)}\right)\right)$$

Where R is the major radius of the torus, r is the minor radius of the torus, and h is the height above the outer rim at R+r.

This analysis first investigates the air pressure at the outer rim, where it will be sea-level air pressure. The air pressure $P_0$ is set to one atmosphere at the outer rim of our stations. Setting h=0 in the equation, the air pressure $P_{h=0}$ is equal to $P_0$ as expected.

For a comparison, consider the air pressure at the center or ceiling of the space station. The center of the cylinder or ellipsoid station is at the height h=R. The air pressure at that height would be:

$$P_{h=R} = P_0 \exp\left(-\frac{Mg_0}{kT}\left(\frac{R}{2}\right)\right)$$

With a large radius R, the air pressure at the center approaches zero. In a torus or dumbbell, the air pressure also decreases from the outer rim to the ceiling. The outer rim of the torus is at R+r (the major radius R plus the minor radius r). The height of the torus ceiling would be twice the minor axis, 2r. The air pressure at the ceiling would be:

$$P_{h=2r} = P_0 \exp\left(-\frac{Mg_0}{kT}\left(\frac{2rR}{R+r}\right)\right)$$

The Denver altitude is our air pressure limit. It is 1609 meters above sea level and has an air pressure of 83,728 Pascals. Using these values in the equations, the center of a cylinder station with a radius of 3212 meters would have the Denver air pressure. The ceiling of a torus station with a major radius of 9237 meters would have the Denver air pressure. Smaller stations would have higher air pressures at the center or ceiling.

Figure 2-3 compares the air pressure for the Earth and a set of rotating cylinder stations. The y-axis shows the air pressure, and the x-axis shows the height above the outer cylinder surface. The chart includes data for the Earth and 5 example cylinder radii. It also includes 6 example locations for reference. All stations have sea-level air pressure at the outer rim by design. The air pressure decreases with height, just like the air pressure on Earth. The data in the chart shows that the air pressure does not decrease below the Denver air pressure for the cylinder stations with R=1000 and R=2000 meters. The chart shows the lowest air pressure for the 1000, 2000, and 5000-meter radius cylinder stations. The lowest air pressure at the center of the 15000-meter radius cylinder is 38053 Pascals, and for the 30000-meter radius cylinder, it is 14291 Pascals. The air pressure in larger cylinder stations decreases with height but not as low of pressure as on the Earth.

## 2.3  Radiation Limitation

Radiation in space comes from galactic cosmic rays (GCR) and solar particle events (SPE) [Dunbar 2019]. Solar flares produce the SPE, x-rays, gamma rays, protons, and electrons. GCRs are relativistic-speed charged particles such as protons and heavier ions. Researchers speculate their source could be supernovae, their remnant superbubbles, and the galactic center.

Earth's magnetic field and thick atmosphere protect people on the surface from this radiation. This protection is missing in outer space. Radiation primarily damages the DNA in cells [Blanchett and Abadie 2018]. Long-term effects from this radiation include cancer, sterility, cataracts, cardiovascular damage, cognitive impairment, and memory deficits. Dosages of radiation are measured in rems or sieverts. An OSHA standard recommends less than 5 rems per year [OSHA 1996]. For context, background radiation on Earth ranges from 0.2 to 0.6 rems per year. The unshielded radiation in deep space is 39 to 88 rems per year [Mewaldt et al. 2005].

Various studies suggest using a regolith layer to protect from cosmic rays. The depth of the regolith ranges from 2.5 meters [O'Neill 2008] to more than 8 meters [Turner and Kunkel 2017] to reduce the radiation to an Earth background level. Researchers analyzed this radiation interacting with the lunar surface [De Angelis et al. 2002]. They found 6 meters of regolith would eliminate the radiation effects; see Figure 2-4. Asteroid regolith is likely to be as protective as the lunar regolith.

Our designs fill the outer shells of the station with regolith. The truss framework of the shell provides most of the structural integrity. Ten-meter walls would be sufficient to provide radiation protection. We prefer solutions with greater thickness on the outer walls. The extra thickness provides additional integrity from potential debris, small meteoroids, and ship collisions. Our designs typically use a shell thickness of 20 meters in our analysis.

## 2.4  Rotational Stability

This subsection begins with the background to introduce rotational stability. It then describes spin-stabilized systems and their relevance to space stations. It then covers our passive rotational stability analysis approach and alternative approaches to stabilize the rotating space station. Our separate paper on Space Station Stability [Jensen 2024s] includes much more detail and analysis on these topics.

Newton's first law of motion is a body in motion tends to remain in motion. The space station will rotate about a central axis to produce centrifugal force as a pseudo-gravity. In the vacuum of space, once the space station is spinning, it will tend to remain spinning. Other forces on a space station could potentially disrupt the spinning stability. Examples of

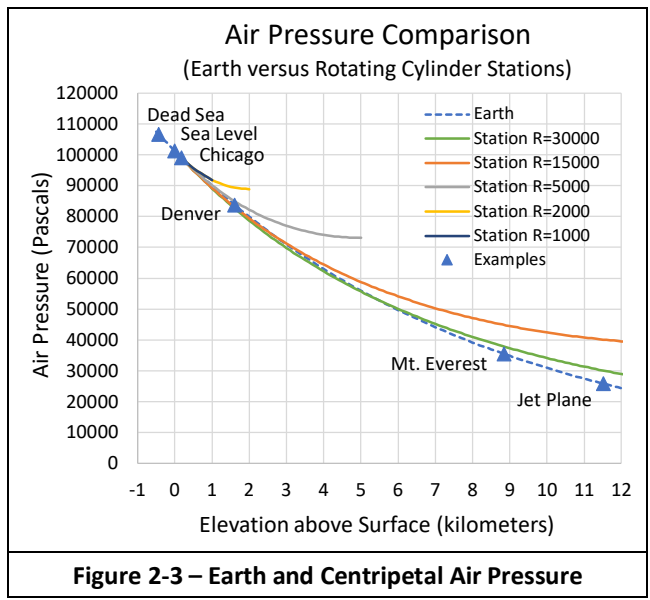

**Figure 2-3 – Earth and Centripetal Air Pressure**

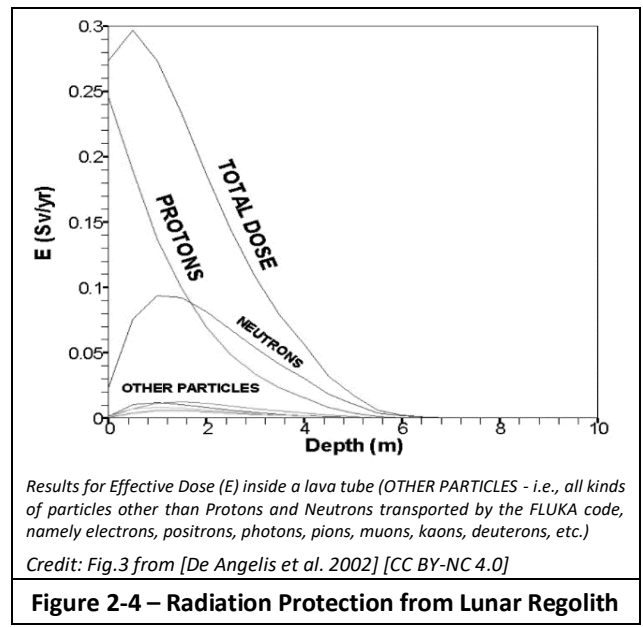

*Results for Effective Dose (E) inside a lava tube (OTHER PARTICLES - i.e., all kinds of particles other than Protons and Neutrons transported by the FLUKA code, namely electrons, positrons, photons, pions, muons, kaons, deuterons, etc.)*

Credit: Fig.3 from [De Angelis et al. 2002] [CC BY-NC 4.0]

**Figure 2-4 – Radiation Protection from Lunar Regolith**

these forces include crafts arriving and leaving the station, objects moving internally, solar wind, and micrometeors.

The stability criterion was recognized and addressed in a recent space station design [Globus et al. 2007]. The team described the Kalpana One Orbital Station: *"In an ideal space environment, any cylinder rotating about its longitudinal axis will continue to do so forever; but in the real space environment perturbations cause rotating systems to eventually rotate about the axis with the greatest angular moment of inertia. If that axis is not along the cylinder length, this introduces a catastrophic failure mode where the settlement gradually changes its rotational axis until it is tumbling end-over-end."* [Globus et al. 2007]

Obviously, we want to prevent a catastrophic instability resulting in a settlement tumbling end-over-end.

### 2.4.1 Spin Stabilized Systems

There is a risk for rotation instabilities with asymmetric geometries. This instability has been named after the Soviet cosmonaut Vladimir Dzhanibekov, who discovered this effect on the MIR space station in 1985 [Kawano et al. 2022]. The cosmonaut spun a T-shaped handle and discovered bi-stable rotation states. A Berkeley website [Kawano et al. 2022] analyzes this rotational behavior and has links to a NASA T-handle video. The abrupt changes in orientation between the two states would not be desirable in a space station.

Even symmetrical designs can be rotationally unstable. The geometry of the space station must be designed properly to avoid this behavior. In the *Elements of Spacecraft Design*, Charles Brown recommends that the desired axis of rotation should have an angular moment of inertia (MOI) at least 1.2 times greater than any other axis to provide rotational stability [Brown 2002]. This constraint is important in developing the station geometry dimensions.

A recent paper investigated the rotational instability of long rotating cylinder space stations [Globus et al. 2007]. That study adjusted the cylinder length to control the rotational stability passively. We extend their rotational stability analysis to our space station geometries. Our paper includes summaries in *§3.7 Single Floor Rotational Stability* and *§3.8 Multiple Component Rotational Stability*. More details and analysis are available in [Jensen 2024s].

Historic station designs typically ignore such stability rules. The Kalpana space station researchers noted that the Bernal Sphere and the O'Neill Cylinder would both be unstable because their desired axis of rotation was not 1.2 times greater than the other axes [Globus et al. 2007]. They applied this guideline to design the Kalpana cylinder design space station. This design constraint provides the spinning station passive control and stability. This stability criterion dramatically changes the original long-and-narrow O'Neill Cylinders [O'Neill 1976] to the short-and-squatty Kalpana One Station [Globus et al. 2007].

### 2.4.2 Stability Metric Analysis

The rotational MOIs around the three station axes are needed to compute the station stability. Generating these moments of inertia uses the MOI equations for various geometry shapes expected in the space station geometries. Section *3.5 Station Components* describes the components of the four geometry stations. The MOI of the station is the sum of the MOIs for all of the component geometries in the station. A table of MOI equations for the various geometry components has been developed and collected; see [Jensen 2024s]. The table includes equations for solid, thin-shell, and thick-shell geometries. This paper uses the same inertia equations.

It is usually possible to create a closed-form equation for the station stability for a single floor geometry with the thin shell (or thick shell) equations. This study uses multiple MOI equations to compute the stability of the geometries using various radius and floor designs. Unlike the single-floor designs, it is tedious or impossible to create closed-form equations for stability using multiple geometry equations. In that case, some of the geometry shapes are further subdivided into many small pieces with well-defined masses and MOIs. Those masses and moments of inertia are summed for the original shape. Often, thin disks rotating about the z-axis are used to achieve this evaluation. This would be a numerical Riemann sum of small parts of the shell. The Newton Raphson numeric algorithm method (goal seek in Excel) adjusts the station geometry dimensions to solve the station stability. This uses the equations of the many components to produce a rotationally stable station.

### 2.4.3 Rotational Stability Alternatives

There are multiple ways to maintain the rotational stability. Table 2-2 shows the advantages and disadvantages of three rotation stabilization techniques. The following paragraphs briefly discuss Passive Spin Stabilization and Active Three-Axis Stabilization.

As the name implies, passive spin stabilization uses spin to stabilize the spacecraft. Spin stabilization is a proven technique originally developed in 1844 by William Hale [Benson 2021]. He patented a method where escaping exhaust struck vanes at the bottom of the rocket and caused the rocket to spin. Spinning creates a gyroscopic action that maintains the rocket's and station's orientation. The rotating space station has angular momentum. The law of conservation of angular momentum implies that the station's spin will remain constant unless acted upon by an external force. A change in the axis of rotation would require an external force. This spinning provides a low-energy approach to keep the spacecraft stable. As explained by NASA in the Basics of Space Flight: *"The gyroscopic action of the rotating spacecraft mass is the stabilizing mechanism. Propulsion system thrusters are fired only occasionally to make desired changes in spin rate, or in the spin-stabilized attitude."* [NASA 2021].

| Table 2-2 – Rotation Stabilization Techniques | | | |
|---|---|---|---|
| | **Spin Stabilization** | **Three Axis Stabilization - Thrusters** | **Three Axis Stabilization – Momentum Wheels** |
| Advantages | Simplicity and reliability. Provides sweeping motion for sensors and instruments. Reduces fuel budget | Flexibility and maneuverability. Low weight. Precise attitude control | Flexibility and maneuverability. Low weight. Reduces fuel budget. Precise attitude control |
| Disadvantages | Limited control of orientation. Increases fuel cost and weight. Complicates pointing of antennas and cameras | Complex implementation and maintenance. Only able to rotate a spacecraft around its center of mass. Increases fuel use and budget | Only able to rotate a spacecraft around its center of mass. Flywheels have a limited mechanical lifetime. Requires periodic corrective maneuvers |

An active three-axis stabilization is an alternative and contrasts the passive spin stabilization. In this active system, the spacecraft's attitude (travel direction) will be continually nudged to maintain a desired orientation. The spacecraft does not need to rotate; as such, cameras and antennas can remain stationary and stay on target. The three-axis stabilization can use small thrusters to maintain the attitude. This requires fuel and increases the weight and cost of the craft. An alternative is to use electrically powered reaction wheels to provide three-axis stabilization. These reaction-wheels or momentum-wheels contain flywheels. Increasing or decreasing the flywheel speed changes the angular momentum of the spacecraft system. This change in the angular momentum causes the spacecraft to counter-rotate to the spin of the reaction wheel [NASA 1997].

Space stations rotate to provide centripetal gravity to the residents. This rotation provides spin stabilization; however, this stabilization will not be perfect. The mass distribution will not be uniform about the rotation axis. Shuttles, materials, equipment, and people will be in motion in the station and further destabilize the station. These movements could cause the equivalent of earthquakes on the station [Globus et al. 2007]. Our large space stations will have large angular momentum, and most movements will have little impact on the stability. Even so, the design will use computers and accelerometer sensors to provide active control and counteract such instabilities. Authors have suggested active control techniques using motors to move large weights on cables [Globus et al. 2007] or using pumps to move water [Lipsett 2005].

## 2.5 Population Limits

The station geometry determines the population. Conversely, the desired size and population determine the geometry selection. This section considers three aspects of the station population. First is the target population for the design. Second is the introduction of multiple floors in the design. Third is the required allocation of floor space.

*2.5.1 Target Population*

Designers define a target population size for their space stations and colonies. This subsection reviews three ranges of population sizes and then explores the station's purpose.

Small spacecraft have been launched for early research. **These spacecraft typically support a crew of less than 20 people.** Examples of this include the Salyut and Skylab spacecrafts with 3 person crews. The International Space Station (ISS) supports a 7-person crew. The Haughton-Mars Project Research station in northern Canada is viewed as a planetary analog of the Moon and Mars. It houses a population of 4 to 13 [Mars Institute 2022].

Multiple programs exist to create lunar bases (China, Russia, ESA, and USA). These bases typically house a few hundred people. Space X aims to send 100 persons at a time to the red planet [Salotti 2022]. Even the first group could be self-sufficient. **Self-sufficient colonies would require a population of about 100 people.** Professor Jean-Marc Salotti defined a quantitative approach to determine the minimum self-sufficient population on another planet or space colony [Salotti 2022]. Salotti defined five required station activities: ecosystem management, energy production, industry, building, and social activities. He notes, "*A greater number of individuals makes it possible to be more efficient through specialization and to implement other industries allowing the use of more efficient tools.*" [Salotti 2022]. His mathematical model found the minimum number needed would be 110 individuals [Pallister 2020].

Rotating space stations with a radius of 200 meters or greater can provide Earth-like gravity. With a radius of 250 meters, the Kalpana space station was planned with a population target of 3,000 residents. With a radius of 830 meters, the Stanford Torus was planned with a population of 10,000. **Larger space stations can support 100s of thousands of people.**

Another aspect of planning the target population is the intended purpose of the station. Early research spacecraft only supported small crews. **University**: Future stations supporting entire universities of researchers, staff, and support could require thousands of people. A state university could have 30,000 students and 2,000 staff. The surrounding city to support that university would have another 30,000 people. **Industry**: Even larger stations with more people could support industry. A typical semiconductor plant employs 2,000 people and might be representative of a space station industry. The surrounding town around that plant would have a population of 40,000. A space colony constructed from an asteroid or lunar material will have a wealth of materials to spawn multiple industries. These industries could produce solar cells, fuel, agriculture, metal, and manufactured goods. **Tourism:** It is possible that supporting tourism might require even larger populations. Walt Disney World is one of the world's largest tourist attractions. It employs 77,000 people and covers 11 million square meters. The two O'Neill Model 3 Cylinders [O'Neill 1974] have over 126 million square meters of floor space and could easily support such a facility. Al

Globus predicted that tourism to a large space station might match international tourism to a city such as Washington DC [Globus 2006]. Over 2 million international tourists visit Washington DC each year, and most of them are likely to fly there [DC 2017]. The number of rocket launches could reach tens of thousands per year - a rate comparable to the early decades of aviation [Globus 2006]. The tourism employees, tourists, and support industry personnel could easily reach several 100 thousand people.

With these later population targets, large stations become necessary. Those stations could be supporting research, industry, and tourism. It would be a complete city with all the associated support personnel. This review suggests a broad range of population targets and a need for large space stations. These larger stations would enable civilization to grow and become sustainable beyond our planet Earth. This begs the question of why humanity would want to become an interplanetary species. *"The answer is complex and multifaceted, but here's the short version: to ensure the long-term continuation of our species and our earthly evolutionary branch."* [Bates 2017].

*2.5.2 Multiple Floors*

Historically, most space station designs use only a single floor on the outer perimeter of the rotating station for living space. Adding multiple floors to a space station greatly increases the available floor space. Figure 1-1 includes a cutaway rendering to help visualize these floors. This torus station was described in detail in [Jensen 2023]. This torus station has a major radius of 2300 meters. The torus has an elliptic cross-section with minor radii of 400 and 1150 meters. The ceiling in the torus is 500 meters above the top floor. This station provides an open vista like a long valley of 6 kilometers over the curved top floor. The top floor in this station would have 33.6 million square meters of floor space. The 18 floors in such a station would have 487 million square meters of floor space. Using 155.2 square meters per person [Johnson and Holbrow 1977], the floors of this station could support over 3 million people [Jensen 2023]. Figure 2-5 includes cross-section diagrams of a similar torus. The inner torus on the left has a vista of 635 meters and a ceiling height of 45 meters. The outer torus on the right has a vista of 2470 meters and a ceiling height of 500 meters.

*2.5.3 Floor Space Allocation*

Everyone in a space station needs room to support their living, working, industry, and agriculture needs. As an example, a 1970s NASA design study created a set of surface usage metrics [Johnson and Holbrow 1977]. These included 49 square meters per person for residential, 10 for open spaces, 12 for transportation, and 61 square meters per person for agriculture. This totals 155.2 square meters per person. NASA used multiple-story structures of various heights for the different space categories (an average height of 11.2 meters). Using these multiple stories results in 67.0 square meters of projected surface area per person.

Figure 2-6 reuses the chart and supporting text from [Jensen 2023]. It uses open areas, support, agriculture, industry, and residential categories. This chart also includes our estimates for different floor spacings. We modernized requirements based on changes from the 1970s to the 2020s. Our study also uses multiple-story structures to support the usage requirements; however, they are constrained between station floors. The vertical distance between floors varies in our analysis. The chart includes floor spacings of 5, 10, and 15 meters. With a fixed floor spacing of 15 meters, the updated space allocation provides each individual 65.5 square meters per person, which is close to the 1970s NASA metric. Figure 2-6 shows different usage metrics for the categories and floor heights. The most efficient volume and area usage in a station with many floors is with the 5-meter floor spacing.

Even with the different floor spacings, many floor usage categories use the same volume; as such, their space requirement scales with the changing floor distance. As an example, the agriculture allocation increases from 13 to 39 square meters per person as the floor height distance decreases from 15

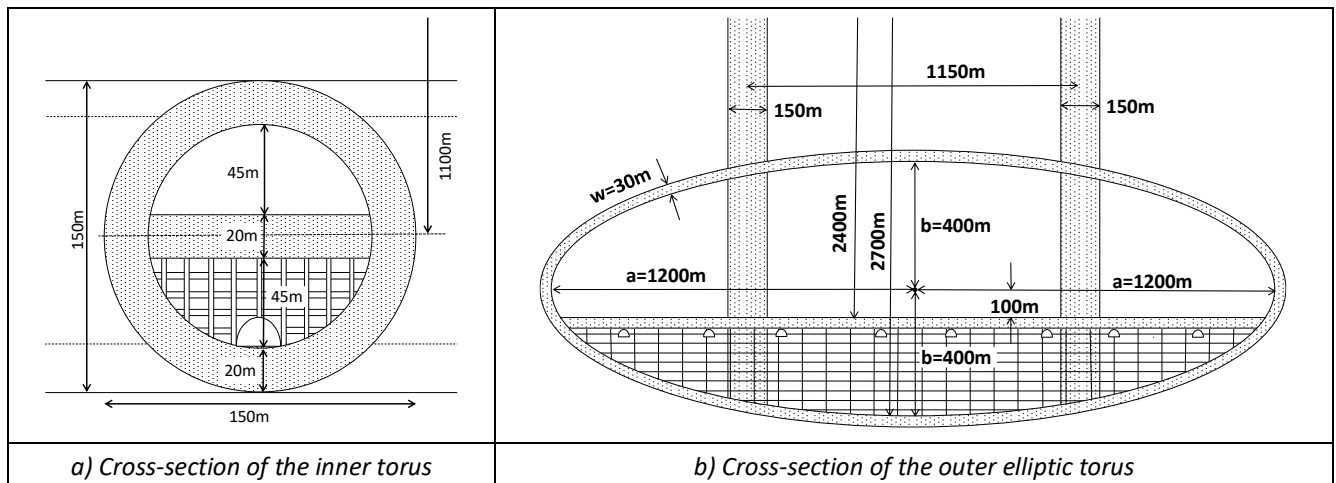

| a) Cross-section of the inner torus | b) Cross-section of the outer elliptic torus |

Credit: Figure 3 13 from [Jensen 2023] [CC BY-SA 4.0]

**Figure 2-5 – Torus Cross Sections – Multiple Floors and Elliptical Design**

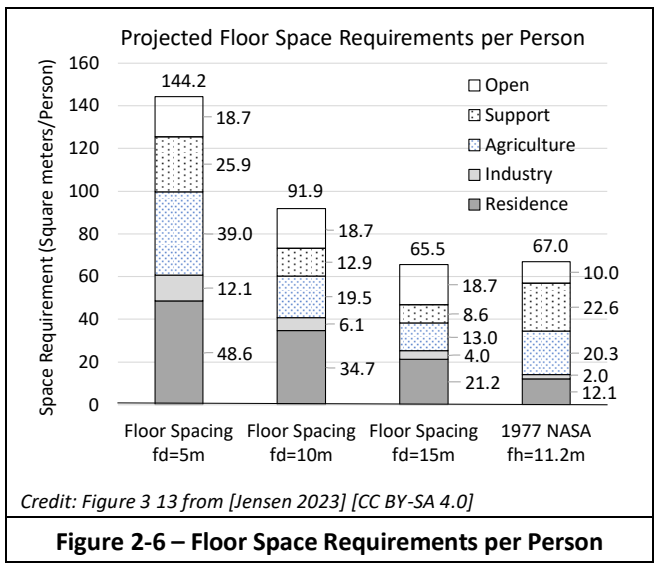

*Credit: Figure 3 13 from [Jensen 2023] [CC BY-SA 4.0]*

**Figure 2-6 – Floor Space Requirements per Person**

to 5 meters. Stacks of plants would fill the floors. The total agriculture volume remains constant at 195 cubic meters per person.

The open space in the NASA study is increased from 10 square meters per person to 18.7 square meters per person. This includes some of the public areas in the support category and adds an extra 5 square meters per person. In large stations, the top floor can often meet the residents' open space requirements. There is also an opportunity to be creative and use low and high gravity regions to create additional open space areas.

## 2.6 Psychological Limits

Living in a space station could cause psychological issues with the inhabitants. NASA studies have considered the oppressive closed-quarters ambience of a space station to be a risk to the colonists' psychological well-being [Johnson and Holbrow 1977] [Keeter 2020]. There is potential for feelings of isolation and confinement. Those feelings can cause reactions such as impaired intellectual functioning, motivational decline, somatic complaints, psychological changes, and social tensions [Connors et al. 1985]. These are typically mild symptoms and not severe psychiatric disturbances [Connors et al. 1985]. This section considers multiple station characteristics to address the psychological limits of a rotating space station. These characteristics include the multiple floors, aesthetics, sunlight, and qualitative assessment.

### 2.6.1 Multiple Floors

Multiple floors in rotating space stations were detailed in [Jensen 2023]. Living space on multiple floors is fairly common on the Earth. Limited space and costs in urban environments promote high-rise living. Adding floors to a structure greatly increases the available floor space. Multiple-floor structures include skyscrapers, cruise ships, submarines, and underground cities. Historic and modern underground cities exist [Garrett 2019]. Entrepreneurs have begun to convert abandoned military missile silos into multiple-floor homes and underground cities [Garrett 2019].

There are issues with creating and using many floors. Studies have found that people living in highrises on the Earth suffer from issues such as more significant mental health problems, higher fear of crime, fewer positive social interactions, and more difficulty with child-rearing [Barr 2018]. The biggest issues for underground cities are not technical but social [Garrett 2019]. NASA studies have considered the psychological risks from the potential oppressive closed-quarters ambience [Johnson and Holbrow 1977] [Keeter 2020]. Fortunately, researchers offer approaches to address these issues and risks using proper planning and space allocation. Recent literature and the lessons learned offer knowledge and experience to address such issues. These lessons support the idea that using many floors is acceptable for space station habitation.

### 2.6.2 Open Vistas Aesthetics

The NASA SP-413 study [Johnson and Holbrow 1977] used a population density metric of 155.2 square meters for each individual. Our asteroid restructuring study [Jensen 2023] used a metric of 547 square meters per individual in a large station with a population of 700,000. Colonists will enjoy wide-open vistas on the main floor. Lower levels could include concourses, atriums, and other large spaces. These additional spaces and vistas should improve the population's psychological well-being.

In a torus, this wide-open vista is available by looking down the tube in the rotation direction. In our designs, the top floor is typically at the center of the torus tube. This vista in the rotation direction can be computed. With a major radius of R and a minor radius of r, the vista would be:

$$\text{Torus Vista} = 2\sqrt{R^2 - (R-r)^2}.$$

This can be a spectacular vista. Unfortunately, with larger radius multiple-floor habitats, much of the square footage is on the lower levels. With large space stations, many of the lowest floors would not be desirable for residential use (only Morlocks would live far underground [Wells 1898]). This affects the station geometry. To reduce the number of floors and maintain large surface areas, our designs evolved the cross-section of several station geometries from circles to ellipses. This reduced the number of floors, increased the surface area, and improved the vista on the main floor.

Ellipsoids, hatbox cylinders, and dumbbells all have similar vista aesthetics. Their best interior vista is the open space overhead. The top floor typically has minimum gravity at height h above the outer rim $R_o$. For the cylinder and ellipsoid, the vista can be computed as:

$$\text{Cylinder and Ellipsoid Vista} = 2(R_o - h).$$

For the dumbbell, the top floor is at the major radius R. The vista would be the radial axis length of the dumbbell node. For a spherical node, this vista would be its radius, r.

For the same population (or construction mass), the torus typically has lower ceilings but longer valley vistas than the other geometries. The cylinder and ellipse have the next best vista, and the dumbbell typically has the smallest vista.

NASA detailed a torus space station design in the 1970s. The station's minor radius was 65 meters, and its major radius was 830 meters. The ceiling was 130 meters above the floor, and a person would have a vista of 892 meters. Figure 2-7 shows this torus interior vista is like a long, narrow valley whose ends curve upwards. Figure 2-8 shows our larger torus example with even a better vista on a valley that is 3.0 kilometers wide, curves upward for 3.0 kilometers, and has a ceiling 500 meters above.

Ideally, the upper-half, low-gravity regions of the space station serve as open space. This open space provides good vistas and beneficial aesthetics for the residents and visitors. The top floor provides the open recreation regions for psychological well-being. Below the top floor, facilities such as courtyards, aquariums, and botanical gardens could create more open areas to provide psychological benefits.

### 2.6.3 Sunlight

Many space station designers insist on natural sunlight in the space station. This provides psychological benefits, aesthetic value, and low-cost illumination. O'Neill wanted his space communities to have normal gravity, normal day and night cycle, natural sunlight, an earthlike appearance, and efficiently use solar power and materials [O'Neill 1974]. He envisioned an outer shell that was one-third to one-half windows; see Figure 2-7. Supporting natural sunlight in the station often adds design complexity by directing the sunlight through windows with a set of mirrors. More recently, designers have not relied on natural sunlight. Solar cells and lights have become much more efficient since the 1970s. For example, LEDs have improved lighting efficiency from 50 lumens per watt in the 1970s [Bock, Lambrou, and Simon 1979] to research efficiencies of 200 lumens per watt. Solar cells are still about 15% efficient; however, their cost has dropped from $20 per watt to 20 cents per watt. The station design becomes much simpler without the complex mirrors and windows.

### 2.6.4 Qualitative Assessment

Significant planning went into the interior description of the Stanford Torus and the O'Neill Cylinders [Johnson and Holbrow 1977] [O'Neill et al. 1979]. Other design criteria from the Stanford study [Johnson and Holbrow 1977] were considered to further evaluate the potential of such a station. The NASA SP-413 study explored physiological, environmental design, and organizational criteria for space stations. They felt a successful space habitat for colonization must meet these qualitative criteria. The large elliptic torus was evaluated with this same qualitative evaluation in [Jensen 2023]. The station has the potential to meet or exceed the design criteria of the Stanford Torus.

Figure 2-8 shows two views inside the large outer torus. A picture is worth a thousand words, and these inside views address qualitative criteria of the environmental design. The left picture is taken near one of eight dividers and includes a two-story house for scale. The divider is an airtight vertical wall thick enough to contain spokes in the design. This wall separates the station into multiple airtight compartments and provides a failsafe feature in case of a catastrophic impact to the torus. In this image, the floors on the divider are spaced 15 meters apart. Figure 2-8 also includes a wide field of view of one of the eight dividers in the station. This view is 3 kilometers wide, and each of the four entryways into the divider is as large as the Arc de Triomphe in Paris. The rendering

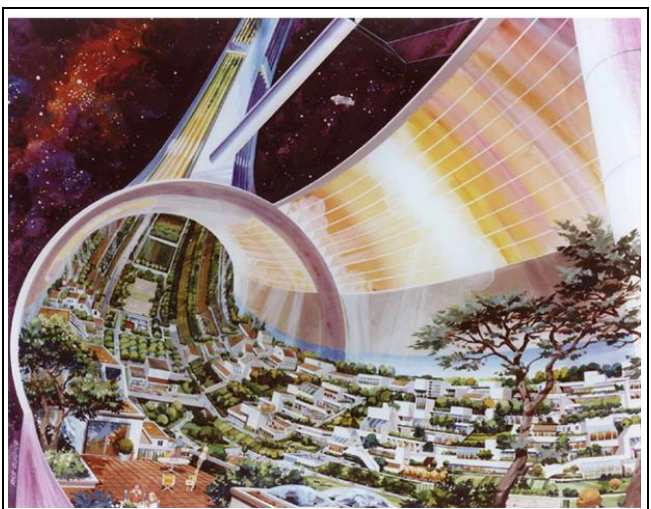

*Credit: NASA Ames Research Center. Artist: Rick Guidice; [NASA Image Public Domain]*

**Figure 2-7 – Open Vista in Torus Space Station**

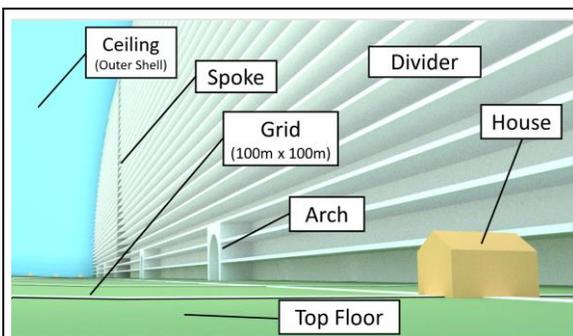
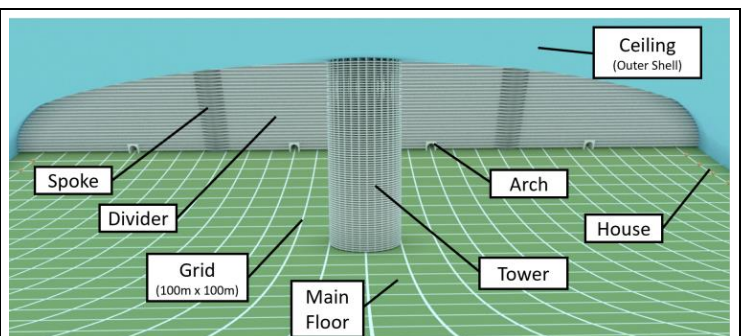

*Credit: Adapted from Figure 5-3 from [Jensen 2023] [CC BY-SA 4.0]* | *Credit: Self-produced using Blender*

**Figure 2-8 – Qualitative Inside Views of Large Torus Station**

tool paints a 100-meter square grid on the main floor – about the size of a city block. Two of the spokes are within the divider. The divider includes 33 floors, each 15 meters in height. Along the side are several of the many two-story homes rendered to help with scale. The scene has a 77-story tower built into one of the spokes. This scene has a fabulous view with a curved floor vista of over 2.6 kilometers and a vertical height of 500 meters.

## 2.7 Human Limits Summary

Human frailty imposes many limits on the design of a space station. This section covered the constraints from gravity, air pressure, radiation, rotational stability, population, and psychological. Our human bodies have evolved in a very narrow range of gravity. Our rotating stations are designed to create centripetal gravity between 0.95g and 1.05g in the inhabited regions. Rotation is kept at less than 1 rpm to prevent motion sickness. This implies a station radius of greater than 900 meters. The station will need an atmosphere in the vacuum of space. Air pressure is kept between a maximum of sea level to a minimum similar to Denver's air pressure. This implies a maximum altitude of about 1600 meters for the top floor inside the station. There is sufficient radiation in space to cause sickness. The radiation comes from galactic cosmic rays (GCR) and from solar particle events (SPE) [Dunbar 2019]. Earth's magnetic field and thick atmosphere protect people on the surface from this radiation. A thick shell of regolith protects our station's residents. There is a risk for rotational instabilities in the space stations. Feedback from such instabilities could cause an abrupt change in the rotating space station orientation. Such a change could be catastrophic for the station and the colonists. The station is designed to provide passive rotational balance.

This section reviewed aspects of the station's human population. The design must consider the station's purpose and the target population. The allocation of floor space to categories such as living, agriculture, industry, and openness creates limits for the population. Multiple floors provide more floor space, more open space, and lower population densities. There are perceived limitations on the station's population in an isolated space station. Psychological issues must be considered. Lower population densities, long vistas, and open spaces should provide pleasing visual aesthetics.

## 3 Station Design Limitations

The previous section introduced various constraints and their effect on humans. Where possible, that section provided background equations to define those constraints. The purpose of this section is to apply those constraints to the station. This will determine physical limits on the construction of large space stations. The design must consider limits from gravity, air pressure, and materials. These limit the height of the top floor and the station population. This section reviews station components and geometry limits for the station design. It also overviews rotational stability concepts and applies them to single-floor and multiple-floor stations.

### 3.1 Habitable Gravity

Gravity plays an important role in regions where humans will live long-term. The previous section introduced centripetal gravity and its benefits and detriments. This section covers gravity in an example large station, defines the range of acceptable gravity, and describes the effects on the station design.

*3.1.1 Example Station Gravity*

Figure 3-1a shows a rendering of a large elliptical torus space station. This paper reuses this example and text from [Jensen 2023] for context and continuity. The station has a large outer torus and a smaller inner torus. For this analysis, the outer torus's major radius and top floor are 2400 meters from the station's center. The major radius of the inner torus is 1100 meters from the center of the station. Rotating the station once every 1.6 minutes produces a sensation of Earth-like gravity on the main floor of the outer elliptic torus. Figure

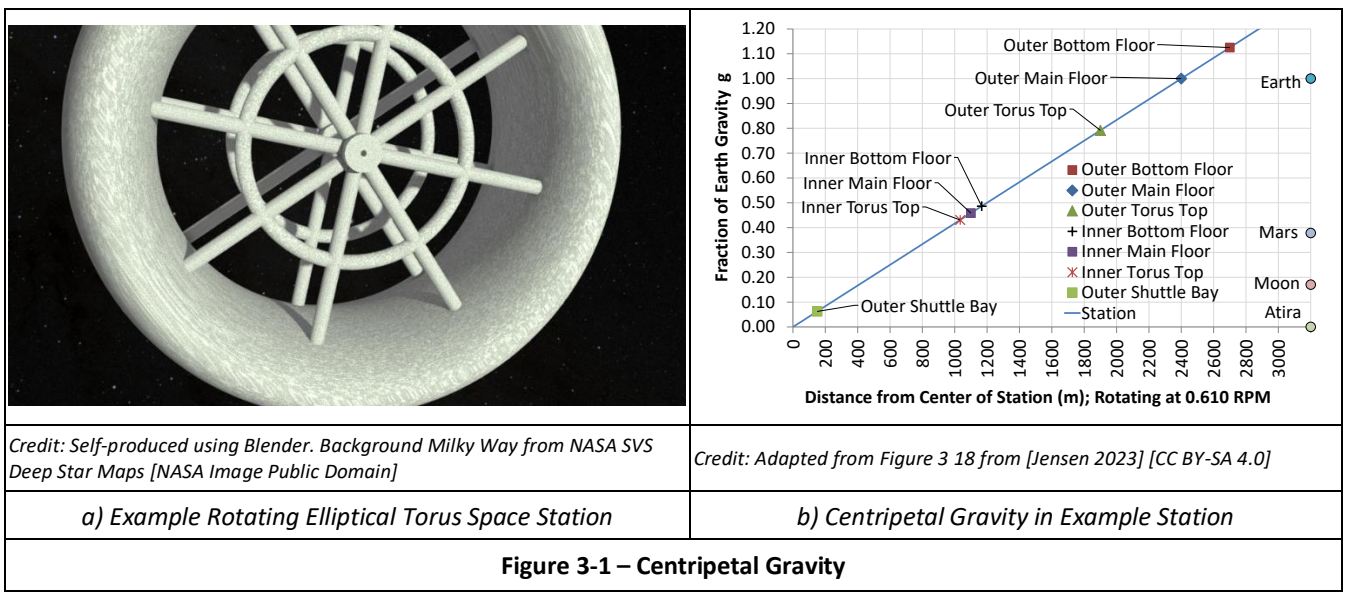

| Credit: Self-produced using Blender. Background Milky Way from NASA SVS Deep Star Maps [NASA Image Public Domain] | Credit: Adapted from Figure 3 18 from [Jensen 2023] [CC BY-SA 4.0] |
|---|---|
| *a) Example Rotating Elliptical Torus Space Station* | *b) Centripetal Gravity in Example Station* |

**Figure 3-1 – Centripetal Gravity**

3-1b shows the centripetal gravity in this rotating space station. For reference, the gravities of the Earth, Moon, Mars, and the asteroid Atira are shown on the right side of the chart. The bottom floor of the outer torus has the highest gravity of 1.12g. In the outer torus, the top floor of the divider and the tower have a gravity of 0.79g. The main floor of this design has a gravity of 1.0g and only increases to 1.05g on the 20[th] floor below the main floor. Most of the 700,000 residents will live and work in this region between 1.0g and 1.05g. Figure 3-1 also shows the inner torus will provide about half of Earth's gravity. The shuttle bay will provide a gravity even less than the Moon.

*3.1.2 Geometry Gravity Ranges*

The large torus station in Figure 3-1a has a gravity range of 1.0g to 1.12g on its floors. Stations can be designed with different radii and rotation rates to vary the gravity values over the station floors. Historically, organizations have used different ranges in their designs. A 1977 NASA study used a gravity range of 0.9g to 1.0g [Johnson and Holbrow 1977]. They also considered a relaxed constraint range of 0.7g to 1.0g. A decade later, NASA authors designed a torus station with a gravity range from 0.97g to 1.03g [Queijo et al. 1988]. They felt this range would not significantly influence human physiology and performance.

The previous section covered gravity constraints for humans. This section uses those limits to define the gravity range over the multiple floors in our large stations. In our studies, the multiple-floor stations are usually designed with a gravity range of 0.95g to 1.1g. The population will spend most of their time on floors with Earth-like gravity. This gravity range should minimize associated health risks.

This subsection reviews details on gravity ranges from [Jensen 2023]. That paper defined two classes of rotating stations. One class rotates about an axis outside the enclosed region and includes the torus and dumbbells. The other class rotates about an axis inside the enclosed region and includes the cylinder and ellipsoid. That paper analysis also varied the number of floors in the station volume. That paper included both a half-filled and a fully-filled volume. This subsection fills the enclosed volume halfway. Figure 3-2 illustrates these half-filled volume classes. These show the floor location for the minimum and maximum gravities for the two classes. The center floor is at the radius R. The outer rim is at radius $R_o$, defined as radius R plus the floor height h. The figure includes equations that define the relationships between the gravities, the radius, and the height.

For a torus or dumbbell, the gravity range can be controlled with the rotation radius R and the minor radius a. The minor radius is the floor height h. The analysis introduces a scale variable m that defines R as equal to m times a. The scale m equals the minimum gravity over the difference between the maximum and minimum gravity. Table 3-1 provides a small table of the habitability scale metrics for a range of minimum and maximum gravities. A broader range of gravities supports more floors and greater populations in the station. The table highlights two values: 9.5 and 6.33. These represent a

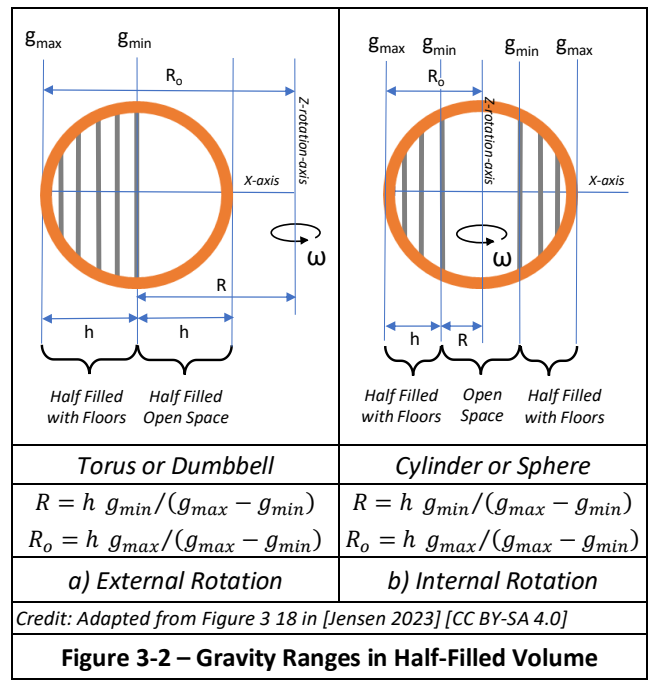

| Torus or Dumbbell | Cylinder or Sphere |
|---|---|
| $R = h\, g_{min}/(g_{max} - g_{min})$ | $R = h\, g_{min}/(g_{max} - g_{min})$ |
| $R_o = h\, g_{max}/(g_{max} - g_{min})$ | $R_o = h\, g_{max}/(g_{max} - g_{min})$ |
| a) External Rotation | b) Internal Rotation |

Credit: Adapted from Figure 3 18 in [Jensen 2023] [CC BY-SA 4.0]

**Figure 3-2 – Gravity Ranges in Half-Filled Volume**

minimum of 0.95g on the center top floor and a maximum of 1.05g or 1.1g on the outer rim. This paper and our earlier research extensively use those values and gravity ranges. A larger table with a broader range is included in [Jensen 2023]. The same scaling factor values work for a cylinder or ellipsoid. The designs can control the gravity range with the rotation radius R and the minor radius a. Using the outer radius Ro to analyze those geometries is often more convenient. The outer radius equations and relationships are in Figure 3-2 and in Table 3-1. The analyses of the dumbbell and torus designs use the minor radius as the floor height h.

## 3.2 Air Pressure Limits

We use a low air pressure limit of 83,728 Pascals, which represents the Denver air pressure at 1609 meters above sea level. Mount Everest, at 8848 meters, has an air pressure of 35,447 Pascals. High air pressures will typically not be a constraint for our design. For example, the Dead Sea is 430 meters below sea level and has an air pressure of 106,624 Pascals. The human body can tolerate a maximum of roughly

| Table 3-1 – Half-Filled Floor Scaling Factor for Various Centripetal Gravity Ranges | | | | |
|---|---|---|---|---|
| Scale (m) | gmin | | | |
| gmax | 0.85g | 0.9g | 0.95g | 1.0g |
| 1.05g | 4.25 | 6.00 | 9.50 | 20.0 |
| 1.1g | 3.40 | 4.50 | 6.33 | 10.0 |
| 1.15g | 2.83 | 3.60 | 4.75 | 6.67 |
| 1.2g | 2.43 | 3.00 | 3.80 | 5.00 |

*Torus (and Dumbbell) Major Radius=R; Minor radius=a; Outer Radius=R+a; R=m×a; m=gmin/(gmax-gmin)*

*Ellipsoid (and Cylinder) Outer Radius=Ro; Top Floor Height=h; Ro = (m+1) h = m̂×h; Top Floor Radius R=Ro-h; m̂ = gmax / (gmax-gmin)*

400,000 Pascals. We use an upper limit of 101,325 Pascals, which represents sea level on Earth. We use the air pressure equations presented in *§2.2.2 Air Pressure Equations*.

*3.2.1 Air Pressure Comparisons*

In the previous section, Figure 2-3 compared pressures of the Earth and multiple station radii. For more detail, Figure 3-3 shows three charts comparing rotating station air pressures to Earth's air pressure. The x-axes show the height above the outer rotating surface. For the Earth data, this measurement is from sea level. The y-axis shows air pressures ranging from 0 to 110,000 Pascals. The charts include dashed lines showing the sea level and Denver air pressure limits.

The charts in Figure 3-3 include the location of the top floor assuming minimum gravity. This location is shown with the vertical dotted line. Using gravity limits, the top floor is set at the 0.95g radius, and the outer radius is at 1.05g. With these cylinder examples, the top floor is at a height of the outer radius over 10.5, producing 333.3 meters, 3,333 meters, and 33,333 meters.

The charts show that with increasing station radii, the air pressure in the rotating stations reduces slower than on Earth. The station air pressure data ends at the center of the cylinder (height equals station radius). Earth's air pressure continues to decrease with additional height. In these examples, the air pressure decreases to less than Denver's air pressure.

The top chart shows a rotating station with a radius of 3500 meters, the middle shows a radius of 35,000 meters, and the bottom shows 350,000 meters. The data of the 3,500-meter radius station shows the air pressure drops below our Denver lower limit at a height of 2242 meters. The air pressure in the 35,000-meter radius station drops below the Denver limit at a height of 1559 meters. The Denver limit would be at 1527 meters in the station with the 350,000-meter radius.

The 3,500-meter radius station has its top floor at 333.3 meters with a habitable air pressure of 97,400 Pascals. This is higher than Denver's air pressure. The 35,000-meter radius station has its top floor at 3,333 meters with an air pressure of 68,250 Pascals. This would be equivalent to an Earth altitude of 3210 meters. Altitude air sickness can occur at 2500 meters on Earth.

The air pressure at any height in small stations will be greater than the Denver air pressure. In large stations, much of the higher altitudes would have low air pressure. The example 350,000-meter radius station would only have air pressures greater than our Denver limit at heights of less than 0.01% of its radius. In very large stations, much of the station volume would be uninhabitable because of low air pressure.

*3.2.2 Top Floor Analysis*

The air pressure equation is used to investigate the top-floor air pressure. An example is to find the station radius and floor height where the top floor would have the Denver air pressure. Consider a cylinder station design with m=10.5 or gmax=1.05g and gmin=0.95g. The top floor would be at a

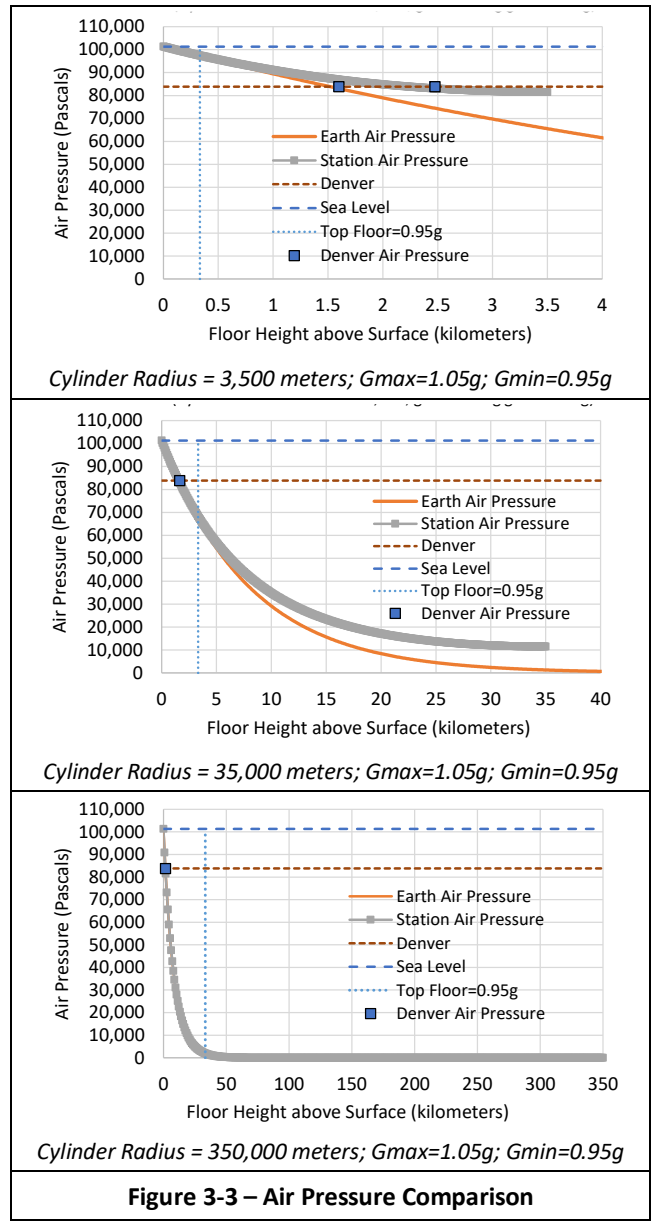

*Cylinder Radius = 3,500 meters; Gmax=1.05g; Gmin=0.95g*

*Cylinder Radius = 35,000 meters; Gmax=1.05g; Gmin=0.95g*

*Cylinder Radius = 350,000 meters; Gmax=1.05g; Gmin=0.95g*

**Figure 3-3 – Air Pressure Comparison**

height of R/10.5 or a radius of 0.9048 R. The pressure at that height would be:

$$P_h = P_0 \, exp\left(-\frac{Mg_0}{kT}\left(h - \frac{h^2}{2R}\right)\right)$$

$$P_{h=R/10.5} = P_0 \, exp\left(-\frac{Mg_0}{kT}\frac{R}{11.025}\right)$$

In this equation, the outer rim gravity $g_0$ is set to 1.05g. The outer rim air pressure $P_0$ is set to sea level, and $P_h$ is set to the Denver pressure of 83728 Pascals. With m=10.5, $Mg_0/(kT)$=1/8033.0. Solving for R:

$$ln\left(\frac{83728}{101325}\right) = -\frac{1}{8033.0}\frac{R}{11.025}$$

0.19076*8033.0 = 1532.37 = R/11.025

R=16894.4 meters

h=R/10.5=1609.0

At the top floor height of 1608.99 meters, the air pressure in the 16894.4-meter radius cylinder station would be the Denver air pressure.

Compare this to the torus station where the outer rim is at R+r, the top floor is at radius R, and the minor radius r is R/9.5 for gravity uniformity. The height h in our equations is the minor radius r. The torus air pressure equation is:

$$P_h = P_0 \, exp\left(-\frac{Mg_0}{kT}\left(h - \frac{h^2}{2(R+r)}\right)\right)$$

Similar to the approach with the cylinder analysis above, the top floor is at the radius R and is height r above the outer rim. The equation becomes:

$$P_h = P_0 \, exp\left(-\frac{Mg_0}{kT}\left(\frac{r(r+2R)}{2(R+r)}\right)\right)$$

$$ln\left(\frac{P_h}{P_0}\right) = ln\left(\frac{83728}{101325}\right) = -\frac{1}{8033.0}\left(\frac{r(r+2R)}{2(R+r)}\right)$$

Again, m=9.5 and R is 9.5r to provide habitable gravity over the torus multiple floors. Solving:

$$1532.37 = \frac{R}{m}\frac{2m+1}{2m+2}$$

1532.37 = R (1/9.5) (20/21) = R / 9.975

R=15,285.5

R=15285.4 meters; r=1609.0 meters; Ro=16894.4 meters

At the torus top floor height of 1609.0 meters, the air pressure in the 15285.4-meter radius torus station would be the Denver air pressure of 83728 pascals. The heights and outer radii match despite the different geometries, air pressure equations, and gravity ratios.

Figure 3-3 included three station sizes and varied the radius and floor height to compute the air pressure. Previous paragraphs provided two top-floor air pressure examples. Figure 3-4 also provides a comparison of air pressures with a range of station radii. Normalizing the height as a percent of the station radius supports the comparison of multiple station radii in a single chart. The x-axis shows the normalized height as a percentage of the total radius. The y-axis shows the rotating station air pressure normalized as a percentage of Earth's air pressure. The legend and data show the difference for four station radii ranging from 350 to 350,000 meters. In all stations, the normalized air pressure is always less than the Earth's air pressure at a given height. The chart data shows air pressures are more similar to Earth in smaller stations. Their air pressure is near Earth's air pressure throughout the station's interior. The air pressure becomes very small at high altitudes in large stations and on Earth. The Earth's air pressure continues to decrease with increasing height, while the station air pressure reaches a small but relatively much larger value at the center of the cylinder station.

### 3.2.3 Air Pressure in Station

The large stations are typically designed to produce a top floor with at least the Denver air pressure. A rough estimate of this height is 1600 meters (or 320 floors). The centripetal air pressure changes at a rate different than on Earth. To be

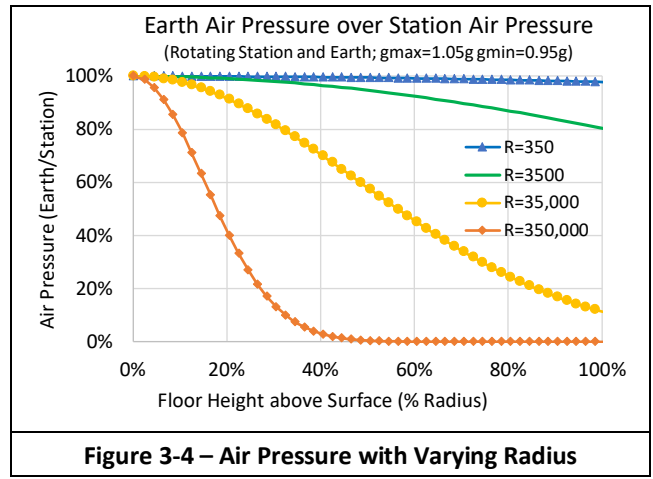

**Figure 3-4 – Air Pressure with Varying Radius**

more precise, consider the cylinder air pressure formula from *§2.2.2 Air Pressure Equations*:

$$P_h = P_0 \, exp\left(-\frac{Mg_0}{kT}\left(h - \frac{h^2}{2R}\right)\right)$$

Rearranging the equation into a quadratic equation supports solving the floor height given the station radius and the Denver air pressure:

$$h^2 - 2Rh - 2R\left[ln\left(\frac{P_h}{P_0}\right)\bigg/\left(\frac{Mg_0}{kT}\right)\right] = 0$$

This solution takes into account centripetal gravity and air density. The gravity $g_0$ is at the outer rim, and we typically assign a value of 1.05g or 1.1g. In a rotating space station, the altitude for this pressure is dependent on the station radius. $P_0$ is set to sea level, and $P_h$ is set to the Denver air pressure. Table 3-2 contains useful constants for this analysis. Using the appropriate constants, solving for the height in this quadratic equation:

$$h^2 - 2R\,h + 3064.75\,R = 0$$

$$h = \left(2R \pm \sqrt{4R^2 - 4*3064.75\,R}\right)/2$$

These equations can evaluate the stations with our gravity and air pressure limits. The following paragraphs consider the air pressures of the top floor and cylinder center.

**Top Floor:** The quadratic pressure equation can find the radius where the top floor would have the Denver Air Pressure. Recall that $P_h$ is the Denver Air Pressure, $P_0$ is the Sea Level Air Pressure, and $g_0$ is the gravity on the outer surface (1.05g). Those constants reduce to 2 $ln(P_h/P_0)/(Mg_0/kT)$, which is equal to -3064.75. A habitable gravity range for the cylinder uses R=10.5h. The top floor would be at 0.95g, and the outer rim would be at 1.05g. In a cylinder, the top floor

| Table 3-2 – Air Pressure Equation Constants | | | | | |
|---|---|---|---|---|---|
| $g_0$ (gmax) | (kT)/ (Mg$_0$) | 2 ln(P$_h$/P$_0$)/ (Mg$_0$/(kT)) | Scale m gmin=0.95g | Cylinder Exp Constant | Torus Exp Constant |
| 1.00g | 8434.66 | -3217.99 | 20 | 20.51 | 19.49 |
| 1.05g | 8033.01 | -3064.75 | 10.5 | 11.025 | 9.975 |
| 1.10g | 7667.87 | -2925.44 | 7.333 | 7.870 | 6.797 |

would be at h=R/10.5. Using R=10.5 h, the quadradic equation becomes:

$$h^2 - 2(10.5\,h)\,h + 3064.75\,(10.5\,h) = 0$$
$$-20\,h^2 + 32179.9\,h = 0$$
$$h = 1609.0 \text{ meters}$$

The radius of the cylinder would be 10.5h:

$$R = 10.5h = 16{,}894.5 \text{ meters}$$

This solution matches the earlier example with the same geometry and uses the air pressure equation directly. This approach adapts to solve other geometries. As an example, for the torus the air pressure equation becomes:

$$P_h = P_0\, exp\left(-\frac{Mg_0}{kT}\left(h - \frac{h^2}{2(R+r)}\right)\right)$$

$$-ln\left(\frac{P_h}{P_0}\right)\bigg/\frac{Mg_0}{kT} = h - \frac{h^2}{2(R+r)}$$

$$h^2 - 2(R+r)\,h - 2(R+r)\left[ln\left(\frac{P_h}{P_0}\right)\bigg/\frac{Mg_0}{kT}\right] = 0$$

This is identical to the cylinder equation by replacing the cylinder radius R with the torus outer rim radius R+r. The gravity limit in the torus sets m=9.5 and R=9.5h to have gmax=1.05g and gmin= 0.95g. The top floor is at the height of h or r. To find h in the torus where the top floor would have an air pressure equivalent to Denver:

$$h^2 - 2(R+r)\,h - (R+r)\,3064.75 = 0$$

This can be solved as a quadratic equation or, in this case, by replacing r with h and R with 9.5h:

$$h^2 - 2(9.5h + h)\,h - 2\,(9.5h + h)\left[ln\left(\frac{P_h}{P_0}\right)\bigg/\frac{Mg_0}{kT}\right] = 0$$

$$h^2 - 21\,h^2 + 3064.75\,(10.5\,h) = 0$$
$$-20\,h^2 + 32179.9\,h = 0$$

The equation is identical to the result with the cylinder. The solution is again h=1609.0 meters and is also the minor radius r of the torus. The major radius of the torus would be:

$$R = 9.5h = 15{,}285.5 \text{ meters}$$

This solution matches the earlier example with the same geometry that directly used the air pressure equation. The outer rim of the torus would be at R+r and equal to:

$$R+r = 16{,}894.5 \text{ meters}$$

This is the same distance as to the outer rim of the cylinder.

**Cylinder Center:** Another interesting evaluation is to find the air pressure at the center of the cylinder station. This varies with the station radius; see Figure 3-3. For very large stations, the air pressure at the center becomes very small; see Figure 2-3. For small stations, the air pressure at the center remains larger than our Denver air pressure limit. The air pressure equation can be used to find the station radius where the center of the station would be at the Denver air pressure limit. In this case, h would be equal to R and:

$$h^2 - 2R\,h + 3064.8\,R = 0$$
$$R^2 - 2R^2 + 3064.8\,R = 0$$

$$R = 3064.8$$

Stations at this radius would have acceptable air pressure from the outer rim (h=0) to the center of the rotating cylinder (h=R). This would support activities in the lower gravity regions above the top floor. One example would be low gravity flying from high floors on the cylinder spoke.

*3.2.4 Station Air Mass Examples*

Eventually, the station will be filled with air. Estimating the required air by assuming a uniform density throughout a space station is straightforward. With centripetal gravity compressing the air towards the outer rim, the air would not be uniform. As stations become larger, the air density varies sufficiently with height to affect the mass and stability. The density decreases with height. The previous air pressure and density equations can be used to compute the mass of the air in the station.

Section *2.2.2 Air Pressure Equations* provides the analysis and air pressure equation. Figure 2-3 provided a comparison of the air pressures for the Earth and the rotating stations. The air pressure in the enclosed station drops slower than on the Earth with height.

Just like on the Earth, the atmosphere becomes thinner with increasing altitude. Figure 3-5 shows the decreasing density with various size rotating stations. The chart shows the air density on the y-axis, ranging from a vacuum to below sea level. The x-axis shows the height above the outer rim of the rotating station. This includes multiple cylinder radii sizes; as such, it normalizes the height as a percentage of the cylinder radius. Section *2.2 Air Pressure Constraints* presents the associated equations. All the rotating stations have the same gravity and air density at the outer rim, but the gravity decreases at different rates. In smaller stations, the air density does not decrease below our minimum Denver constraint (shown as a green dashed line at 1.01 kilograms per meter cubed.

Figure 3-6 shows the air mass required to fill a cylinder station. The figure shows the air mass as a percent of the fully

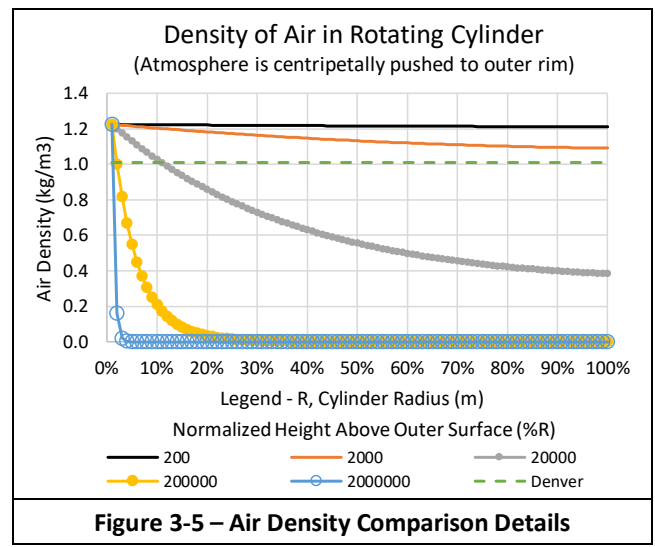

**Figure 3-5 – Air Density Comparison Details**

filled station and the total mass value. The chart includes a logarithmic x-axis showing the cylinder radius ranging from 100 meters to 2 million meters. The left y-axis shows the required air mass labeled as a percentage of mass to fill the fully-filled uniform-density station. The right logarithmic y-axis shows the required air mass in kilograms ranging from 10e3 kilograms to 10e24 kilograms.

Figure 3-6 also includes solid lines showing the mass of the air in the station. These are measured on the logarithmic right-hand axis. The solid gold line shows the mass to fully fill the station with uniform-density air. The other solid red line with triangle markers shows the mass to fill the station with the air centripetally pushed to the outer rim. Significant effects do not appear until the rotating station radius exceeds 20,000 meters. The two values are nearly equal for small stations and diverge with larger stations. Note that the right y-axis is logarithmic, and each unit is a factor of 100. As reflected in the percentages, the total mass to fill the very large rotating station is much less than the fully filled station.

The fully filled uniform density air mass has a volume relationship to the radius with a power trendline fit as *mass=5.0 $R^{3.0}$*. The mass of the air centripetally pushed to the outer rim is less than the fully filled. Until a radius of 20,000 meters, the fully filled and centripetally filled air masses are quite similar. Beyond 20,000 meters, the centripetal mass follows a power trendline fit as *mass=27847 $R^{2.09}$* and is an area relationship.

Figure 3-6 includes a dashed blue line to show the percent of the fully filled station air mass. With small stations, this shows near 100% because the air density is relatively uniform from the rim to the center; see Figure 3-4. Larger stations show that a smaller fraction of the air is required to fill them. It requires less than 20% of the fully filled uniform-density air to fill a large station with a radius of 100,000 meters. The percentage decreases with larger stations. Most of the air is near the outer rim. The air is pressed to the outer rim and becomes thinner (less dense) with increasing height. The chart clearly shows the reduced air mass.

It has been noted in large stations that much of the cylinder station volume would be uninhabitable because of low air pressure. With a 35,000-meter station, only the first 1600 meters above the outer rim would have livable air pressure. Constrained by gravity, the top floor could be almost 3500 meters above the outer rim. The situation is worse with larger stations. In a very large station with a 350,000-meter radius, and using the gravity constraint, the top floor would be at about 30,000 meters. The livable height would still only be 1600 meters because of air pressure limits. A later section, *§4.2 Very Large Space Stations*, offers a means to use more of that space between the air pressure limit and the gravity limit in large stations.

*3.2.5 Air Pressure and Station Size*

Figure 3-7 illustrates the relationship of a station's air pressure as a function of the station radius. The chart shows the cylinder radius on the logarithmic x-axis ranging from 100 meters to 2 million meters. The logarithmic left y-axis shows height in the multiple-floor station ranging from 1 to 1 million meters. The linear right y-axis shows air pressures ranging from 0 to 120,000 Pascals.

The chart includes a solid blue line to show the position of the top floor height when limited by a gravity range. This chart assumes gmax=1.05g, gmin=0.95g, and the gravity range scale is 10.5. The height of the top floors would be the cylinder radius R divided by that scaling factor of 10.5. This top floor would have a gravity of 0.95g. The chart shows that the top floor height increases from 19 meters to almost 200,000 meters as the station radius increases from 200 to nearly 2 million meters.

The air pressure decreases with the increasing height of the top floor. The outer rim (bottom floor) is designed to have a sea-level air pressure of 101,325 Pascals. The solid yellow line shows the air pressure of the gravity-limited top floor. The air pressure at the center of the cylinder is an orange line with triangle markers, which is included for reference. The top floor height line reaches the Denver Earth height of 1609.2 meters when the station radius is 16,897 meters. The top floor of larger stations would be higher than this Denver altitude, and the air pressure would not be acceptable. The chart includes a yellow line with circles to show the livable air pressure. This yellow line initially tracks the top floor air pressure with increasing station size. The livable air pressure line does not decrease below the Denver air pressure.

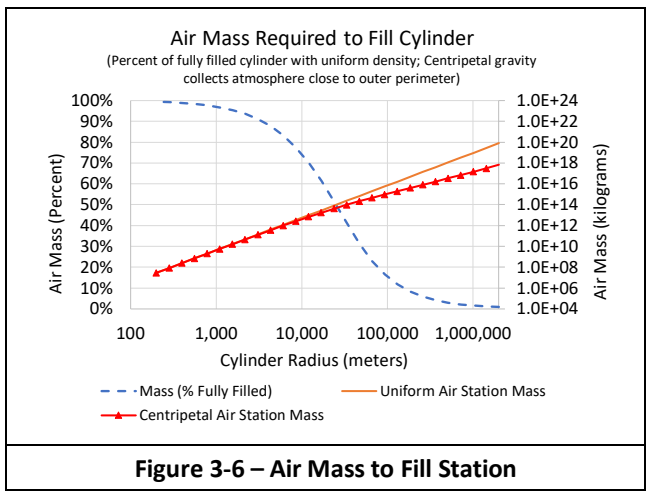

**Figure 3-6 – Air Mass to Fill Station**

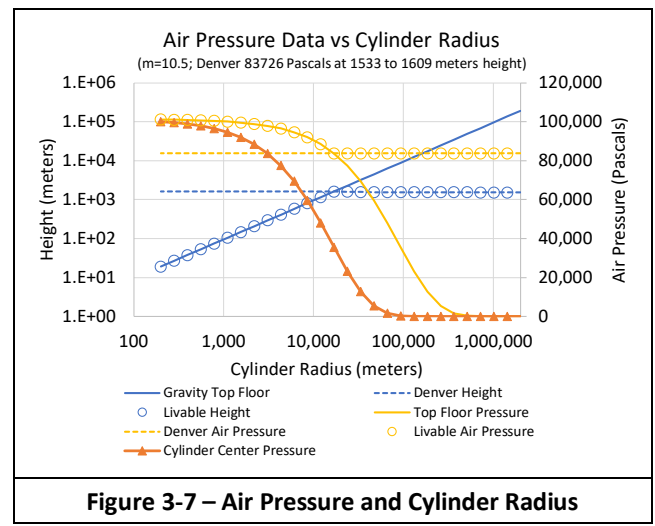

**Figure 3-7 – Air Pressure and Cylinder Radius**

The Figure 3-7 chart also includes a blue line with circles representing the livable height limited by gravity and air pressure. This line tracks the top floor height until the station radius reaches 16,897 meters. At this station radius, the top floor is at 1609 meters altitude and does not increase beyond this height because of the minimum Denver air pressure limit. Further increasing the station radius causes the top floor to drop slightly to 1533 meters because of the changing station rotation rate and air pressure. Recall the top floor height solution is a quadratic equation and was solved as:

$$h = \left(2R \pm \sqrt{4R^2 - 4*3064.75\,R}\right)/2$$

Using Wolfram Alpha, the top floor height reaches 1532.38 meters in the limit as the radius approaches infinity.

### 3.2.6 Air Pressure for Gravity Range

This analysis considers an alternative approach to specifying the station gravity range using gravity and air pressure. Our original approach set the outer rim at 1.05g or 1.1g and the top floor at 0.95g. This approach is used extensively throughout this document. The new alternative maintains the center floor of the gravity range at 1.0g and derives the outer rim and top floor gravities using the gravity and air pressure limits. This is a recent result and is included for completeness and future investigations.

We developed equations that can find the height of the top floor with 1.0g in the center of the gravity range. The key was to set the $g_0$ at the outer rim in the equation scaled from the Earth gravity at the center. Using the approach from the previous subsection:

$$h^2 - 2Rh - 2R\left[\ln\left(\frac{P_h}{P_0}\right)\bigg/\left(\frac{Mg_0}{kT}\right)\right] = 0$$

$$g_0 = g_E \frac{R}{R - h/2}$$

$$h^2 - 2Rh - 2R\ln\left(\frac{P_h}{P_0}\right)\left(\frac{kT}{Mg_E}\right)\frac{R - h/2}{R} = 0$$

$$h^2 - (2R + 1609)\,h + 3218\,R = 0$$

Which compares somewhat to the previous top-floor equation with the outer rim at 1.05g:

$$h^2 - 2R\,h + 3065.13\,R = 0$$

These two equations generate floor heights over the example gravity ranges. Our original approach sets the outer rim and the upper gravity range to 1.05g. The alternative sets the center of the multiple floors to 1g. Two charts in Figure 3-8 illustrate the air pressure constraint on the gravity range.

The graph in Figure 3-8a shows the gravity range difference for these two approaches. The x-axis shows the rotation radius ranging from 0 to 200,000 meters. The y-axis shows gravity in 1g (Earth gravity=9.86065 meters per second squared) ranging from 0.94g to 1.06g. The red dots show the 1g Earth gravity for all the radii. The yellow lines and dots show the original gravity range approach setting the outer rim (gmax) to 1.05g. The top floor is 0.95g for radii less than 16,897 meters. For radii greater than 16,897 meters, habitable air pressure limits the top floor to approximately 1600 meters.

Consider the original approach and setting the outer rim to 1.05g. The graph in Figure 3-8a shows that the top-floor centripetal gravity increases with the increasing rotation radius. It reaches 1g at a rotation radius of about 35,000 meters. For even larger stations, the top-floor gravity continues to increase. All floors of this large station have a gravity greater than 1g. This is not desirable and motivated the approach introduced in this subsection.

The blue lines and squares show the new gravity range approach with the center floor at 1g. Using this approach with increasing radius, the center floor is maintained at 1g, the top floor gravity is increased, and the outer floor gravity is decreased. This produces a more desirable range of gravity with extremely large stations.

The graph in Figure 3-8b shows another effect from this gravity range approach. This approach shows that the top floor height remains constant at 1609 meters. Previous sections and analyses have shown the top floor height peaking at the air pressure limit and then decreasing slightly with the increasing rotation radius. The increasing radius, changing rotation rate, and changing air pressure cause the top floor height to decrease somewhat after reaching the maximum. Figure 3-8b has an x-axis that shows the rotation radius ranging from 0 to 200,000 meters. The y-axis shows the top floor height ranging from 1520 to 1620 meters. The yellow solid squares show the top floor height as a function of the rotation radius with the outer floor set to 1.05g (original approach). This decrease corroborates the data in earlier figures, such as Figure 3-9b. The hollow blue squares show the top floor height with the center floor set to 1g (new approach). With this approach, the top floor remains at 1609 meters.

This centered gravity range has significant advantages and is recommended for future investigations. The most significant benefit is that the habitable gravity range is centered about Earth gravity for large stations. The range would be more comfortable for the inhabitants. This also avoids the situation where both the top floor and the outer rim are greater than Earth gravity; see Figure 3-8a. This approach results in the top floor being slightly higher with the gravity and air pressure changes. This produces a larger multiple-floor region. Figure 3-8b shows the multiple floors would increase by about 50 meters and add 10 floors. The larger multiple-floor region would increase the station floor surface area and the population. Note that the main floor would be at a smaller radius and would be slightly smaller. These changes would also change the MOIs and the stability. The station geometry would also change slightly to maintain the stability ratio of Iz/Ix=1.2. These MOI changes would be small and typically less than 5%. The 5% differences only happen on the largest stations. Figure 3-8a shows the outer rim gravity is less for the 1g center floor approach. This lower gravity at the same radius implies the station would rotate slower and produce

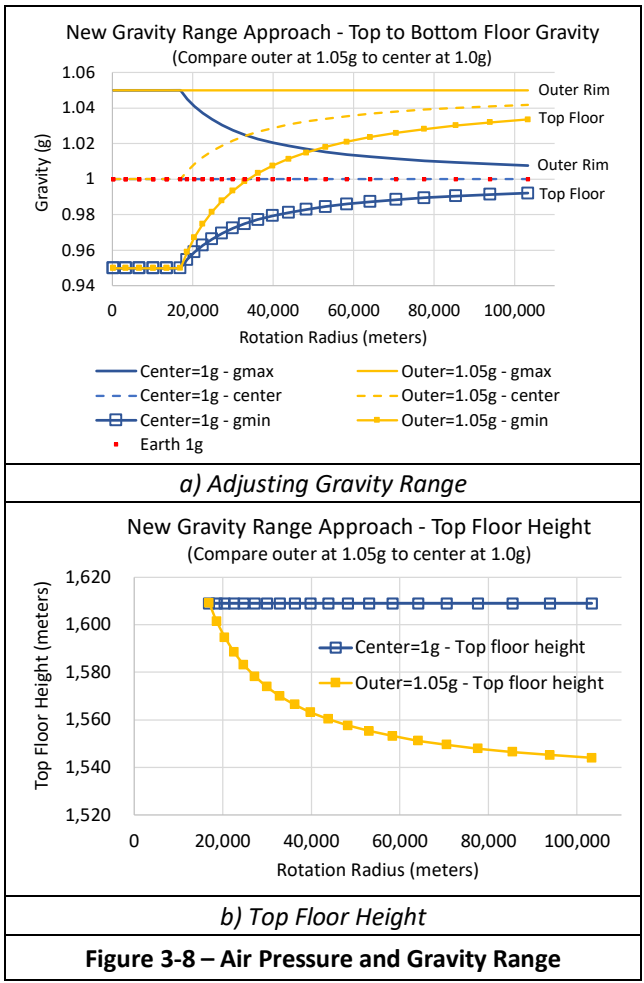

*a) Adjusting Gravity Range*

*b) Top Floor Height*

**Figure 3-8 – Air Pressure and Gravity Range**

less stress on the outer structures. At this time, we are reluctant to make this global change throughout our analysis. This paper continues to assume the outer rim has a centripetal gravity of 1.05g in the remaining sections.

### 3.2.7 Station Air Pressure Summary

This section has provided a comparison of Earth's air pressure to that of rotating stations. The air pressure in rotating space stations constrains the habitable floors and station design. Similarly, the air pressure constrains the population and the livable regions. The centrifugal force in the rotating station pushes the atmosphere towards the outer rim of the station. This ultimately defines the amount (mass) of atmosphere required in the station. Air pressure in large rotating stations also defines the station size and can define the gravity range.

## 3.3 Population Limits

This section offers more population details about the single and multiple-floor designs. It presents these details for the four station geometries. The population increases with the size of the station. These concepts were introduced and detailed in [Jensen 2023]. This section reuses some graphs and text and extends the details. This information provides context and continuity for this population limits section.

### 3.3.1 Station Radius and Floor Counts

Figure 3-9 illustrates the gravity and air pressure constraint on the number of floors in a cylinder station. This, of course, affects the population and cylinder design. The x-axis shows the cylinder radius on a logarithmic x-axis ranging from 100 meters to 200,000 meters. The linear right y-axis shows the number of floors in the station ranging from 0 to 350. The left y-axis shows height in meters on a logarithmic scale ranging from 1 to 10 million meters.

Floors are spaced 5 meters apart. The number of floors increases linearly with the cylinder radius until 16,897 meters. The top floor is at the minimum of 0.95g in the habitable gravity range. At the radius of 16,897 meters, the number of floors has reached a maximum. At that radius and on the top floor, the air pressure has reached the minimum Denver air pressure, and the gravity has reached the minimum 0.95g. With radii less than 16,897, both the radius and the number of floors increase. Greater than 16,897, only the radius increases. The length of the cylinder continues to increase with the radius. The number of floors decreases slightly because of the changing rotation rate and air pressure. Even so, the population increases slower after 16,897 feet. The cylinder floors' circumference length and surface area continue to increase.

To provide additional insight, Figure 3-9 includes a habitable height curve. The habitable height above the outer rotating rim of the cylinder is constrained by gravity and air pressure. The left y-axis shows height on a logarithmic scale ranging from 1 to 100 million meters. For small stations, the gravity constrains the livable height. The minimum gravity is 0.95g on the top floor. The livable height reaches the maximum at the top floor height of 1609.2 meters with the station radius of 16,897 feet. At 1609.2 meters, the station's top floor has the minimum air pressure (Denver). The dotted line showing the maximum height to the center continues to increase with the radius. The Denver air pressure limit continues to constrain the top floor height. Because of the increasing radius and the changing rotation rate and air pressure, the top floor height decreases slightly after reaching the maximum at 1609.2 meters. On the logarithmic scale, it is difficult to see the height decrease from 1609 meters at the radius of 16,897 to 1538 meters at 200 thousand meters. The right y-axis shows floor count on a linear scale ranging from 0 to 400 floors. The floor count is calculated from the top floor and shows the peak at the station radius of 16,897 meters. The floor count initially increases linearly with the livable top floor height. It then shows the slight decrease with larger stations after reaching the Denver air pressure. Again, this is because of the increasing radius and the changing rotation rate and air pressure.

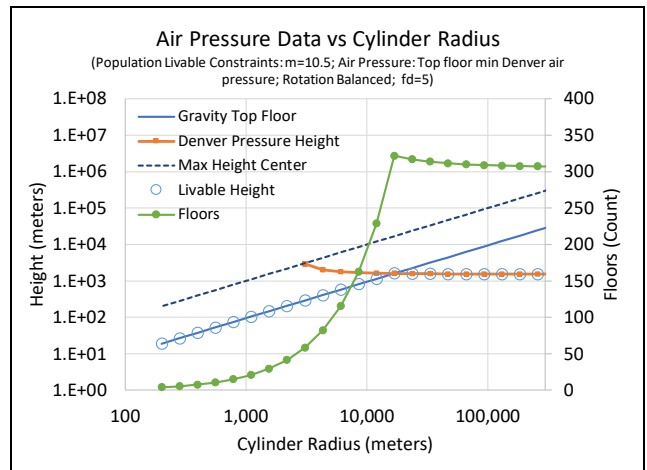

**Figure 3-9 – Floor Height and Cylinder Radius**

In Figure 3-9 the gap between the gravity-limited floor height line and the air pressure-limited line represents an unused space in the station with acceptable gravity. This open area provides an aesthetically pleasing vista for the inhabitants but can also be considered an inefficient use of space. It is possible to create airtight floors or layers of floors in the station and fill the entire gravity-limited region with floors. These airtight layers are not required until stations become very large.

*3.3.2 Top Floor Limits*

Previous sections have shown how air pressure and gravity affect the top-floor position in multiple-floor station designs. There are limits on the number of floors in our multiple-floor station designs. The viable gravity range is one of those limits; see *§3.1.2 Geometry Gravity Ranges*. The designs will have a habitable gravity range. The floors typically range from 0.95g to 1.05g. A viable air pressure is another limit in the design; see *§3.2 Air Pressure Limits*. With multiple airtight layers, the entire gravity region could be habitable. The following paragraphs offer more detail on geometry's top floor limits.

**Cylinder Top Floor Limits:** Figure 3-10a illustrates the gravity and air pressure limits in the cylinder station. The graph shows top floor heights as a function of the station radius. The y-axis shows the height above the outer rim of the cylinder and ranges from 0 to 5000 meters. The x-axis shows the station radius on a logarithmic scale ranging from 100 to 50,000 meters. The station rotates and provides a gravity range of 1.05g to 0.95g from the outer rim to the top floor. This gravity limit for the top floor is shown as a solid blue line with round markers. For this gravity range, the line is at the radius R over the value 10.5. This chart includes an air pressure limit on the top floor equal to Denver's air pressure. That air pressure is a function of the station radius and the top floor height. An orange line with diamond markers represents this air pressure limit. At a station radius of about 16,900 meters, the air pressure limit and the gravity limit intersect at a height of 1609 meters. The habitable limit follows the gravity limit with small stations and transitions to the air pressure limit at this intersection point. A solid grey line with hollow circle markers in Figure 3-10a represents this habitable height. This habitable height excludes the use of the multiple airtight layers concept; see *§3.3.4 Airtight Layers of Floors*. The chart also shows the maximum height with a dashed grey line. This represents the height to the center of the cylinder from the outer rim. It represents the position in the station with a minimum air pressure and where the centripetal gravity is zero.

**Elliptical Torus Top Floor Limits:** Figure 3-10b illustrates the gravity and air pressure limits in the elliptical torus station. The results are shown using the same axes as with the cylinder results. In the torus, the gravity limit sets the minor radius a to the major radius R over 9.5 in this station. This height is shown as the blue line in Figure 3-10b. The 9.5 ratio creates a minimum gravity of 0.95g and a maximum gravity of 1.05g. The top floor is at the center of the torus tube at the major radius R. Figure 3-10b shows the top floor air pressure height with an orange line and diamond markers. Until a station radius of 8045 meters, the air pressure in the entire torus tube is above our Denver limit of 83,728 Pascals. At this size, the ceiling is at the Denver air pressure. The gravity limits the top floor height for stations smaller than 15285 meters. At this radius, the air pressure height limit is the same as the gravity height limit. The top floor is at 1609 meters when this occurs. With larger stations, the gravity limit height continues to increase, but floors at those heights would have low air pressure. The habitable air pressure limit height slowly decreases because of the increasing radius and the changing rotation rate and air pressure. The minimum gravity and air pressure limits represent the habitable height of the torus. Figure 3-10b illustrates this habitable height with a solid grey line and hollow circle markers. This habitable height does not include the multiple airtight layers concept; see *§3.3.4 Airtight Layers of Floors*. The dashed grey line represents the ceiling height of the torus from the outer rim. Above that ceiling (outer shell) is the vacuum of outer space.

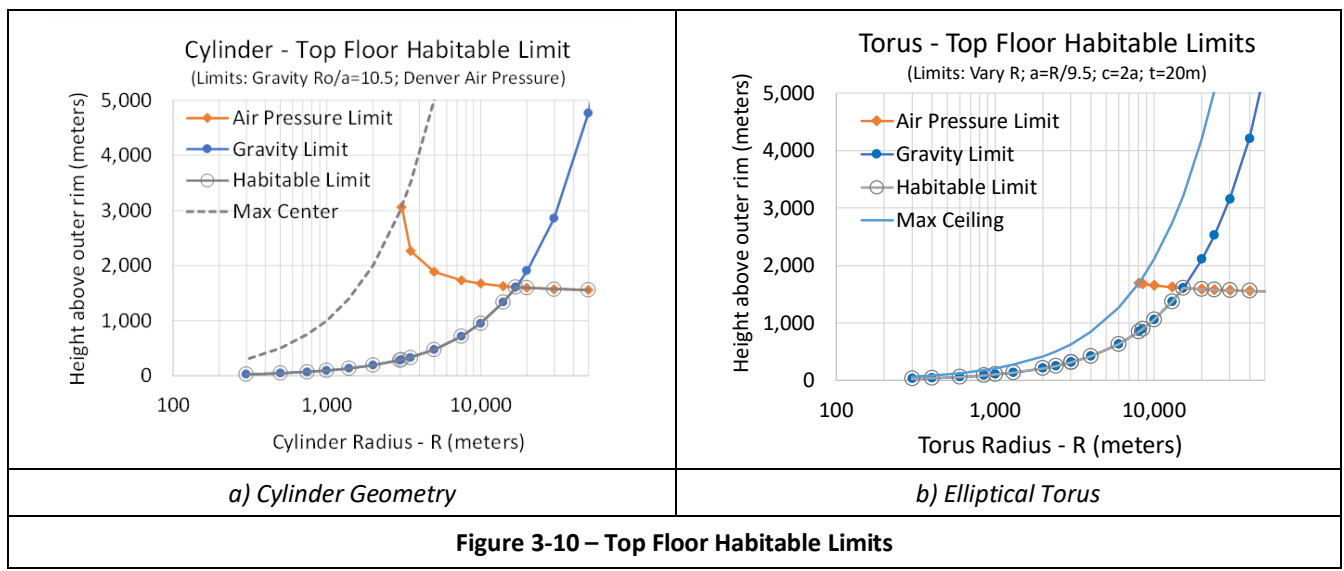

| a) Cylinder Geometry | b) Elliptical Torus |

**Figure 3-10 – Top Floor Habitable Limits**

**Top Floor Summary:** Table 3-3 shows the largest rotating station design using the air pressure limit. These results use the same set of previously presented metrics. These include the sea level air pressure of 101325 Pascals, the Denver air pressure of 83726 Pascals, and the Denver altitude of 1609 meters above sea level. It includes the cylinder and the torus geometries; the ellipsoid and dumbbells would have the same corresponding gravity and air pressure limits. Table 3-3 shows two gravity ranges for the cylinder and the torus. The minimum gravity would be 0.95g and be at the top floor of the geometry. The maximum gravity of 1.05g or 1.1g would be at the outer rim of the geometry. The Denver air pressure is a second limit on the top floor. Table 3-3 includes the top floor height that reaches the gravity and air pressure limits. Figure 3-9b showed an example where the cylinder station radius is 16,897 meters, and the top floor air pressure and gravity limits are 1,609.2 meters.

### 3.3.3 Single and Multiple Floors Population

The literature contains many types of space station geometries including spheres, torus, dumbbells, and cylinders. This subsection focuses exclusively on the torus and the cylinder. One has the rotation axis inside the station shell, and the other has the rotation axis outside the station shell. Later sections consider and evaluate all four space station geometries. Figure 3-11 shows the populations for some of the literature's cylinder and torus geometry space stations [Johnson and Holbrow 1977]. This subsection first describes aspects common to both charts. It then describes specific details for the single-floor charts and the multiple-floor charts.

| Table 3-3 – Largest Design for Top Floor at Denver Air Pressure | | | | |
|---|---|---|---|---|
| **Geometry** | **Torus** | **Cylinder** | **Torus** | **Cylinder** |
| Gravity Range | min/max 0.95g/1.05g | min/max 0.95g/1.05g | min/max 0.95g/1.1g | min/max 0.95g/1.1g |
| Scale m=(R/a) | 9.5 | 10.5 | 6.33 | 7.33 |
| Station Radius (R) | 15,287 meters | 16,897 meters | 9,938.1 meters | 11,508 meters |
| Top Floor Height (h) | 1,609.2 meters | 1,609.2 meters | 1,570.0 meters | 1,570.0 meters |

The space station geometry determines the available surface area. For example, a torus's main floor surface area is the floor width (twice the minor radius) times the station circumference (two pi times the major radius). The vertical axis of the charts in Figure 3-11 show the population on a logarithmic scale ranging from 10 to 1 billion. It shows the major radius of the station along the horizontal axis. It is a logarithmic scale and ranges from 100 meters to 40,000 meters.

Both charts include population lines assuming specific geometry relations. The population lines also show the relationship between the station radius and the population. For the cylinders, this line shows populations when the length of the cylinder is 10 times the radius and produces a long, narrow cylinder. This is a historic geometry ratio. For the torus, this line shows populations when the major radius is 10 times the minor radius. The single floor only uses the internal projected surface area of the station geometry. The population is proportional to the radius squared for both the cylinder and the torus.

Figure 3-11 includes population estimates from various reports [Johnson and Holbrow 1977] [O'Neill et al. 1979] [Globus et al. 2007] [Brody 2013] [Jensen 2023]. The populations in both charts have been normalized to a projected surface area allocation of 67 square meters per person, which comes from a NASA study [Johnson and Holbrow 1977].

#### 3.3.3.1 Single Floor Population

Historically, most designs use only projected floor values using only a single floor on the outer perimeter of the station for living space.

**Toruses:** Figure 3-11a includes the Stanford Torus and a Tiny Torus from a NASA study [Johnson and Holbrow 1977]. A 1974 Stanford study detailed a rotating torus design with a minor radius of 65 meters and a major radius of 830 meters [Johnson and Holbrow 1977]. The Tiny torus and Stanford torus fall along the r=R/10 torus line. Figure 3-11a also includes various other torus designs for comparison. It uses several tori from [Jensen 2023]. These results are for the single main floor of those multiple-floor tori. The Atira torus has a major radius (R) of 2116 meters and minor radii of 1003 meters and 334 meters (r). Its population lies above the torus r/R line because of its elliptic cross-section. The Atira torus is similar to the tori shown in Figure 1-1 and Figure 2-5. The chart also includes the station from the movie Elysium [Brody 2013]. Elysium is a torus with a major radius of 30,000 meters and a minor radius of 1500 meters [Brody 2013]. For consistency, the Elysium population in Figure 3-11 uses the same population density metric (67 meters squared per person).

**Cylinders:** Figure 3-11a includes four O'Neill cylinder models, shown as C-1 through C-4 [Johnson and Holbrow 1977]. The lengths of these cylinders are 10 times the radius. Figure 3-11 includes the Kalpana One station, which is a rotating cylinder with a radius of 250 meters and a length of 325 meters [Globus et al. 2007]. This cylinder is not along the cylinder line in the chart because the R/L for Kalpana

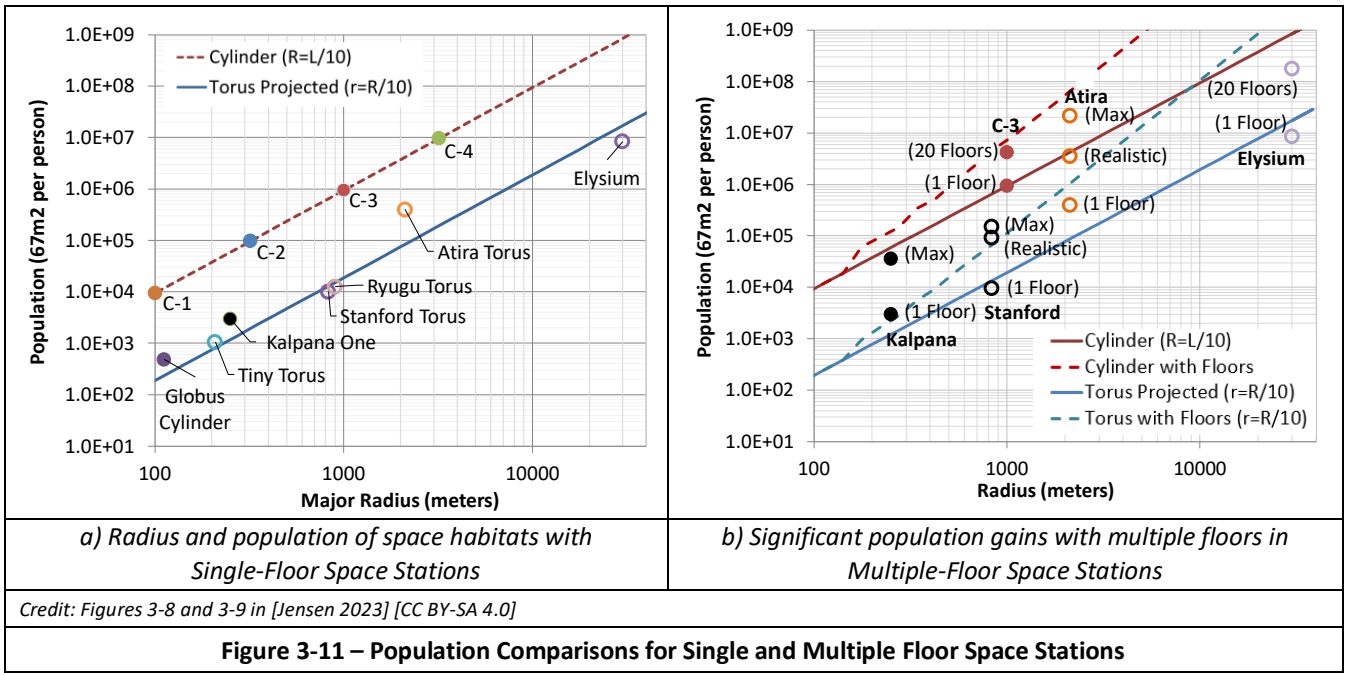

*a) Radius and population of space habitats with Single-Floor Space Stations*

*b) Significant population gains with multiple floors in Multiple-Floor Space Stations*

Credit: Figures 3-8 and 3-9 in [Jensen 2023] [CC BY-SA 4.0]

**Figure 3-11 – Population Comparisons for Single and Multiple Floor Space Stations**

One is 250/325 or about 3/4 instead of 1/10. The O'Neill cylinders use two counter-rotating cylinders to reduce precession and improve stability. The designers of Kalpana One used only one cylinder for their analysis and reduced the length for stability [Globus et al. 2007].

3.3.3.2 Multiple Floors Population

Figure 3-11b compares the population of stations with a single floor and with multiple floors. This chart includes the potential populations for some of literature's cylinder and torus geometry space stations [Johnson and Holbrow 1977] [Globus et al. 2007].

The multiple-floor chart shows the same cylinder and torus geometries population lines. The chart also includes analogous lines to show the population increase from using multiple floors. Adding floors dramatically increases the square footage and larger populations. The population lines also show the relationship between the multiple-floor station radius and the population. For a single floor, the population is proportional to the radius squared. For both the cylinder and the torus, the multiple-floor population is nearly proportional to the radius cubed. Multiple floors take advantage of the internal volume of the station geometry.

Figure 3-11b includes the results of the analysis of five stations. These are a subset of the stations in Figure 3-11a. These stations include two cylinder and three torus stations. The stations include population estimates for a single floor, a maximum number of floors, and, in some cases, a realistic number of floors. We don't expect colonists to live in the deep "bowels" of very large stations; as such, we reduce the number of inhabited floors to a more realistic number. The analysis in Figure 3-11b also limits the number of cylinder floors to one-tenth the radius to provide better gravity range uniformity. It includes one of the O'Neill cylinder models (C-3) with floors. It includes the Kalpana One cylinder with floors. The Kalpana One population increases from a single floor supporting 3357 people to over 12,160 people with multiple floors.

The charts in Figure 3-11 also include various torus designs for comparisons. The torus stations assume the outer half of the torus is filled with floors. The Stanford torus is from the Stanford study [Johnson and Holbrow 1977], and its single-floor population falls along the r=R/10 torus line in Figure 3-11 The maximum population for a multiple-floor Stanford torus increases from about 10,000 people for the single floor to a realistic 92,000 people using 12 floors.

This chart shows the Atira station with multiple floors. The top floor can accommodate almost 400,000 people using 67 square meters per person. It can support a maximum population of 21.5 million and a realistic population of 3.5 million.

The movie Elysium Station provides 1131 square meters of space for each individual with its population of 500,000. Using only 20 of the multiple floors possible, Elysium would increase its floor area from 565 square kilometers to 11,319 square kilometers. With 67 square meters per person, the 20 floors of Elysium would support 176.7 million people. Using the generous 1131 square meters per person, it would still support over 10 million people.

*3.3.4 Airtight Layers of Floors*

The previous *§3.2.3 Air Pressure* covered the constraining effect of air pressure on the livable regions of rotating stations. An air pressure at 1600 meters would be like Denver, Colorado, and is used as our minimum air pressure. The previous *§3.1.2 Geometry Gravity Ranges* introduced the gravity range constraint of 1.05g at the outer rim and 0.95g on the top floor.

Figure 3-9 and Figure 3-10 showed the air pressure and gravity constraints for the cylinder and the torus stations. Those figures showed an uninhabitable low air pressure zone between the air pressure limit and the gravity limit beyond the cylinder radius of 16,900 meters. There is a straightforward approach to using that "uninhabitable" zone and increasing the top floor in large stations above that minimum air pressure height of 1600 meters. That top floor could be made airtight. Immediately below that top floor, the air pressure would be at the Denver limit. Being airtight, immediately above that top floor, the air pressure could be raised to the sea level maximum. Additional floors could be constructed above that original top floor. Another 1600 meters of floors could be added before reaching that Denver air pressure limit again.

Figure 3-12 illustrates these limits in a cylinder station with a radius of 35,000 meters. The chart shows the air density of the station along the y-axis, which ranges from 0 to 1.3 kilograms per meter cubed. The x-axis shows the height in the station. The chart shows the gravity floor limit as a yellow dashed line at 3333 meters and the air pressure limits as dashed blue lines at 1567 and 3127 meters. It shows the minimum Denver air density at 1.01 kilograms per meter cubed as a yellow dashed line.

The chart shows three air density models. The first model is without airtight layers (open floors). The red line shows this model with the air density monotonically decreasing (using air density equations). The second model is the airtight layers of floors followed by open air to the center. The blue line shows saw tooths varying between the sea level and minimum air pressure limits, followed by a decreasing air density to the center. The third model is a uniform density (sea level) shown by a green line across the top of the chart. A fourth model would be airtight floors. Every floor would be airtight and have sea-level air density. This model is not shown but would be a combination of the uniform density (green line) from the outer rim to the top floor gravity limit followed by the decreasing airtight air density (blue line) to the station center.

Figure 3-12 includes a design with two airtight layers at heights of 1567 and 3127. The blue line showing the airtight layer density follows the same air density equations; however, the height resets to zero, and the gravity constant changes at each of the layers. Once the height is above the gravity limit of 0.95g, the airtight layer air density drops monotonically to the center of the station.

The regions under the curves of the four models reflect their associated air mass. The open-floors model uses the least air, the airtight-layers then open-floors model uses more, the airtight floors model uses more, and the uniform-density model uses the most. Table 3-4 shows the four models and their associated mass for this 35000-radius cylinder station.

There are more floors in the station design with the airtight models. In Figure 3-12, the open-floors design has 314 floors, while the airtight design has 663 floors. Using the surface area of all those floors, the open-floors design could support a population of 25 billion, while the airtight design could support over 50 billion people using 144.2 square meters per person.

*3.3.5 Population for Example Stations*

This subsection describes the population for several example stations. It compares the population of stations with single and multiple floors, shows the effect of limiting the top floor location in our stations, and reviews the populations of single and double dumbbell stations.

3.3.5.1  Single and Multiple Floors

Figure 3-13 shows populations of torus space habitats. The estimates use 67 square meters per person. It shows the population along the vertical axis based on available surface area. The horizontal axis shows the major radius of the torus. The chart shows torus geometries where the minor radius is one-tenth the major radius. A 1979 NASA Study [O'Neill et al. 1979] also used this ratio. Figure 3-13 considers three population allocations for the torus surface area. The lowest population results from using a single floor (1 Floor) projected across the minor diameter. This surface area is twice the minor radius times the major radius circumference. This population is proportional to the radius squared. Multiple floors are created under a center main floor; see Figure 1-1. As the minor radius increases, the space under the center diameter floor increases. This space can be used for additional floors. For stations with a small radius, Figure 3-13 shows little increase in population by adding floors. As the radius increases, the benefits from adding floors appear. A torus

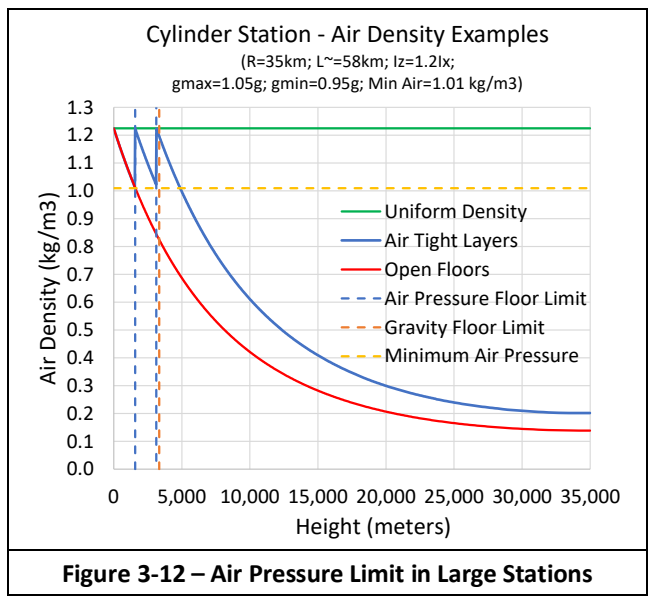

**Figure 3-12 – Air Pressure Limit in Large Stations**

| Table 3-4 – Masses of Air Modelling Examples | |
|---|---|
| **Air Density Modelling** | **Mass (kilograms)** |
| Uniform | 2.72e14 |
| Airtight Floors | 1.42e14 |
| Airtight Layers then Open Floor | 1.39e14 |
| Open Floors | 1.08e14 |
| Cylinder Station; R=35 km; L≈58 km; Stable Design | |

station with a major radius of 1000 meters and a minor radius of 100 meters could have nearly 20 floors under the main floor. Each floor extends entirely around the major radius of the torus. The population with multiple floors is nearly proportional to the radius cubed. There will be open vistas on the main floor for colonists to enjoy. Unfortunately, with larger radius habitats, most of the square footage is on the lower levels. Some of that lower-level square footage could be used for open spaces.

3.3.5.2  Limiting the Number of Floors

The lowest floors might be psychologically undesirable as living quarters to many people. The lowest floors would also have higher centripetal gravities than desired for residential usage. To compensate for these undesirable issues, Figure 3-13 includes a population estimate using only the top 20 floors. The lower floors would be for industrial, research, ventilation, transportation, storage, and agriculture. This 20-floor constraint results in a population that is first proportional to the radius cubed. Once there are more than 20 floors (at a radius of about 1000 meters), the population is only proportional to the radius squared. This 20-floor allocation is only relevant for large stations; normally, all the floors will be used. For reference, the chart includes several tori located at their published major radii and population. It may be reasonable to only use a limited number of floors in the large stations. Although there could be over 300 floors in a station with a radius of 15,000 meters, psychological limits may reduce that number to 20 to 30 floors.

Previous sections have analyzed constraints from gravity and air pressure in the station. These two constraints provide direct methods to determine the station's top floor. The viable gravity range and air pressure limit the top floor position in the station, which in turn limits the number of floors in our multiple-floor station designs. This subsection included the psychological limit of 20 floors on the population. Figure 3-14 is included to show the effect of these limits on the station's population (floor surface area). One graph shows the effects on a cylinder station and the other on a torus station.

The graphs show the population as a function of the station radius. These charts assume a gravity range of 0.95g to 1.05g. They also use an air pressure range from sea level to a height representing Denver, Colorado air pressure (1600 meters).

The chart in Figure 3-14a shows the population supported by the multiple floors above the outer rim of the cylinder. The chart in Figure 3-14b shows the population supported by the multiple floors above the outer rim of the torus. The y-axes shows the population supported by the habitable floors, assuming 144.2 square meters per person. The y-axes show the population and use a logarithmic scale ranging from 1 to 100 trillion people. The x-axes show the station radius and use a logarithmic scale ranging from 100 to 500 thousand meters.

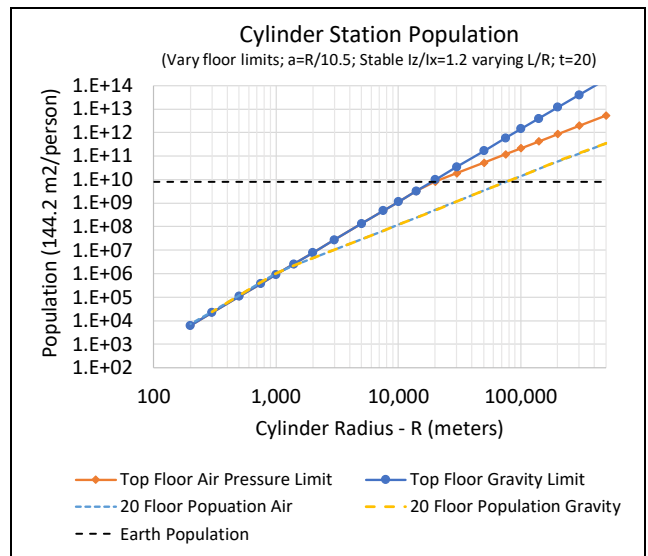

*a) Cylinder Geometry Station*

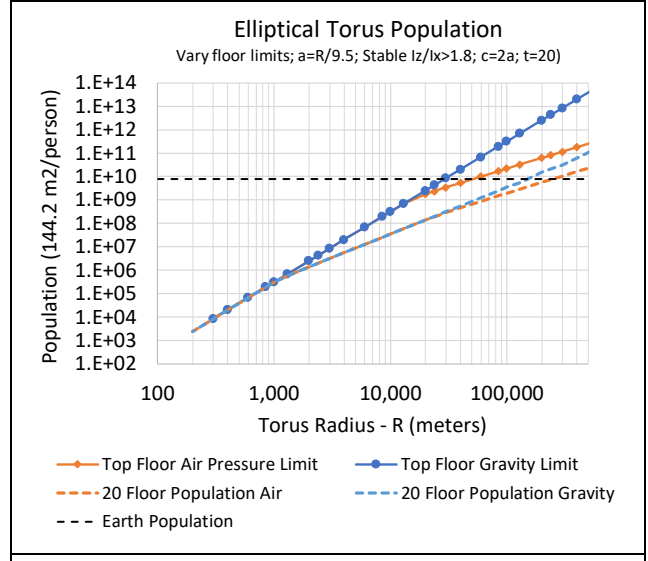

*b) Elliptical Torus Geometry Station*

**Figure 3-14 – Air Pressure and Gravity Limits on Station Population**

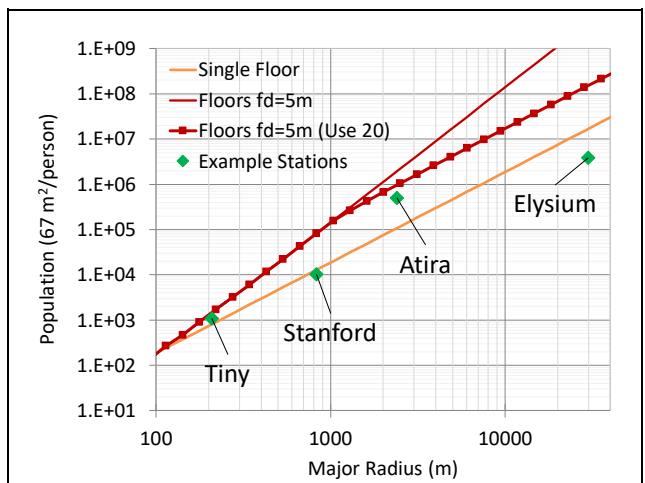

**Figure 3-13 – Torus Space Station Populations**

The charts show the population with the top floor constrained by gravity, air pressure, and psychology (20 floors). Gravity again limits the top floor height and the number of floors for smaller stations. At about a radius of 2000 meters, the 20-floor limit begins to take effect. The charts show that at a station radius of about 16,900 meters, the air pressure limit and the gravity limit intersect. With larger stations, the air pressure is more restrictive, and the top floor height remains at about 1600 meters. The population continues to increase because the major radius and minor radii continue to increase and provide more floor space.

The gap between the gravity-limited population line and the air pressure-limited population line represents an unused space in the station that has acceptable gravity. This space can be allocated as open space for aesthetics and the inhabitants' well-being. The gap can be considered inefficient and is a motivation not to build such large stations. Finally, this space can be exploited for larger populations by using airtight layers, and floors could reach the gravity floor limit. It becomes possible to support gravity-limited populations in large stations by using airtight layers of floors.

Without using multiple airtight layers, the air pressure limit in larger stations appears to be restrictive. In Figure 3-14, the air pressure-limited population is much less than the gravity-limited population. Restrictive is a relative term; these stations could hold the entire population of the Earth. With a cylinder radius of about 19,800 meters and limited by the air pressure, the station's top floor would be at 1600 meters. The station would have over 300 floors and a total floor surface area of 1.16e12 square meters. Assuming 144.2 square meters per person, this station would support 8 billion people. Even using only the top 20 floors, this cylinder station would support 394 million people. In comparison, today the Earth has about 8 billion people and 43.3e12 square meters of habitable space. On Earth, this provides over 5400 square meters of space per person.

3.3.5.3  Dumbbell Populations

This subsection offers a brief discussion on dumbbell populations. For reference, Figure 3-21 provides an illustration of the double dumbbell. We have found that a single dumbbell is not rotationally stable. Doubling the dumbbell creates a rotationally stable geometry. Later sections describe this stability issue.

Our stability paper [Jensen 2024s] provides more detail on the dumbbell stability and population. We include the following text from that paper. For a specific radius and gravity range, the node sizes are identical for the single and double dumbbell geometries. There are twice the number of nodes in the double dumbbell design; as such, the double dumbbell design supports twice the population at a specific radius. For a specific mass, the double dumbbell population is consistently about 0.75 times the single dumbbell population. Figure 3-15 shows the node radius for the single and double dumbbells as a function of the construction mass. It shows the 0.75 ratio found from the two radii. For a given mass, the double dumbbell population may be smaller than the single dumbbell; however, the double dumbbell does not have the instability wobble. To avoid active rotational balance, the double dumbbell appears to be the better choice.

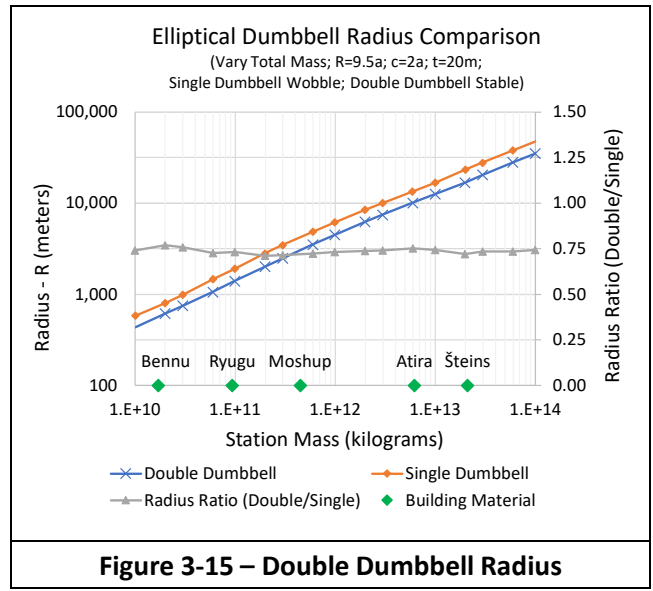

**Figure 3-15 – Double Dumbbell Radius**

*Section 3.10 Station Constraints and Population* provides more details and comparison of the dumbbells with the other geometries. Data shows the dumbbell geometries support at least twice the populations of the other three geometries. For the smaller masses, the double dumbbell population is about 0.75 times the single dumbbell population.

The single dumbbell often will provide the most population for a given mass. The single dumbbell has multiple issues that preclude it from being the ideal geometry choice. It is not rotationally stable. Other geometries have better vistas. The population is divided into at least two nodes, which creates potential social issues. Tether stresses will limit the size of the station. Dumbbells are expected to be less stable in the event of a severe impact.

*3.3.6 Population Summary*

This section reviewed station population limits. Additional details are covered in [Jensen 2023]. The analysis showed that multiple floors significantly improved the supported population for all four geometries. For a single floor, the population is proportional to the radius squared. For multiple floor geometries, the population is nearly proportional to the radius cubed. Multiple floors take advantage of the internal volume of the station geometry.

## 3.4  Material Limits

The materials used in the construction of a rotating space station must be able to withstand a variety of stresses and forces, including mechanical, thermal, and radiation. Our focus is on the mechanical framework of the rotating station. This rotation provides centripetal gravity. The thick shell provides radiation protection. This subsection covers the desired material, material availability, material strength, station stress, and radiation protection. Minimal details are provided on thermal stresses.

*3.4.1 Desired Construction Material*

The construction materials for rotating space stations must withstand the harsh conditions of space, including radiation, extreme temperatures, and vacuum. These materials should also be lightweight and durable enough to support the station's structure and function. A report on Space Structures and Support Systems [Bell and Hines 2012] focused on glass, basalt, metals, concrete, and anhydrous glass. They presented that glass, basalt, and metals have high compressive strength. They presented that anhydrous glass has the greatest tensile strength. They suggested basalt and concrete might be used for relatively thick-walled pressurized structures.

Spacecrafts today commonly use metals for their structure and exterior. Titanium and aluminum are both lightweight and strong. Designers today use Kevlar and carbon composites for spacecraft design because they are resistant to corrosion, temperatures, and radiation. Glass and composites can withstand temperature extremes and are resistant to corrosion and radiation. The idea of using anhydrous glass for much of the space station construction was presented in [Jensen 2023]. The oxides to produce anhydrous glass are common on the lunar and asteroid surfaces. If necessary, it may be possible to reinforce glass structural beams with asteroidal nickel-iron steel to withstand both tension and compression forces [Prado and Fraser 2018]. This complexity may not be necessary with the additional strength produced with the anhydrous, vacuum-produced glass [Blacic 1985].

*3.4.2 Construction Material Availability*

There are many sources of construction material for a space station. These include the Earth, other planets, our Moon, Near Earth Objects, Main Belt Asteroids, other moons, and dwarf planets. Figure 3-16 shows various asteroid, comet, and moon sources being considered as construction material. The x-axis shows the semimajor axis of these bodies. The y-axis shows the mass of the bodies. The size of the bubbles represents the diameter of the bodies. Their color represents their spectral composition. The following subsections describe sources of the construction material.

3.4.2.1   Earth

Constructing a space station requires a variety of materials, including metals, polymers, composites, and electronic components, as well as various life support systems and consumables. All the materials can be found on Earth. The Earth can be considered a nearly unlimited source of material. The space station construction also requires construction equipment in addition to the construction materials. Generally, it would be cost-prohibitive to launch the equipment and materials into space using rockets. The cost of launching material into space is high because it requires a large amount of energy, specialized equipment and expertise, and extensive safety and regulatory requirements.

In the early days of space exploration, the cost of launching one kilogram into space was extremely high. For example, the cost per kilogram of the 1969 Apollo 11 mission was estimated to be around $50,000 (in 1969 dollars), equivalent to over $350,000 in 2023 dollars. Over the decades, the cost of rocket launches to Low Earth Orbit (LEO) has dropped from $20,000 per kilogram for the United Launch Alliance Atlas V rocket in 2015 to $1410 per kilogram for the Space X Falcon Heavy rocket in 2019 [de Selding 2016] [SpaceX 2018]. Even at these lower costs, launching all the materials and equipment to build a space station would be very expensive.

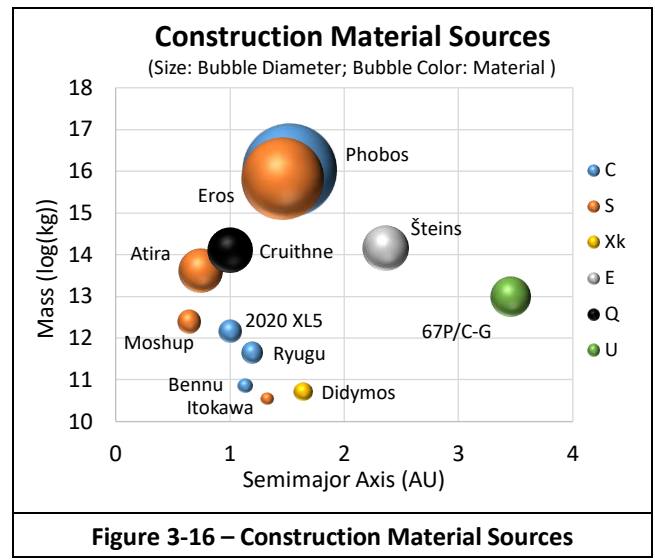

**Figure 3-16 – Construction Material Sources**

3.4.2.2   Lunar Building Material

The moon has materials that would be valuable for space station construction [Williams et al. 1979] [Schmitt 2005]. These materials include regolith, helium-3, water ice, metals, and solar power. The moon has a mass of 7.3e22 kilograms, and its surface has over 38 million square kilometers. It could be considered a nearly unlimited source of material. O'Neill envisioned using lunar material to build his space stations in the 1970s [O'Neill 1976]. The availability of these resources on the Moon could make it an attractive location for the construction material of a space station, as it could reduce the cost and logistical challenges of launching materials from Earth.

3.4.2.3   Asteroid Building Material

There are many large near-Earth asteroids with enough material to construct space stations where each could house one million people. There are over 22,000 near-Earth asteroids, and over 5000 have a lower-cost orbit. The asteroids of the solar system have sufficient material to build enough space stations to house more than the world's population. Asteroids represent another nearly unlimited resource that can be used to develop many large space stations.

3.4.2.4   Asteroid Composition

In the article *Processing of Asteroidal Material* [Mazanek et al. 2014], the author identified three main types of asteroidal materials for In-Situ Resource Utilization (ISRU): volatiles, free metals, and bulk dirt. Volatiles are simple compounds of hydrogen, oxygen, carbon, sulfur, and nitrogen. Free metals are iron, nickel, magnesium, and platinum group metals. Many of these metals are in the form of oxides or sulfides. Bulk dirt is the oxides, sulfides, and carbon materials.

The paper [Jensen 2023] reviewed key points from the asteroid composition analysis. Figure 3-17 provides the estimated abundance and composition for the three major asteroid categories. S-Type (Stony or Siliceous) and C-Type (Carbonaceous) asteroids have similar composition of materials [Ross 2001]. The M-Type (Metal) asteroids are quite different in composition. Key takeaways from Figure 3-17 are that C-Type or S-Type are most common in the solar system, and those asteroids are comprised primarily of oxides.

Oxides of asteroids are used to produce anhydrous glass tiles for our examples and analysis. Glass made in the absence of hydrogen or water has significantly better mechanical properties [Prado and Fraser 2018]. They note this glass would be competitive or superior to metals [Carsley, Blacic, and Pletka 1992]. Rapidly cooling the glass should increase its strength [Yale 2013]. It may be possible to substitute this glass for structural metals [Blacic 1985].

3.4.2.5  Example Processing

Most asteroid mining studies focus on retrieving and processing metals and/or volatiles. Supporting research provides optimism that asteroid oxides can be used as building materials. It appears that these oxides could be formed into anhydrous glass.

Table 3-5 contains a set of high-level process steps to convert asteroid regolith to building material. It accounts for asteroid porosity, processing losses, and asteroid composition. This is for an S-Type asteroid called Atira. Its volume is estimated at (2.19e10 cubic meters) and its mass at 41 trillion kilograms. Using the mass and volume, analysis found the asteroid has a porosity of 48.3% and assumes there is a 24.4% loss when processing the oxide. Table 3-5 contains these processing values and metrics. The table also includes the amount of material required to build an example Atira space

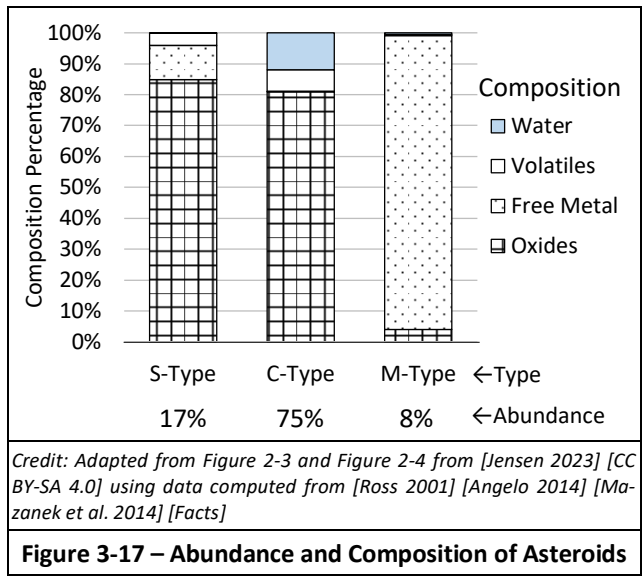

Credit: Adapted from Figure 2-3 and Figure 2-4 from [Jensen 2023] [CC BY-SA 4.0] using data computed from [Ross 2001] [Angelo 2014] [Mazanek et al. 2014] [Facts] station. Harvesting only the top 14% of the asteroid would provide enough building material for the station. This volume of harvested material represents 30% of the total volume of the Atira asteroid. Again, additional details on this approach are in [Jensen 2023].

**Figure 3-17 – Abundance and Composition of Asteroids**

| Table 3-5 – Asteroid Material Processing Example | | |
|---|---|---|
| **Material** | **Metric** | **Volume (m³ millions)** |
| Atira | 1928 meters radius | 21,900 |
| Harvested | 269 meters deep | 6,572 |
| Packed | 51.7% | 3,397 |
| Loss | 24.4% | 829 |
| Regolith | Processed | 2,569 |
| Building | 84.9% | 2,181 |
| Metals | 11.1% | 285 |
| Volatiles | 4.0% | 103 |
| Station | Calculated | 2,181 |

3.4.2.6  Summary

Construction material for a space station could come from multiple sources. Earth material is the most accessible today but expensive to launch into space. Multiple researchers have recommended using lunar material for constructing space stations. Asteroids have a similar composition to the moon and can also be used for constructing space stations. Each source has its own set of advantages and detriments. Our research focuses on using asteroids for the construction material.

*3.4.3 Material Strength*

Historically, space station designs have ranged from tiny cans to huge habitats. With the near-unlimited resources of a large asteroid, one wonders how large to make the space station. Table 3-6 replicates maximum station radius values from [O'Neill 1974] and [McKendree 1995]. The table extends their results with additional materials, including anhydrous glass.

The exterior station shell could be over 20 kilometers in radius using melted asteroid material (basalt rods). The table includes anhydrous glass data. Multiple sources are available to review this anhydrous glass [Blacic 1985] [Carsley, Blacic, and Pletka 1992] [Bell and Hines 2012] [Soilleux 2019]. The tensile strength of anhydrous glass suggests a station of over 120 kilometers could be feasible. Again, more background and details on these materials are in [Jensen 2023]. Realistically, the solar-powered, somewhat primitive manufacturing techniques will not produce perfect results. Such anhydrous glass might only produce a station radius of 26.8 kilometers. Our estimates include a filled shell structure using anhydrous trusses filled with crushed regolith. The structure does have lower tensile strength; however, the low density increases the station radius to 21.1 kilometers in Table 3-6. Initially, it seems prudent to aim for a much smaller radius, closer to 3 or 4 kilometers. One could reinforce these structures by using cables created from surplus metal or basalt fiber.

| Table 3-6 – Materials and Space Habitat Radius | | | |
|---|---|---|---|
| Material | Tensile Strength (MPa) | Density (g/cm3) | Radius (km) |
| Molecular Nanotechnology | 50,000 | 3.51 | 343.9 |
| Anhydrous Glass (max) | 13,800 | 2.70 | 123.4 |
| Basalt fiber | 3,000 | 2.67 | 27.1 |
| Anhydrous Glass | 3,000 | 2.70 | 26.8 |
| Basalt rod (7mm) | 2,471 | 2.79 | 21.4 |
| Filled Structure | 1,500 | 1.72 | 21.1 |
| O'Neill Future | 2,068 | 3.12 | 16.0 |
| Titanium | 1,450 | 4.50 | 7.8 |
| Steel | 1,240 | 7.80 | 3.8 |
| Aluminum | 352 | 2.65 | 3.2 |
| Iron | 275 | 7.20 | 0.9 |
| Glass | 7.0 | 2.50 | 0.1 |

*Credit: Self-produced in [Jensen 2023] using data from [O'Neill 1974] [McKendree 1995] [Bell and Hines 2012]; [Facts].*

This section focused on the station's mechanical stresses. The rotating space station uses anhydrous glass for its panels and trusses, which provides strength and thermal and radiation resistance. Table 3-6 shows the strength of anhydrous glass. Unlike many other materials, when exposed to radiation, anhydrous glass generally maintains its structural integrity and does not break down.

### 3.4.4 Radiation Protection

The outer shell has a thick layer of regolith to provide shielding from radiation and debris. Historically, researchers noted that shielding results in a substantial structural shell weight penalty. As a reference, a 1979 NASA paper considered radiation shielding and offered requirements and analysis for the various geometries [Bock, Lambrou, and Simon 1979]. They considered both attached shielding, which is integral to the design, and unattached stationary shielding, which has a separate living area rotating inside. As another reference, a different author stated that the material strength of attached shielding would not support a rotation rate greater than 1 revolution per minute [Graem 2006]. This implies that stations must be larger than 900 meters in radius to have an integrated attached shield. Even for thin-shell single-floor designs, one quickly finds that shielding mass dominates the station structural mass [Bock, Lambrou, and Simon 1979]. The shielding is 100 to 1000 times greater for smaller stations than the habitat interior structural mass. For large stations, the structural mass of the shielding is only 10 times greater than the station's habitat portion (multiple floors). The external shell and shielding typically dominate the mass for modest-sized, multiple-floor stations. With very large multiple-floor stations, other components of the station begin to equal or exceed the outer shell (shield) mass.

### 3.4.5 Station Stress

The stresses that space station materials must endure include mechanical, thermal, and radiation stresses. Mechanical stresses are caused by the weight of the station and the forces that act upon it. The station materials must be strong enough to support the weight of the station and its inhabitants, the centrifugal forces generated by rotation, and the forces when the station is maneuvered or docked with another spacecraft. Thermal stresses result from the extreme temperature variations in space. Radiation stresses come from the exposure to cosmic radiation and solar flares.

The stresses produced in the rotating space station were covered in [Jensen 2023]. For context, this subsection reviews that material. Stresses are produced from air pressure and from centripetal forces. Figure 3-18 compares these stresses in a rotating cylinder. The logarithmic x-axis shows the rotation radius from 100 meters to 20,000 meters. The logarithmic y-axis shows the stress ranging from 1 kilopascal to 1 gigapascal. The cylinder shell is 20 meters thick. The shell rotates at a speed that produces one Earth gravity on the outer rotation radius. The shell centripetal hoop stress, $\sigma_c$, is the largest and is equal to $g\rho R$ where g is the centripetal gravity, $\rho$ is the material density, and R is the rotation radius. In this design, the circumferential air pressure, $\sigma_a$, is the next largest and is equal to $P R/t$ where P is the internal air pressure, t is the shell thickness, and R is the rotation radius. The chart uses the furnishing stress values from the 1977 NASA study [Johnson and Holbrow 1977]. The radial centripetal force and the radial air pressure force are both minimal in the overall stresses in the cylinder; see Figure 3-18. A truss-like structure made of anhydrous glass filled with crushed regolith is assumed. The tensile strength of this structure is defined as 1500 MPa, and the density is set to 1720 kilograms per cubic meter [Jensen 2023].

Figure 3-19 presents the working stress for various materials. Again, these stresses were covered in [Jensen 2023]. For completeness, this subsection includes text and graphs from

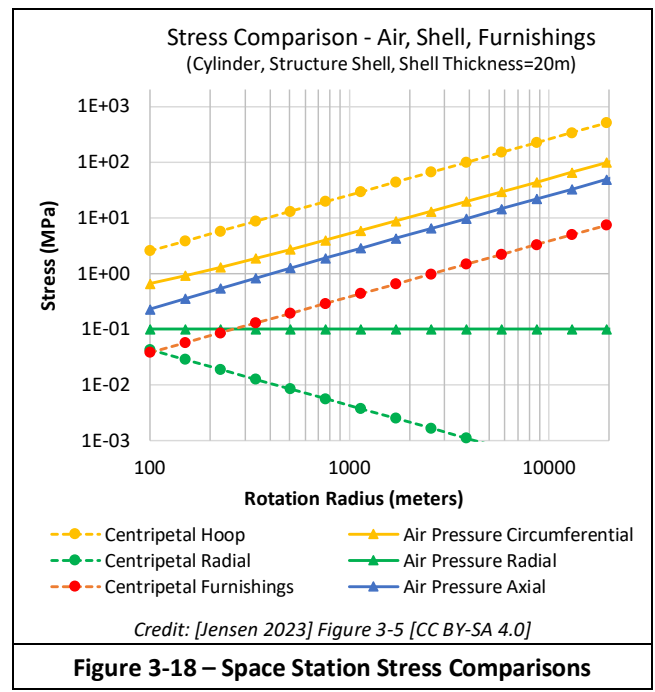

*Credit: [Jensen 2023] Figure 3-5 [CC BY-SA 4.0]*

**Figure 3-18 – Space Station Stress Comparisons**

that review. The working stress is the sum of the stresses introduced in Figure 3-18 and includes the stresses from the air pressure, the centripetal forces on the shell, and the centripetal forces on the internal structures and furnishings. It shows the working stresses in megapascals ranging from 1 to 20,000 on the logarithmic y-axis. The x-axis shows the outer rim radius of the rotating station, ranging from 100 to 100,000 meters. The station is rotating at a speed to produce 1g at the outer rim. The chart includes the material stresses for four materials. The anhydrous glass has the largest tensile strength, and aluminum has the smallest. Steel, with the highest density, creates the largest working stress. With the lowest density, the filled structure creates the smallest working stress. All the materials support their working stress below the rotation radius of 10,000 meters. This radius is used as a maximum value in many parts of this paper because of this material strength analysis. Considering the previous background and these results, structures have the potential to increase this rotation radius to 60,000 meters, and anhydrous glass could increase the radius to 100,000 meters.

As originally developed in [Jensen 2023] and shown in Figure 3-19, the material densities and tensile strengths affect the working stress. Figure 3-20 shows the effect of the shell thickness on the station stresses. It shows the two largest stresses from Figure 3-18 – the shell centripetal stress and the air pressure stress. The chart shows the stresses on the y-axis and the shell thickness on the x-axis. The chart includes data from a torus station and a cylinder station. The torus is like the Stanford Torus, with a major radius of 830 meters and a minor radius of 65 meters. The cylinder is like the O'Neill C-3 design, with a radius of 1000 meters. With a higher ceiling and thicker atmosphere, the cylinder air pressure stress is greater than the torus air pressure stress.

The cylinder has standard sea-level air pressure at the outer rim. In both designs, the air pressure stress decreases with the increasing shell thickness ($\sigma_{air}$ is equal to P R/t). The Stanford torus was designed to use ½ standard sea level air pressure; as such, its air pressure marker is lower than the torus air pressure line. The Stanford Torus has a thin shell of 1.68 centimeters [Johnson and Holbrow 1977]. The C-3 cylinder is designed with a thick shell of 20 meters. The centripetal shell stress is slightly greater for the torus than the cylinder. This is because the cylinder uses a filled structure shell, and the torus uses an aluminum shell. The centripetal stress is proportional to the density, and aluminum is denser than the anhydrous structure. Additional analysis is recommended; however, this brief review of structure stress suggests that large stations with thick shells are viable.

*3.4.6 Material Limitations Summary*

This subsection overviewed limitations on material for the space station. It covered the availability, strength, and desired characteristics. Materials to build a space station are available from the Earth, the Moon, other planets, and asteroids. The cost to launch from Earth makes other sources more desirable. A simple analysis of the material strengths shows that large stations can be built. Stations with a radius of 10,000 meters seem structurally feasible. It may be possible to create stations with a radius of 60,000 meters. The desired construction materials for rotating space stations must be strong, lightweight, durable, and resistant to the harsh conditions of space. Given the background research and these results, we advocate that anhydrous glass from asteroid oxides is an obvious choice for constructing large stations.

## 3.5 Station Components

This section introduces the major components of our stations. These components are essential parts of the space station. They do not represent a completed station; however, they provide a more accurate evaluation than just a thin shell.

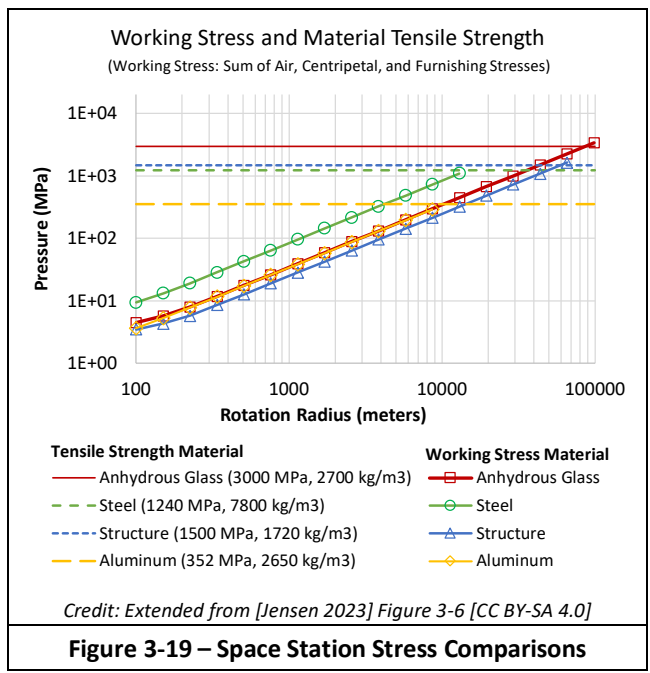

**Figure 3-19 – Space Station Stress Comparisons**

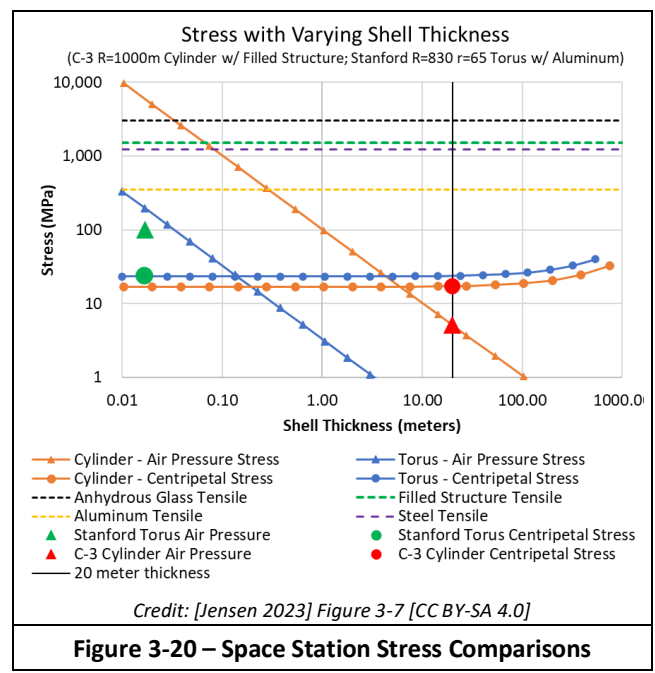

**Figure 3-20 – Space Station Stress Comparisons**

Simple geometries represent these components and are combined to model the space stations. This section reviews the refined station geometries, introduces the station components, and previews some differences between small and large stations.

*3.5.1 Refined Space Station Geometries*

The previous section and subsections have reviewed and extended the features of large space stations. These features included station geometries, multiple floors, mass, rotational stability, and population. They also reviewed features of the station such as the surface area allocation to living quarters, public areas, industry, and agriculture.

The set of typical space station geometries to address rotational stability and gravity issues has been refined. Figure 3-21 introduces those refined geometries and shows line drawings of them. This is an extension and refinement of the earlier work in [Jensen 2023]. All the geometries include multiple floors and components in their design. Spherical stations become ellipsoidal stations; long cylindrical stations become short hatbox stations; circular cross-section torus stations become elliptical cross-section torus stations; and dumbbell structures are doubled. Each of these refinements addresses specific limitations, and they are introduced in Figure 3-21. The following subsections provide the components and mass equations for these refined geometries to support our analysis.

*3.5.2 Station Components*

Each of the space stations in Figure 3-21 are composed of multiple simple geometries for modeling. Table 3-7 includes equations for seven component geometries. These equations compute the volume of those components. The mass of these components can be calculated with these equations and the density values from Table 3-8. The densities of structures were computed using their composite materials and dimensions. For example, the floor densities assume anhydrous glass beams, trusses, and panels with air-filled spacings of 5, 10, and 15 meters. Most of the components exclude the air density. The air volume is computed inside the station shell. The air density varies with height above the rotating outer floor. Including it in the other components would double count that mass.

The following paragraphs describe each of the four refined space station geometries and their major components as an overview. The station stability analysis in [Jensen 2024s] includes more details, illustrations, and equations.

**Hatbox Cylinder Station**: The cylinder station is decomposed into 7 components: outer shell, end caps, air, floors, main floor, spokes, and shuttle bay. These components are modeled with multiple geometric primitives from Table 3-7. As an example, the outer shell of the cylinder in Figure 3-21 has a radius of R and a length of L.

**Oblate Ellipsoid Station:** The ellipsoid station is decomposed into 6 components: the outer shell, air, floors, main floor, spokes, and shuttle bay. These components are modeled with multiple geometric primitives from Table 3-7. As an example, the outer shell of the ellipsoid in Figure 3-21 has an axis of rotation with length c. The radial axes distances are equal and have lengths a and b.

**Elliptical Torus Station:** The elliptical torus station is decomposed into 7 components: the outer shell, air, floors, main floor, dividers, spokes, and shuttle bay. The torus has a rotation radius of R, and the elliptical tube has a cross-section radial axis of length a and an axial axis length of c. The torus in Figure 3-21 does not show the spokes nor the shuttle bay.

**Double Dumbbell:** The double dumbbell station is decomposed into 6 components: outer shells, air, floors, main floor, spokes, and shuttle bay. Again, these components are modeled with multiple geometric primitives from Table 3-7. Figure 3-21 shows a line drawing of a double dumbbell. The main floor of the dumbbell node is at the major radius R. The axial axis c of the dumbbell's prolate ellipsoid node is parallel to the axis of rotation. The other radial axes, a and b, are equal in length and perpendicular to the axis of rotation. The double dumbbell provides rotational stability, whereas the single dumbbell is rotationally unstable [Jensen 2023s]. The double dumbbell has four nodes, two sets of spokes, and one shuttle bay.

| HATBOX CYLINDER | OBLATE ELLIPSOID | ELLIPTICAL TORUS | DOUBLE DUMBBELL |
|---|---|---|---|
| 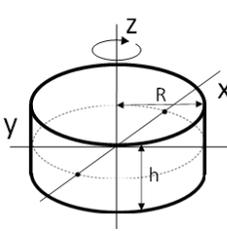 | 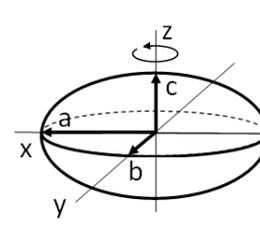 | 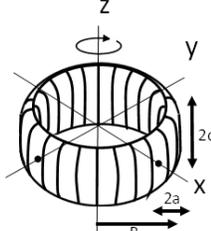 | 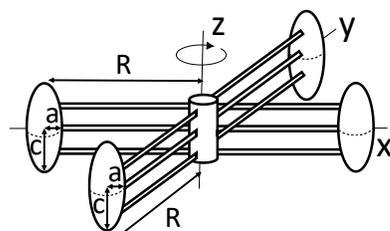 |
| Shortened to provide passive rotational stability | Sphere reduced to provide passive rotational stability | Spherical cross-section expanded to increase population | Single dumbbells can be doubled to provide passive rotational stability |

**Figure 3-21 – Refined Space Station Geometries**

| Table 3-7 – Volume Equations for Geometries | | |
|---|---|---|
| Geometry | Volume for Solid Geometry | Volume for Thick Shell Geometry |
| Cylinder<br>Length: L Radius: r | $V = L\,pi\,R^2$ | $V = L\,pi\,(r_o^2 - r_i^2)$ |
| Ellipsoid<br>Axes: a, b, c | $V = \frac{4}{3}\,pi\,a\,b\,c$ | $V = \frac{4}{3}\,pi\,(a_o\,b_o\,c_o - a_i\,b_i\,c_i)$ |
| Oblate Ellipsoid<br>Polar Axis: c<br>Other Axes: a=b | $V = \frac{4}{3}\,pi\,c\,a^2$ | $V = \frac{4}{3}\,pi\,(c_o a_o^2 - c_i a_i^2)$ |
| Sphere<br>Radius: r | $V = \frac{4}{3}\,pi\,r^3$ | $V = \frac{4}{3}\,pi\,(r_o^3 - r_i^3)$ |
| Torus<br>Major Axis: R<br>Minor Axis: r | $V = 2\,pi\,R\,pi\,r^2$ | $V = 2\,pi\,R\,pi\,(r_o^2 - r_i^2)$ |
| Elliptical Torus<br>Major Axis: R<br>Minor Axes: a,c | $V = 2\,pi\,R\,pi\,a\,c$ | $V = 2\,pi\,R\,pi\,(a_o\,c_o - a_i\,c_i)$ |
| Rod<br>Length: L Radius: r | $V = L\,pi\,r^2$ | $V = L\,pi\,(r_o^2 - r_i^2)$ |

*Inner dimensions are labeled with subscript "i". Outer dimensions are labeled with subscript "o". Outer dimension – Inner dimension = "t" (thickness)*

| Table 3-8 – Densities of Station Components | | | |
|---|---|---|---|
| Density | Value (kg/m3) | Density | Value (kg/m3) |
| $\rho_{steel}$ | 7850 | $\rho_{fill}$ | 1721 |
| $\rho_{basalt}$ | 2790 | $\rho_{spoke}$ | 337.4 |
| $\rho_{rod}$ | 2790 | $\rho_{floor5}$ | 32.2 |
| $\rho_{panels}$ | 2790 | $\rho_{floor10}$ | 17.0 |
| $\rho_{aluminum}$ | 2650 | $\rho_{floor15}$ | 11.2 |
| $\rho_{bay}$ | 2291 | $\rho_{air}$ | 1.5 |

### *3.5.3 Station Design*

This subsection previews details on the components and their models. These models are used to compute the station mass, and their masses are used to compute the component moments of inertia (MOIs). The MOIs are used to compute the station stability. The mass and stability analysis are introduced in the following sections. Again, additional details and illustrations of the components are in our Space Station Stability paper [Jensen 2024s].

#### 3.5.3.1 Shell

Each geometry station has its own unique exterior shell. The shells typically use a shell thickness of 20 meters. The outer shell has a thick layer of regolith fill to protect residents from radiation and debris. A truss framework of the shell provides most of the structural integrity. Ten-meter walls would be sufficient to provide radiation protection. We prefer designs with greater shell thicknesses to provide additional rotational inertia and collision integrity. The 20-meter-thick shell is the default thickness in our designs. Thinner shells are appropriate and often used for smaller radius stations.

#### 3.5.3.2 Shuttle Bay

A shuttle bay is included along the rotation axis in each station geometries. They would serve as the arrival and departure area for products, personnel, and tourists. The shuttle bays have multiple docking bays, service areas, and multiple floors. Analysis at this level of detail supported the density analysis. The shell and endcaps of the shuttle are modeled as filled cylinders and disks. To support multiple shuttles, the shuttle bay can become quite large. Its standard size is a 360-meter diameter and 400-meter height cylinder. The shell is 20 meters thick. Smaller stations use smaller dimensions and thicknesses in their shuttle bays. These smaller-scaled shuttle bays use less mass. That saved mass can be used for increased station living quarters. Larger stations reach the standard-size shuttle bay dimensions. The shuttle bay opening is accessible to shuttles incoming on the rotation axis.

The shuttle bay is divided in half, with each half placed near the outer polar poles. A spoke connects the two halves of the shuttle bay for stability, structural strength, and other functions.

#### 3.5.3.3 Spokes

Spokes connect the shuttle bay to the outer rim of the rotating station. These spokes serve multiple purposes including structural integrity, space for elevators, living space, and air circulation. The modeled spokes of the station use an outer filled thick shell cylinder and an inner hollow structure. The computed mass of each thick shell cylinder uses $m_{spoke} = \rho_{spoke}\,pi\,L\,(r_o^2 - r_i^2)$. The exterior cylinder is a filled structure, and the interior cylinder typically contains space trusses. The dimensions vary with the different geometries. The spoke radii are set to a fraction of the axial station length. The spokes have a minimum radius of 10 meters and a maximum of 100 meters. The spoke thicknesses are typically 5% of the spoke radius.

#### 3.5.3.4 Main Floor

The main floor is the top floor of a multiple-floor station. The main floor is positioned to create a habitable gravity range; see *§3.1.2 Geometry Gravity Ranges*. In large stations, the top floor is also limited by the habitable air pressure; see *§3.2 Air Pressure Limits*. The top floor is thicker than other floors in the station. This top floor provides open aesthetics for psychological well-being. This thick floor would include topsoil on small hills and valleys. The main floor thickness is set to 1/20th the rotation radius of the station shell. The main floor thickness is set to a maximum of 20 meters for large stations. The main floor thickness is set to a minimum of 2 meters for small stations. The thick main floor model computes the mass using $m_{main} = \rho\,pi\,L\,(r_o^2 - r_i^2)$ where $r_o$ is the outer radius of the main floor, and $r_i$ is the inner radius. In the cylinder model of the main floor, the main floor length is represented by the variable L and is equal to the shell width.

#### 3.5.3.5 Multiple Floors

The multiple-floors component is divided into many cylinders. The multiple floors are designed at 5 meters apart. The multiple-floors component is modeled with a series of concentric cylinders that are thinner and closer together. They extend between the main floor and the outer rim. The mass and MOIs of the floors are computed using these concentric cylinders. Like the previous main floor model, each cylinder would have a mass of $m_{floor5} = \rho_{floor5}\, pi\, L_n\, (r_o^2 - r_i^2)$. The lengths of the cylinders, $L_n$, are bound by the station shell. The length, $L_n$, would be a constant in a cylinder station. The lengths would vary with height in the other geometries. The density of the multiple floors, $\rho_{floor5}$, is in Table 3-8.

3.5.3.6 Air

The air component MOI of the station is computed as a low-density solid. The mass and MOIs of the air component use a series of concentric cylinders. Like the previous multiple-floor model, each cylinder of air would have a mass of $m_{air} = \rho_h\, pi\, L_n\, (r_o^2 - r_i^2)$. Again, the length of the cylinder, $L_n$, is bound by the station shell. The length, $L_n$, would be a constant in a cylinder station and would vary with height in the other geometries. The air density, $\rho_h$, decreases with increasing height; see *§2.2 Air Pressure Constraints*. The air density equations are used to compute the air density for each concentric cylinder. Summing the thin concentric cylinders produces accurate estimates even with varying lengths of the air cylinders. The station is pressurized so the outer rim has a sea-level air density of 1.225 kg/m3.

3.5.3.7 Divider

Only the torus design has dividers. The dividers separate the torus tube into airtight compartments as a part of a fail-safe design. A divider is comprised of two vertical walls with a floor structure between them. A divider is placed at each of the station radial spoke locations. In case of a catastrophic impact and loss of atmosphere, only one section of the torus tube would be lost. The divider would have a center interior structure built from trusses and panels. The interior includes floors with a spacing of 5 meters. This divider has two vertical walls filled with regolith and are 5 meters thick each. The thickness of the interior structure and the exterior walls sum to the diameter of the station spoke. In the example torus design, the divider also includes another 10 meters of floor structure on its exterior walls. The mass of the divider is computed using different densities for the three types of walls. The mass of the interior divider structure would be the divider ellipse area (pi a c) times the structure thickness times the multiple floor density. The wall masses are summed to compute the divider mass.

*3.5.4 Small and Large Stations*

Previous subsections described our standard dimensions for the station components. We take engineering liberties and adjust those dimensions in very small and large stations.

Figure 3-22 and Table 3-9 capture some of the design changes between small and large dumbbells. One of the two graphs shows a dumbbell with a radius of 200 meters, and the other shows a dumbbell with a radius of 2000 meters. The axes and scales of the charts are different to better illustrate the station details. The illustrated differences help justify the need to adjust the dimension changes.

Both dumbbells use a shell with a thickness of 20 meters. The shell appears quite thin in the 2000-meter radius station and excessively thick in the 200-meter radius station. The internal space appears much more constrained in the small station. The small station would only hold 4 floors spaced 5 meters apart, and the large station would hold 40 floors. Both stations provide 0.95g on the main floor and 1.05g at the outer rim of the station. The small station would rotate at 2 rpm, and the large station at 0.65 rpm. For most people, neither would cause nausea from rotation effects on the inner ear.

Table 3-9 summarizes these metrics for the two dumbbells in Figure 3-22. Both dumbbells use the same shell thickness. The large dumbbell creates more stress on the spokes. The number, radius, and thickness of the spokes are increased in large-radius dumbbells. Without reducing the node mass, the spokes begin to dominate the total station mass and significantly reduce the size and population of the station. The bay is also reduced in size for the smaller stations. With a smaller station and population, a smaller bay supporting fewer

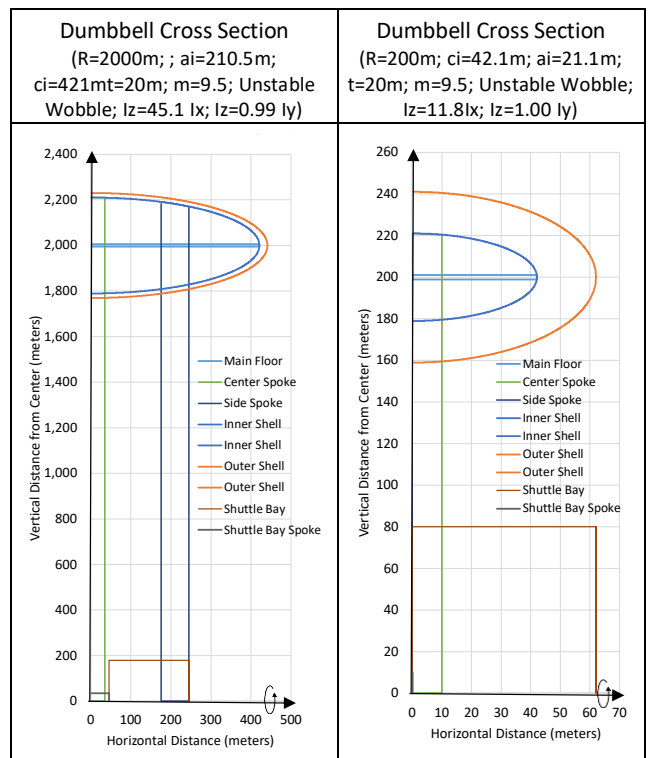

**Figure 3-22 – Large and Small Dumbbell**

| Table 3-9 – Dumbbell Size Comparison | | |
|---|---|---|
| **Size** | **Large Dumbbell** | **Small Dumbbell** |
| Radius | 2000 meters | 200 meters |
| Minor A | 210.5 meters | 21.1 meters |
| Minor C | 421.1 meters | 42.1 meters |
| Shell Thickness | 20 meters | 20 meters |
| Spokes | 3 | 1 |
| Spoke Thickness | 3.5 meters | 1.0 meters |
| Population | 107,605 people | 101 people |
| Mass | 1.2e11 kilograms | 2.6e9 kilograms |
| Half Bay | 90R x 200L meters | 62R x 124L meters |
| Rotation | 0.65 rpm | 2.06 rpm |

shuttles would be reasonable. Even with these reductions, the smaller bay in Figure 3-22 still appears to dominate the station structure visually.

A 10-meter-thick shell might be more appropriate for the small station. This would provide more material to increase the size of the nodes and the population. Using 8-meter-thick shells, the same mass of the small dumbbell (2.18e9 kg) would support a larger 251-meter radius dumbbell and increase the population to 202 people.

We adjust nearly all our component sizes for very small and very large stations. A more refined analysis of these structures is recommended. In the future, material strength analysis and finite element stress analysis will refine these station and component dimensions.

## 3.6 Station Mass

This section introduces the station and component masses. The following subsections provide examples of cylinder and torus stations. This analysis builds on the station and component analysis in the previous subsections. Mass details and trends are presented for the components with different station sizes.

### 3.6.1 Station Mass Background

The station mass uses the volumes of the station components. The masses of those components are derived from material densities and volumes of individual components. Different densities are used for the components and materials; see Table 3-8. The anhydrous tile density is assumed to be 2,790 kilograms per meter cubed. The asteroid densities vary depending on composition and porosity and range from 1,192 to 2,566 kilograms per meter cubed. The fill material uses that asteroid material, is packed, and includes melted regions for stability. The fill material density is set to 1,721 kilograms per meter cubed. This is in the density range of gravel. This is used as our fill structure density too. The densities of complete structures in the station are computed for some analyses. Table 3-8 contains example densities of the shuttle bay, spokes, and different floor spacings. The station's components use the sum of the volume and mass of their constituent pieces. Figure 2-5 illustrates the level of detail for this analysis. The mass includes the material needed to construct the trusses, to fill the exterior walls with regolith, and to cover the exterior and floors with panels.

### 3.6.2 Component Mass Introduction

Figure 3-23 provides the station components and their masses for several torus and cylinder stations. Both charts include masses for tori and cylinders with radii of 300, 1000, and 10,000 meters. They include the masses of the major components of the stations: shell, main floor, air, multiple floors, spokes, shuttle bay, and dividers (for the torus).

The y-axis of Figure 3-23a shows the mass on a logarithmic scale ranging from 1000 kilograms to 1 quadrillion kilograms. The x-axis shows the two station geometries and the three radii in six groups. The columns in each group represent the mass of the 6 or 7 station components. Figure 3-23b presents the same data organized with the x-axis showing seven groups of the station components. The columns in each group represent the masses of six stations. Figure 3-23b also includes the stations' total mass. Of course, the mass increases with the station radii. The cylinder mass is consistently greater than the torus mass.

The data in the charts presents several trends. With all 6 stations, the outer shell comprises much of the station mass. The shell mass scales with the radius of the station. In the smaller stations, the shuttle bay component is a significant part of the station mass. The shuttle bay has a fixed geometry and slightly scales with the largest stations. The air in nearly all the stations is the smallest mass component. The multiple floors and the air both scale as a volume function with the radius of the station. The mass of the multiple floors increases from a minor contributor to the total mass to a major contributor with the increasing radius.

The logarithmic scales in Figure 3-23 tends to obscure some of the details of the component masses. Figure 3-24 includes some of the same data and has a linear scale for the mass. The x-axes of the charts show the 6 and 7 major components of the cylinder and torus stations. Columns are included for each of the geometries and components. The charts show the torus and cylinder data for the 200-meter and 10,000-meter radii. With the linear scale, it is obvious that the mass of the air is very small compared to the mass of the other components. The charts also show that shells are the largest mass component in both these small and large station examples. The multiple floors contribute to the large cylinder station mass. The linear axes also show the significant contribution of the shuttle bay in small stations and its insignificant contribution in large stations. The two charts show that the total cylinder mass grows faster than the total torus mass.

### 3.6.3 Component Mass Preview

As the station dimensions become larger, the masses of the station components obviously increase. Before delving into the specific station geometries, this subsection introduces the changing mass allocation of the station components. This section previews the mass allocations of the cylinder station.

Figure 3-25 shows the component masses in a rotating cylinder station. The station has 6 components: a thick shell with endcaps, the main floor, multiple lower floors, air, spokes, and shuttle bay. The x-axis shows the radius using a logarithmic scale and ranges from 100 to 400 thousand meters. In the top chart, the y-axis shows the mass of those components in kilograms on a logarithmic scale. In the bottom charts of Figure 3-25, the y-axis shows the mass of those components as percentages of the total system mass. The charts on the left show results for a station with uniform air mass. The charts on the right show results with varying air density from the centripetal gravity.

The models for these mass simulations include three constraints: gravity, rotationally balanced, and air pressure. The gravity limits the floors to about the outer 14% of the radius. To maintain the rotational balance, the length of the cylinder varies from 1.2 to 1.9 times the radius length. The air pressure limits the floors to regions below the Denver air pressure (about 1600 meters). The data of Figure 3-25a assumes the station is fully filled with uniform-density air, but the top floor is still set to the Denver height. In actuality, a uniform air density does not limit the position of the top floor like the centripetal air density.

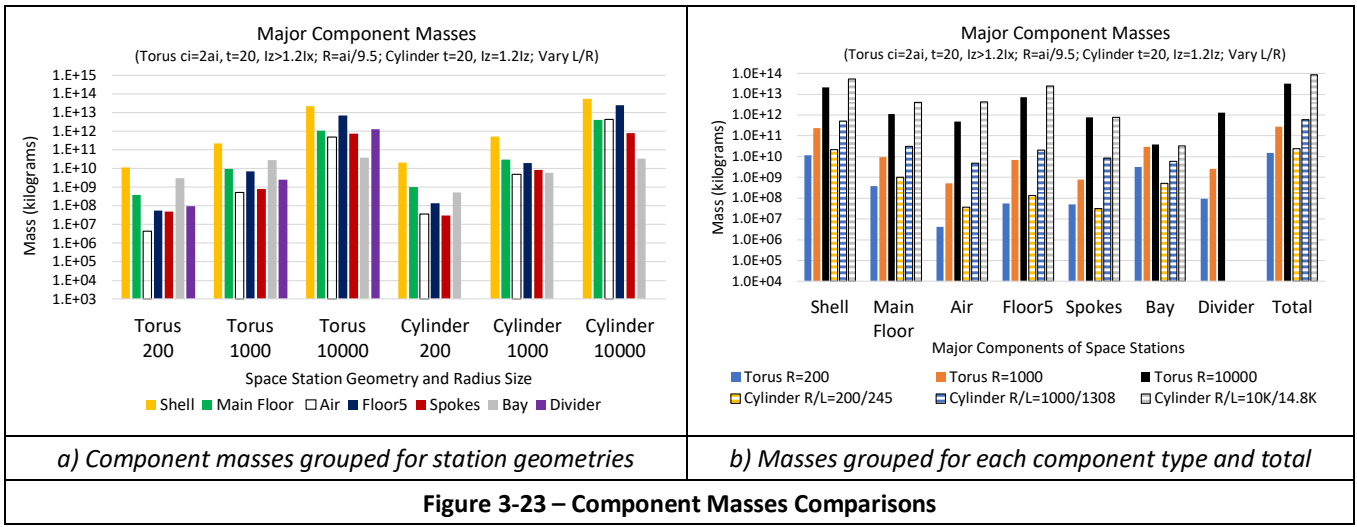

| a) Component masses grouped for station geometries | b) Masses grouped for each component type and total |

Figure 3-23 – Component Masses Comparisons

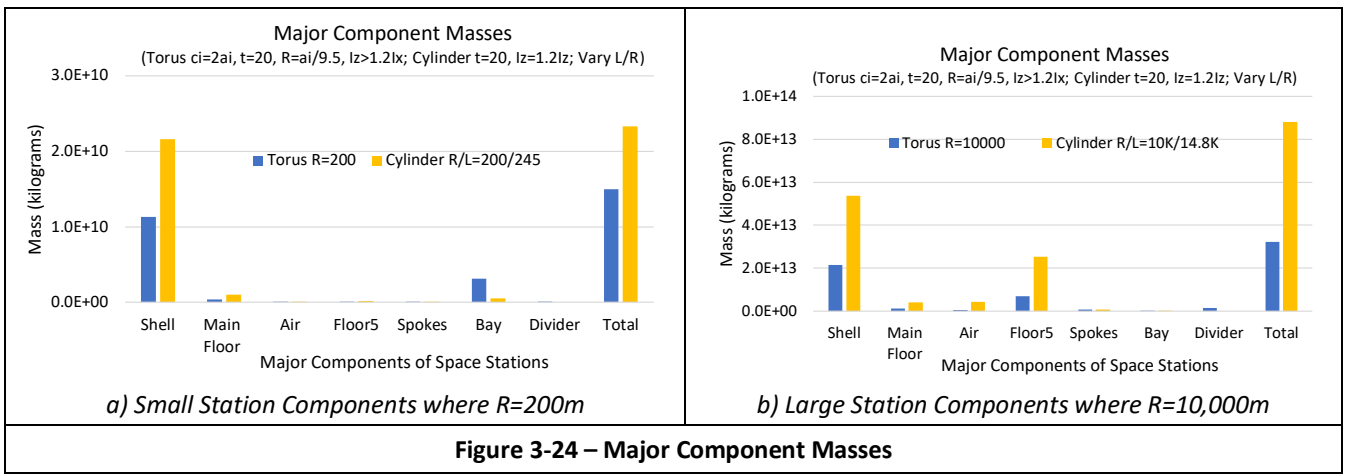

| a) Small Station Components where R=200m | b) Large Station Components where R=10,000m |

Figure 3-24 – Major Component Masses

The shell is the heaviest component in small stations. With increasing radius, the masses of the multiple floors and the air increase faster than the shell mass. The shell surface area of the station is proportional to the radius squared. The air and multiple floors are proportional to the radius cubed. The gravity limit constrains the multiple floors. They only extend over the outer 14% of the radius. Once constrained at 1600 meters (Denver air pressure), the multiple-floor mass is proportional to the station radius instead of the station volume. In contrast, the uniform air density component increases as a volume. Despite its low density, the uniform-density atmosphere becomes the largest mass in Figure 3-25a.

Multiple issues exist with using the uniform density of air as our model in the station. It does not represent the expected centripetal force pushing the air to the outer rim. It creates much more mass (and cost and inertia) than the centripetally pushed air. There is no air pressure limit to the floor location when assuming uniform air density. Given those issues, most of our investigation uses the air model where the air is centripetally pushed to the outer rim.

Figure 3-25b shows the results for the same station components as Figure 3-25a. In this chart, centripetal gravity pushes the air towards the outer rim. The axes and chart formats are the same in both charts in Figure 3-25. Only the air mass and the total mass significantly change in this chart compared to Figure 3-25a. As the air is pressed and accumulated on the outer shell of the cylinder, the chart shows that the atmosphere mass does not increase as it did in the previous chart. The air has a decreasing density and, with large radii, becomes proportional only to the inner surface area of the cylinder shell. The air becomes only a maximum of 8.5% of the total mass instead of 74% as shown in Figure 3-25a. The multiple floors only extend to the Denver air pressure limit, and their mass is again proportional to the radius. The lower multiple floors become the largest percentage of mass in the station at about 39%. This more accurately represents the large rotating station's expected air mass and floor count.

The charts in Figure 3-25 illustrate the changing components with increasing station radius. There are different constraints on the system. These constraints significantly affect the mass of the lower floors and the air. Habitable gravity range and air pressure both limit the floors to a small outer radius of the large stations. The centripetal gravity pushes the air to the outer rim. This limits the height of the habitable floors in the station. These results were generated for a cylinder space station; however, all the geometries rotate and have the same constraints on their components. Details for all the geometries are in [Jensen 2024s].

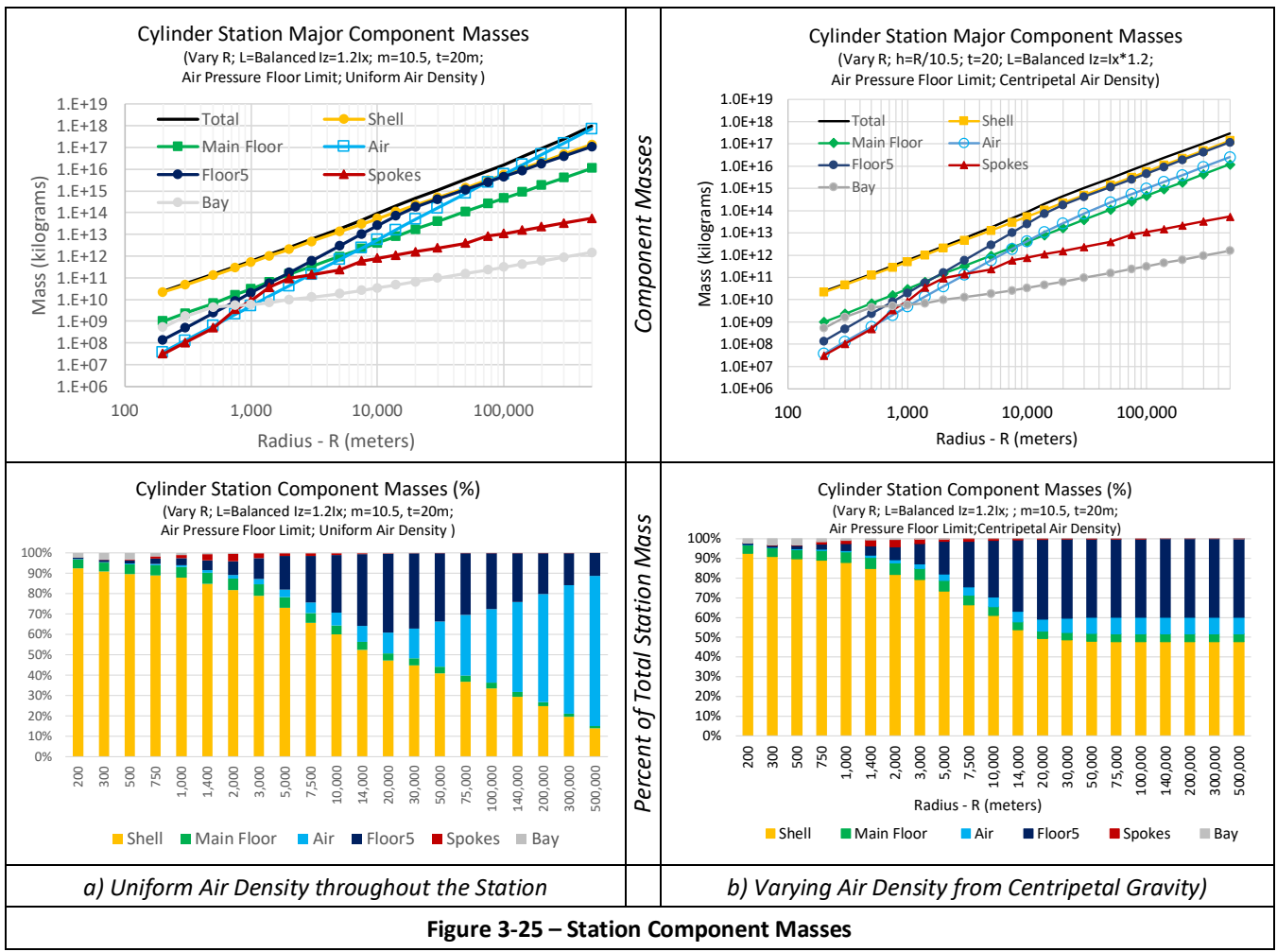

Figure 3-25 – Station Component Masses

## 3.7 Single Floor Rotational Stability

Rotational stability was introduced in *§2.4 Rotational Stability*. This section briefly overviews rotational stability to provide context relevant to large stations. A full analytic study on rotational stability is in [Jensen 2024s].

Stations rotate to produce an Earthlike gravity for the residents. This is important for human long-term health. Perturbations will cause rotating systems in space to eventually rotate about the axis with the greatest angular moment of inertia [Globus et al. 2007]. There is the risk that the changing axis of rotation would cause the station to tumble end-over-end catastrophically. Globus and his co-authors designed a cylinder station that would not have this risk [Globus et al. 2007]. The desired axis of rotation should be 1.2 times greater than the other rotation axes [Brown 2002]. This stability rule provides passive rotational stability. Other stability rules regarding the principle moments of inertia (MOIs) are in [Jensen 2024s]. We have extended the Globus cylinder approach [Globus et al. 2007] to the other geometries. This includes analysis of stations with thin and thick shells. Like their analysis, this analysis only considers the outer shell.

### 3.7.1 Assumptions

The stations are assumed to have a habitable gravity range. On the cylinder geometry, the single-floor model is on the outer shell and has a gravity of 1.0g. With the other geometries, the single floor ranges from 0.95g to 1.05g and limits the usable distance up the curved outer shell. As an example, as previously shown, the rotating radius will be 9.5 times the length of the habitable region to support that gravity range.

The single-floor model stability analysis assumed homogeneous outer shell densities to compute the MOI of the station. The MOI equations are used as an engineering estimate for analytic analysis; see [Jensen 2024s]. The analysis produced a set of closed-form equations to evaluate the stability. The analysis ignored other station components. This single-floor approach provides a quick approximation for initial stability analysis.

### 3.7.2 Single Floor Stability Results

This analysis aims to determine the geometry dimensions to provide passive rotational stability. Again, the MOI of the desired axis of rotation should be 1.2 times greater than the other rotation axes [Brown 2002]. The moment of inertia about that preferred axis of rotation is labeled Iz. The MOIs of the other two axes are equal for the cylinder, sphere, and ellipsoid station geometries. The other two axes are designed to have smaller MOIs than the z-axis. With dumbbells, the MOI of the rotation axis Iz is the same as Iy. Even though Ix and Iy are not equal, these matching Iz and Iy MOIs introduce an unsolvable passive stability problem.

Figure 3-21 illustrated our refined set of station geometries. Table 3-10 contains a summary of our single-floor stabilities from [Jensen 2024s]. The table includes stability equations for each of the four station geometry types. The following presents a very brief summary of that analysis. **Cylinders:** The shell consists of the exterior cylinder and flat endcaps. Like the analysis in [Globus et al. 2007], hatbox cylinders can be stable. **Ellipsoids:** The rotating shell of a sphere station is analytically unstable. Oblate ellipsoids can be stable. **Elliptical Torus:** Our analysis found that elliptical tori are stable for our design space where R≥6.33a and c=2a or c=3a. The torus shell did not include the inner docking station nor spokes. **Dumbbell:** Equal MOIs on rotation and radial axes cause this geometry to be unstable. We contacted an expert [Fitzpatrick 2023] and found that the instability will grow algebraically, and the rotational axis will wobble. A wobble will remain an open concern for the single dumbbell system.

### 3.7.3 Single Floor Stability Summary

Table 3-10 contains a summary of our current stability equations for single-floor analysis of the four station geometry types. These are stability metrics for the single-floor design with a thin shell. These equations represent moments of inertia for thin-shell hollow geometries. An introduction and this table were adapted from [Jensen 2023]. More analysis with results are in [Jensen 2024s].

## 3.8 Multiple Component Rotational Stability

*Section 2.4 Rotational Stability* introduced rotational stability. We extend the single-floor analysis to include multiple components. Previous sections have defined and analyzed the station components. This includes the geometry and mass of station components such as floors, shell, spokes, and atmosphere. The single-floor rotational stability analysis considered only the outer shell and ignored the other components. We extend the stability analysis to include more of the station components. The major components in the station have different densities, geometries, and dimensions. These components include the outer shell, multiple floors, main (top) floor, spokes, and air.

With these complex multiple-component designs, we first tried to use numeric analysis to compute the station inertia values. Closed-form equations were used for the mass and inertia. These became unyielding for some of the more complex components and motivated another approach to compute the rotational moments of inertia. Those complex components are decomposed into many small volumes, and the rotational MOIs and masses of all those individual volumes are summed. This can be considered a Riemann sum of the MOIs and masses of those volumes for those more complex components. Where possible, results were compared to analytic equation results.

Our space station stability paper [Jensen 2024s] provides more details on this multiple-floor analysis approach. The following subsections provide sufficient information to provide context and help understand the constraints from the multiple-floor stability on our large stations.

### 3.8.1 Multiple Component Stability Results

This subsection provides a preview to introduce and provide context for the multiple component stability limits on large stations. This includes the effect of air density on large station stability. Additional subsections show the effects from the component mass and the large station mass. Another subsection evaluates the shell thickness and its effect on the rotational stability. A summary of these results concludes this subsection.

### 3.8.2 Station Stability Preview

A set of guidelines and rules have been organized based on [Fitzpatrick 2011] [Globus et al. 2007] [Brown 2002] and [Birse 2000]. A key stability rule is that the moment of inertia (MOI) about the rotation axis should be greater than 1.2 times the other axes [Brown 2002]. Our analysis assigns the z-axis as the rotation axis and assumes the system would be stable when Iz≥1.2 Ix. The cylinder geometry is used as an example of this relationship.

Figure 3-26 shows the stability when including the multiple components and varying the cylinder geometry. The x-axis

| Table 3-10 – Geometries and Rotational Stability for Thin Shell Geometries ||||
|---|---|---|---|
| **Geometry** | **Key Stability Factor** | **Rotational Stability** | **Notes** |
| Cylinder | L < 1.3 R | Hatbox cylinders can be stable | Flat endcaps |
| Ellipsoids | c < 0.8165 a | Oblate ellipsoids can be stable | Sphere stations are not stable |
| Elliptical Torus | $R^2 > 1.5 (c^2-a^2)$ | Elliptical tori are stable where R>=6.33a and c=2a or c=3a | Torus only – inner docking station and spokes not included |
| Dumbbell | Iz/Ix=1.2 when $R^2 \geq \frac{1}{3}(1.2\ c^2 - 0.8\ a^2)$  Iz/Iy is less than 1.2 for all R. | Instability from equal MOIs on rotation and radial axes grows algebraically and the rotational axis will wobble. | A wobble will remain an open concern for the single dumbbell system. |
| *Credit: Adapted from Table 3-3 [Jensen 2023] [CC BY-SA 4.0] and by extending cylinder concepts from [Globus et al. 2007] [Facts]* ||||

shows the cylinder rotation radius on a logarithmic scale ranging from 100 to over 100,000 meters. The y-axis shows the stability as the ratio of the z moment of inertia (Iz) over the x moment of inertia (Ix). The minimum stability limit is shown as a dotted black line at 1.2 on the stability axis. Geometries with stability values greater than 1.2 are rotationally stable, and values less than 1.2 are unstable.

Figure 3-26 shows the stability of the cylinder geometry with multiple station components including the outer shell, endcaps, spokes, multiple floors, air, and a center shuttle bay. With the changing radius in our analysis, the geometry ratio (L/R) is varied to attain the stability ratio of 1.2. In the single-floor analysis, the thin shell analysis required L/R=1.3 for the cylinder to be stable (Iz/Ix=1.2). Given the thin shell L/R=1.3 and using multiple components, the chart data shows the stability of that L/R=1.3 station crosses below the 1.2 limit. This shows instability with the multiple components when the radius is less than 1000 meters and stable for larger radii.

The station geometry is varied to analyze the station's rotational stability. Geometries with longer lengths tend to be less rotationally stable. As an example, the geometry ratio of L/R=2.0 in Figure 3-26 has no cylinder radius sizes that would be stable. Geometries with shorter lengths (hatbox) are more rotationally stable. Figure 3-26 shows that all cylinders would be stable with the geometry ratio of L/R≥1.0.

### 3.8.3 Air Density and Station Stability

The varying air density affects the station's inertia and stability. It is possible to use a geometric model for the air inertia on the rotation axis (Iz) and other axes (Ix). The fully filled uniform air mass could be considered a solid (low-density cylinder), while the centripetal air mass could be considered a hollow cylinder. The MOIs of the two models are different and not necessarily accurate. Instead of using a single geometry model, one can compute the MOIs of multiple thin-layer cylinders. Each cylinder's density is calculated using the previous section's air pressure and density equations. The MOIs of the multiple thin cylinders are summed to obtain the air component MOI.

Figure 3-27 illustrates the effect of the centripetal air density on the station's stability. It shows results for a rotating cylinder station. The thickness of the shell is 20 meters. The x-axis shows the radius using a logarithmic scale and ranges from 100 to 1 million meters. The y-axes show the ratio stability metric. The y-axis shows a detailed range of the stability ratio from 1.0 to 1.6. Stability can be controlled by the ratio of cylinder length over radius (L/R). Consider assigning the L/R ratio to 1.3. In the thin shell model, this provides the stability of Iz/Ix=1.2 ratio (or better). For a fully filled station with uniform air density, the stability reaches a maximum at a radius of about 30,000 meters and then decreases. With the atmosphere centripetally gathered near the outer rim, the ratios of the inertias continue to increase with increasing radius. The aforementioned hollow cylinder model has a higher MOI than a solid cylinder model. In very large stations, the "hollow cylinder" of air improves the station's stability and is more representative. Figure 3-27 includes two main floor positions: one at the gravity floor limit and the other at the air pressure floor limit. It shows that the stability improves with more floors and mass on the outer edge.

### 3.8.4 Component Stabilities

Figure 3-28 is included to introduce the component stabilities in the system. This graph's data is from the components of an elliptical torus. The x-axis of the chart shows the rotation radius on a logarithmic scale ranging from 100 to 400,000 meters. The y-axes in Figure 3-28 show the components' stability ratio (Iz/Ix). The left axis shows most components' ratios and ranges from 1.75 to 2.00. The right axis shows the ratios for the total stability and the bay component. It ranges from 0.0 to 3.5. The two y-axes provide more detail for the different components.

As previously discussed, the dimensions and number of components vary with the station radius. These increases cause non-linear changes in the component stabilities. Each component has a different mass, so the total station stability ratio is not simply the sum of these individual component stabilities. Our stability analysis paper includes a mass-normalized comparison of these stability factors [Jensen 2024s].

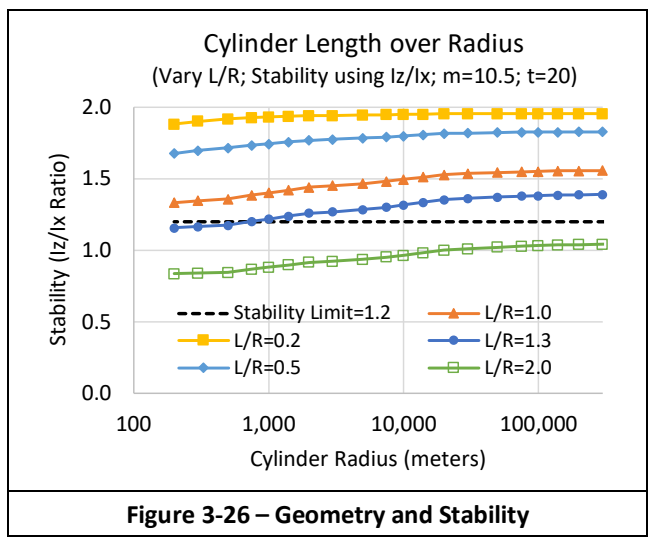

**Figure 3-26 – Geometry and Stability**

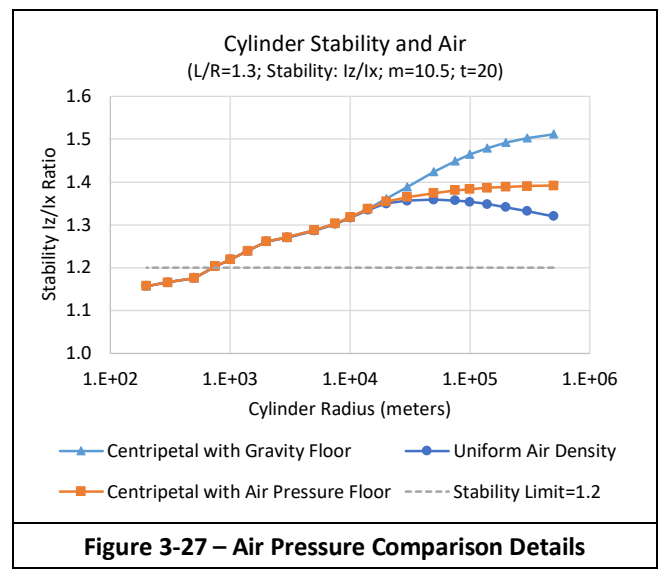

**Figure 3-27 – Air Pressure Comparison Details**

The shuttle bay component stability quickly drops as the two 200-meter length bays are separated on the rotation axis with a connecting shuttle bay spoke. With increasing station radius, more spokes are added to the station. They are added off the center and increase the x-axis inertia, Ix. The number of spokes increases at station radii of 8500 and 85,000 meters. The increased number of spokes causes the step function in the spoke stability. The spoke dimensions also increase with increasing station radius to provide more structural strength. The main floor, the air, and the multiple floors all have a stability Iz/Ix of about 1.95 until radii greater than 10,000 meters. Past the 10,000 meter radius, the air pressure top floor limit takes effect, and the stability of the floors becomes larger. With very large stations, there is an increase in the stability of the air component as more of the air is centripetally located towards the outer edge. The greater distance from the rotation center increases the rotational inertia and the stability. This summary of the individual component stability is covered in more detail in [Jensen 2024s].

*3.8.5 Shell Thickness and Stability*

Figure 3-29 shows the effect on stable cylinder geometries from varying the shell thickness. The legend shows the cylinder radii ranging from 300 meters to 10,000 meters. The y-axis shows the geometry ratio of length over radius (L/R). The changing ratio represents the changing cylinder length for each of the 5 station sizes. The logarithmic x-axis shows the shell thickness. The length L of the cylinder is set to provide passive rotational balance with Iz=1.2 Ix. For the cylinders, the length L ranges from 0.7R to 1.9R over the range of thickness and cylinder radii.

With a thin shell, the station geometry ratio approaches the stability of all the other components. Small radius stations require smaller length-to-radius ratios to be rotationally balanced. The shuttle bay at the station's center degrades the station's stability. Larger radius stations can be rotationally balanced with larger length-to-radius ratios. With thin shells, the large relative mass of the multiple floors far from the rotation axis improves the station's stability. In large stations, the multiple floors are the dominant mass and are on the outer edge of the rotating station. With very thick shells, the geometry ratio begins to decrease. The cylinder shell and endcaps cover the entire station (from the outer edge to the top poles). With very thick thicknesses, its stability counters the improvement from the floors' stabilities, and reducing the length restores that stability reduction.

*3.8.6 Stability Summary*

This section reviewed rotational stability. Our Station Stability paper [Jensen 2024s] covers more details on the single-floor and multiple-component rotational stability for all the station geometries. Each geometry has a design ratio that could be modified to produce a station with passive rotational stability. For context in this large station analysis, the following paragraphs provide summaries from that paper to describe each of those station stability ratios.

**Cylinder:** The stability of the cylinder is controlled by changing the cylinder length (L) and the cylinder radius (R). The thin shell analysis duplicated the Globus ratio L<1.29R [Globus et al. 2007]. The ratio decreases from L<1.29R for thick shells as its thickness increases. Shells with thicknesses up to 1% of the radius require a length-to-radius ratio (L/R) of 1.29. Shells with thicknesses greater than 1% of the radius permit longer-length cylinder sides. A shell with a thickness of 10% of the radius permits a ratio of L/R up to 1.35. With the multiple components, the cylinder stability was evaluated with radii between 200 and 500,000 meters. For those radii, the cylinder is rotationally stable with L/R between 1.2 and 1.9 compared to the 1.3 ratio for thin or thick shells.

**Ellipsoid:** The oblate ellipsoid station has two major axes that are the rotation radius length (a=b) and a third shorter polar minor axis with a length (c). The stability is controlled by changing the length of the polar axis or the rotation radius length. The thin shell ellipsoid is stable with Iz>1.2 Ix when c<0.8165a. The thick shell ellipsoid is stable when $c_o<0.775a_o$. Our analysis of the ellipsoid with multiple components also controls the stability by varying the c/a ratio.

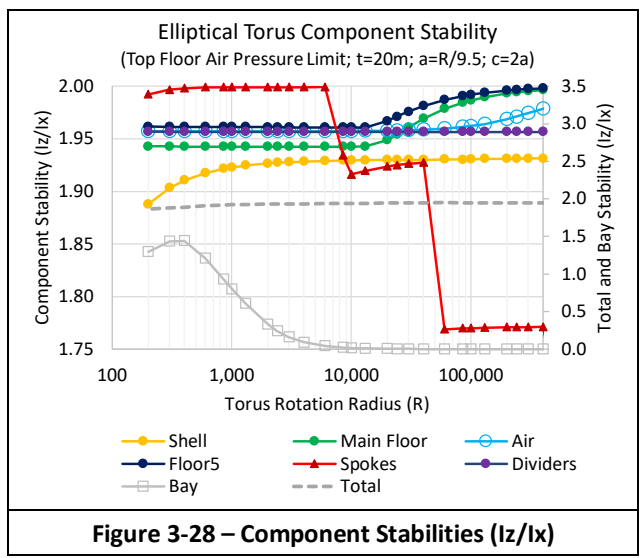

**Figure 3-28 – Component Stabilities (Iz/Ix)**

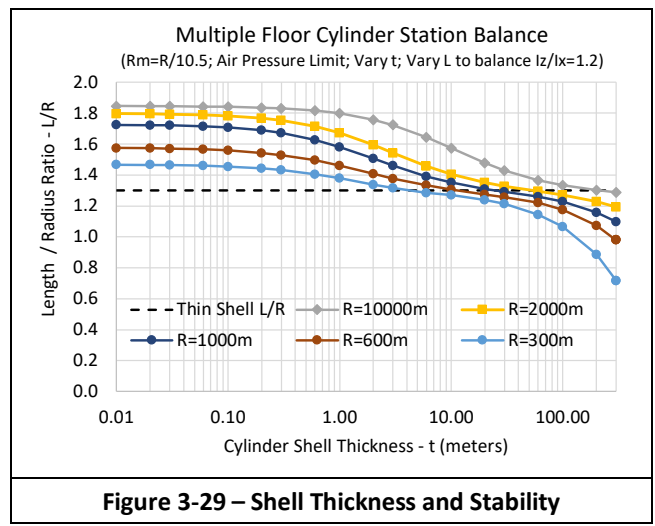

**Figure 3-29 – Shell Thickness and Stability**

With increasing station dimensions, the c/a ratio increases from 0.75 to over 2.25 to produce a rotationally stable station. Some of these ellipsoid stations can be spherical (a=c or ratio=1.0) and be rotationally stable.

**Elliptical Torus:** This torus has an elliptical cross-section. The torus has a major radius of R, and the elliptical cross-section has a radial axis a=R/m and a perpendicular axis c=2a or 3a. The m gravity scaling value defines the gravity range in the torus tube and is typically 9.5 or 6.33; see *§3.1.2 Geometry Gravity Ranges*. The stability is controlled by changing the cross-section or the gravity scaling ratio. Thin and thick shell tori are rotationally stable for a broad range of cross-section and gravity scaling ratios. With all components and a thick shell, the torus is also rotationally stable ($I_z/I_x>1.2$) for a broad range of cross-section and gravity scaling ratios. The cross-section ratio can vary from c=1a to c=6a. The gravity scaling ratio can be our R=6.33a or R=9.5a, and the station remains rotationally balanced.

**Dumbbell:** Dumbbell stations are not rotationally stable. Analysis found the stability ratio $I_z/I_x$ is much greater than 1.2 and would represent a stable system. The $I_z/I_x$ stability ratio ranges from 1.83 to 1.88, with the gravity range scaling set to a=R/6.33 and a=R/9.5. Unfortunately, the station's stability $I_z/I_y$ was found to be approximately 1.0, and this represents an unstable rotation. An expert explained that *"the instability with two equal moments of inertia is algebraic rather than exponential"* and that this unstable rotation *"will settle into a limit cycle in which the rotational axis wobbles slightly [Fitzpatrick 2023]."* This wobble will remain an open analysis concern for the single dumbbell system.

**Double Dumbbell:** The double dumbbell is rotationally stable. Spherical and ellipsoid nodes were considered. Analysis of the thin and thick shell double dumbbell models found that the stability ratios of $I_z/I_x$ and $I_z/I_y$ were nearly identical and greater than 1.7. This is significantly greater than the recommended stability limit of 1.2 [Brown 2002]. Double dumbbells are rotationally stable for our design space using multiple components. With increasing rotation radius, the stability ratio range is consistently greater than the 1.2 stability limit.

Table 3-11 is duplicated from [Jensen 2024s]. This summarizes the design ratios for the four geometries, which all have a broad design space for rotational stability. Additional details and analysis on the stability ratios and the geometries are included in [Jensen 2024s].

## 3.9 Station Constraints Results

This section has covered multiple constraints limiting the size and geometry of a rotating space station. This included limits from gravity, population, materials, geometry, air pressure, and rotational stability. Geometry limits for the station designs were reviewed. This section details the station mass and air pressure for those geometries. This section also overviewed rotational stability concepts. The following subsections present the results of those constraint effects on the station mass.

| Table 3-11 – Typical Stability Ratios | | | | |
|---|---|---|---|---|
| Geometry Analysis | Cylinder $I_z/I_x=1.2$ | Ellipsoid $I_z/I_x=1.2$ | Elliptical Torus R/a=9.5 | Double Dumbbell R/a=9.5 |
| Single Floor Thin Shell | L/R=1.29 | a/c=1.225 | $I_z/I_x$=1.92 c=2a | $I_z/I_x$=1.95 c=a |
| Single Floor Thick Shell | L/R=1.29 to 1.35 | a/c=1.29 | $I_z/I_x$=1.93 c=2a | $I_z/I_x$=1.98 c=a |
| Multiple Components | L/R=1.2 to 1.9 | a/c=0.75 to 2.25 | $I_z/I_x$= 1.86 to 1.96 c=2a | $I_z/I_x$=1.7 to 1.95 c=2a |

### 3.9.1 Station Radius to Mass

Only three station radii were presented in *§3.6.3 Component Mass Preview* in Figure 3-23 and Figure 3-24. Figure 3-30 presents the mass results as a function of the station radius. The x-axes on both charts show the radius in meters ranging from 100 to 40,000 meters. The y-axes on both charts show the masses of the components and the total station in kilograms ranging from 1e6 to 1e16. Both axes are logarithmic. Figure 3-30a shows the results for a cylinder station with the radius ranging from 200 to 40,000 meters. Figure 3-30b shows the results for a torus station with the radius ranging from 300 to 40,000 meters. They again show the mass of the 6 (or 7) station components and include the total station mass data.

The component masses increase with the radius of the station. The shell and bay components have the largest masses at the smallest station radius. As the radius increases, the shell mass increases rapidly, but the bay mass changes little. The cylinder shell and endcap thicknesses are set to a specific thickness value. Most of this paper's analysis uses a thickness of 20 meters. The torus and cylinder shells represent their outer surface area and both increase with the radius. The shuttle bay is a fixed-size cylinder model. As the radius increases, the cylinder is split into two parts and connected with a spoke-like structure. The bays in the cylinder reach the end caps of the cylinder station. The torus bays reach the outer edge of the side spokes. The air density varies with the height above the outer rim. In general, in the same radius cylinder and torus stations, the torus has less air than the cylinder. The torus tube encloses the air while the cylinder is filled to the rotation center.

There are multiple constraints on the main floor and the multiple floors. The air pressure floor limit limits the main floor and the multiple floors in the torus and cylinder. The cylinder length increases with the radius and maintains the rotational stability ($I_z/I_x$=1.2). The length ranges from 1.13R to 1.66R. The torus minor axis, a, is set to R/9.5 to provide habitable gravity from the outer edge to the center of the torus tube. The torus minor axis, c, remains at 2 times the minor axis, a.

Figure 3-30 shows that for small stations, the masses of the main floors and multiple floors of the torus and cylinder are similar; however, in large stations, the cylinder station masses become much larger than those of the torus. The torus top floor length is typically smaller than the cylinder by design. In addition, the curved sides reduce the effective width, area, and volume of the multiple floors. These differences reduce the mass of the torus main floor and multiple floors compared to the cylinder station.

At the 30,000 meter radius in Figure 3-30, there is a 5x difference between the combined masses of the main floor and multiple floors for the cylinder and torus. The cylinder has a top floor length of 48,142 meters, while the torus has a top floor length of 10,917 meters. In addition, the cylinder has more floor surface area (and volume and mass) with its flat endcaps compared to the curved shell of the torus.

The spokes of the cylinder and torus follow similar increases in mass with their increasing radius. The radius of the spokes range from 10 meters to 100 meters. Larger stations have multiple spokes. The shells are designed with materials to support their hoop stresses. The spokes provide space for elevators, housing, air circulation, and some additional structural strength.

### 3.9.2 Non-linear Mass Increases

Figure 3-30 showed that all of the components' masses increased with the station's radius. The mass of many components does not increase linearly from the smallest radius to the largest radius. It is not linear because our analysis varies the dimensions of structures (such as shells and floors) with the changing radius. These dimension changes provide additional support, strength, or radiation protection. Additional structures (such as spokes and towers) are also added to strengthen the large stations.

Figure 3-31 is included to illustrate some of these non-linear changes. The non-linear changes are most noticeable with the spokes and shuttle bay. This chart shows the changing dimensions of the shuttle bay and spokes in an elliptical torus. Other components exhibit less obvious changes. Previously Figure 3-28 showed similar non-linear changes with the stability (Iz/Ix) of the components. Figure 3-31 shows dimensions on the right and left y-axes. It shows the torus rotation radius on the x-axis ranging from 100 to 400,000 meters on a logarithmic scale. The radii and length measures are on the left y-axis, and the associated data is shown with dotted lines with circle markers. Thicknesses and counts are shown on the right y-axis, and the associated data are shown with solid lines and cross markers. The varying radius affects these components. The chart shows the changing count as extra spokes are included. The number of spokes shows two increases from 4 to 12 and from 12 to 20. There are 4 spokes (diameter length) at 45-degree angles about the rotation z-axis. One set of spokes is at the center, and two sets are typically equal and balanced sets on each side of the center in the z-direction. The component dimensions increase until they reach a maximum value. To provide additional strength, the spoke structures increase in dimensions. The additional spokes with increased thickness cause small step increases in mass around radii of 10,000 and 50,000 meters in Figure 3-30. Figure 3-31 also shows smaller shuttle bay dimensions for the smaller stations. Fewer shuttles need to be supported in smaller stations. Figure 3-30 shows an increase in the shuttle bay mass at the smallest radius and nearly no change in mass with the increasing radius. The shuttle bay

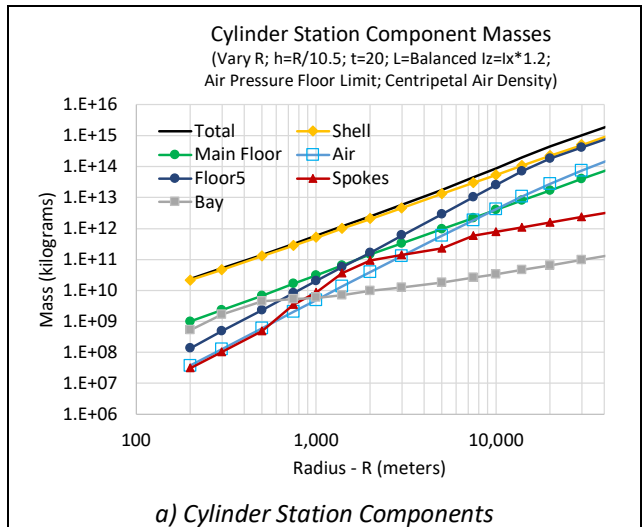

*a) Cylinder Station Components*

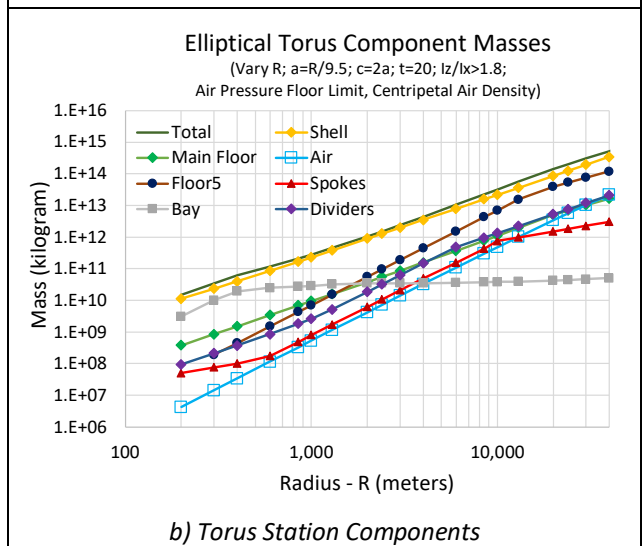

*b) Torus Station Components*

**Figure 3-30 – Component Mass Details**

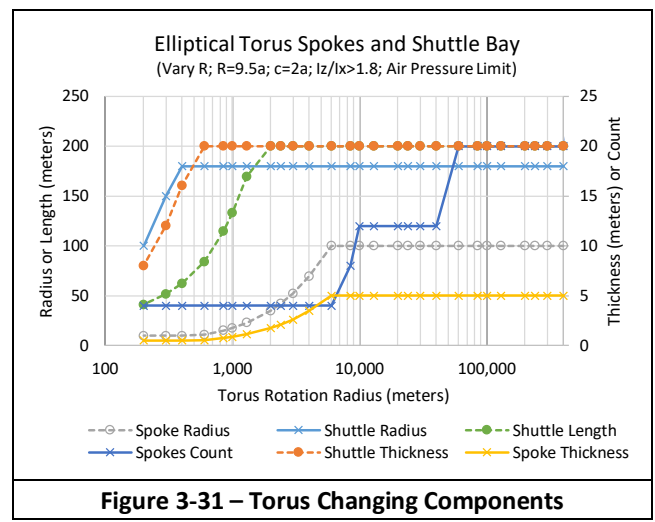

**Figure 3-31 – Torus Changing Components**

connecting-spoke increases in length and visibly increases the mass with the largest radii.

Other components in Figure 3-30 exhibit less visible non-linear mass change. The main floor and the multiple floors show subtle non-linear changes. These floors are constrained when the top floor reaches a maximum height because of air pressure limits. Even less visible, there is a change in the increasing air mass in the station at much larger radii. The centripetal gravity in the station pulls more of the air towards the outer rim and less air mass is required to fill the station.

The size and strength of the components change with the increasing radius. Changing materials would change the size of the station components. Finite element analysis and additional details would further change the size of the components. Additional analysis and refinements could eliminate some of these step functions and non-linearities. The general relationships presented in this subsection should remain valid with modest changes to the size and strengths of the components.

*3.9.3 Radius to Mass – All Geometries*

Figure 3-32 shows the summed masses for all 5 geometries. This shows the building mass to construct the various geometry stations at specific radii. This uses all the station components and design limits from gravity, air pressure, and rotational stability. All five geometries use 20-meter-thick shells and multiple floors spaced 5 meters apart. The stations are rotationally stable with their geometry ratio adjusted to obtain $I_z/I_x >= 1.2$. The stations have a centripetal gravity of 1.05g at the outer rim. A gravity of 0.95g limits the position of the top floor on smaller stations. For larger stations, the air pressure limits the position of the top floor.

The y-axis in Figure 3-32a shows the total mass of the stations on a logarithmic scale of kilograms. The x-axis shows the radii of the stations on a logarithmic scale in meters. There is close to a linear relationship between the radius and the mass; however, even the total mass shows some non-linear behaviors as described in the previous subsection.

The five geometries segregate into two groups. The dumbbells require smaller masses than the other geometries. Of course, the dumbbells also support smaller populations at a given radius. In the other group, the required mass increases from the lowest for the torus geometry, then the ellipsoid, and finally the cylinder geometry. The cylinder uses the most mass at a given radius. The other geometries have curved shell edges that tend to reduce the total mass compared to the flat endcaps of the cylinder station.

With the logarithmic scales in Figure 3-32a it is difficult to discern some trends. Figure 3-32b shows normalized mass values from Figure 3-32a. It shows the same geometries, constraints, and x-axes. The y-axes show the mass and population normalized to the single dumbbell values. This normalization more clearly shows the mass differences.

Figure 3-32b shows the masses of the other geometries are about 10 times the mass of the single dumbbell. It also shows in larger stations that the cylinder and ellipsoid require more

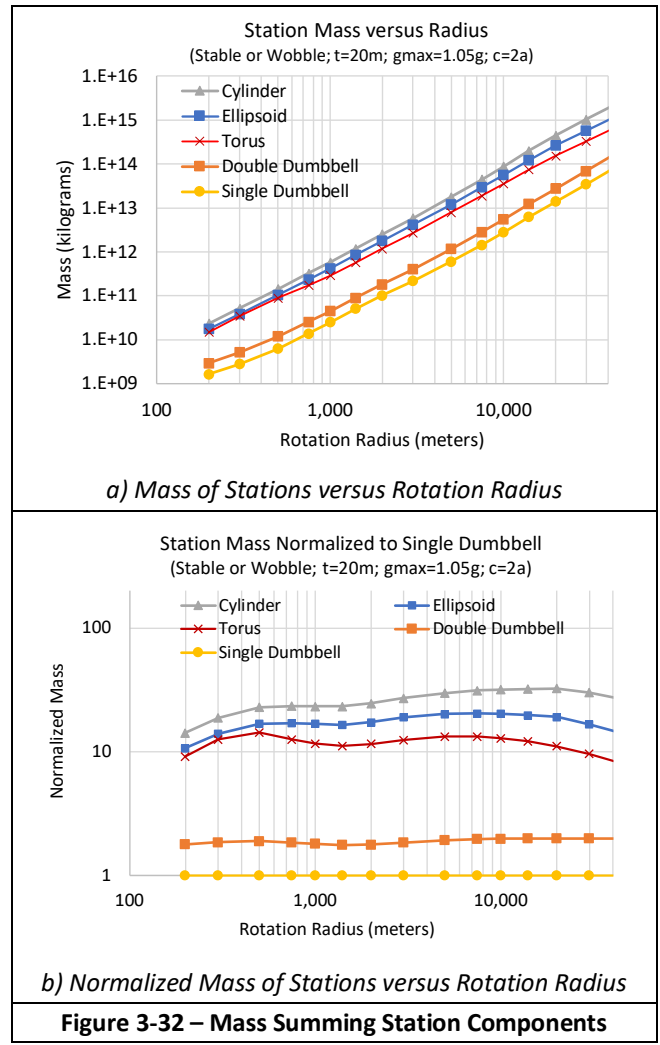

*a) Mass of Stations versus Rotation Radius*

*b) Normalized Mass of Stations versus Rotation Radius*

**Figure 3-32 – Mass Summing Station Components**

mass than the torus for the same radius. The normalized mass relationships remain fairly unchanged over the entire radii range. Reduced dimensions in the small stations cause the increase seen at radii less than 500 meters. As previously discussed, in small stations we reduce the dimensions of their shell thickness, spokes, and shuttle bay; see *§3.5.4 Small and Large Stations*. The relative masses and populations do not change once there is sufficient material to increase all the station component dimensions.

*3.9.4 Mass to Radius – All Geometries*

Figure 3-33 shows the radius of the same five station geometries based on their required construction material. The x-axis shows the amount of building material in kilograms. This is the processed oxide material from the asteroids. It shows the 5 example asteroids selected in [Jensen 2023] along the x-axis at their station building material mass. The y-axis shows the station radius that can be constructed.

The chart shows a line for the minimum radius at 224 meters, which, when rotating at 2 revolutions per minute, produces 1g of Earthlike gravity. Figure 3-33 includes a line at the maximum radius (10 kilometers). Given material density and strength, the maximum rotating station radius can be computed [McKendree 1995]. O'Neill envisioned stations up to 16 kilometers using future materials [O'Neill 1974]. Structures using anhydrous glass tiles and beams are envisioned to support even larger station radii. Table 3-6 contains station radii and properties for these different materials. We conservatively set 10 kilometers as a maximum radius on this graph. Earlier analysis found that most of the considered materials in *§3.4.3 Material Strength* can support our station working stresses up to this radius.

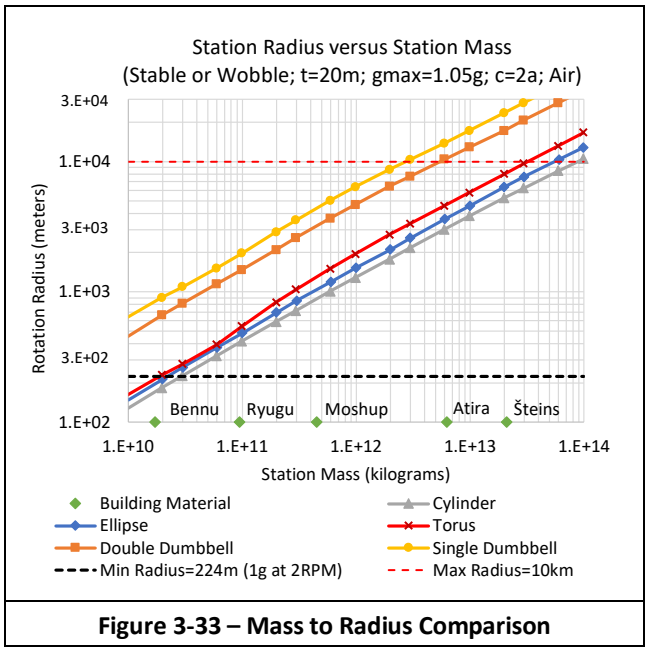

**Figure 3-33 – Mass to Radius Comparison**

The chart shows that the dumbbell station has a much larger radius given the same amount of material. This makes sense because the other geometries encircle the center of rotation. The dumbbell has two nodes (or four nodes) at the radius distance from the center. Given these lower and upper limits, Figure 3-33 shows that dumbbell geometries are viable for small to medium-sized asteroids. Dumbbell geometries are not viable for large asteroids. To support the large node masses and stresses, the cross-sections of tether structures become increasingly large as the station radius increases. A simple analysis suggests that large radii could be supported. We continue to limit the dumbbell rotation radius with our maximum 10-kilometer material strength limit. The building material from the asteroid Šteins would create a huge dumbbell station with a large radius exceeding those metrics. Figure 3-33 also shows that cylinders, tori, and ellipsoids are not viable for small asteroids because their radius would be smaller than our minimum 224-meter radius.

### 3.9.5 Station Shell Thickness

A thick outer shell on our stations provides protection from radiation and debris. The stations often use a 20-meter-thick shell. Figure 3-34 shows the effect of varying the shell thickness with a small and large cylinder station. The logarithmic y-axes show the mass in kilograms. The logarithmic x-axes show the shell thickness. These charts show the mass of the 6 station components and the station's total mass. The length L of the cylinder is computed to provide passive rotational balance using $I_z=1.2\ I_x$. For the large cylinder, the length L ranges from 1.85R to 1.25R with increasing shell thickness. For the small cylinder, the length L ranges from 1.53R down to 0.72R. The 200-meter radius cylinder with a 200-meter thick shell only has a length of 143 meters. The length to maintain the station stability is greatly affected by the thickness.

The mass of the shell increases with the shell thickness. The other components vary less with the changing shell thickness. Some components have dimensions based on the shell thickness but include range limits on those dimensions. An example is the main floor. It has a thick layer of soil to support an Earth-like environment with forests, streams, hills, and valleys. The soil layer is 2 meters thick. The underlying structure thickness is proportionate to the shell thickness. The design has a minimum of 2 meters and increases with the shell thickness to a maximum of 20 meters in the stations. The constant thickness of the higher-density soil layer tends to dominate the main floor mass.

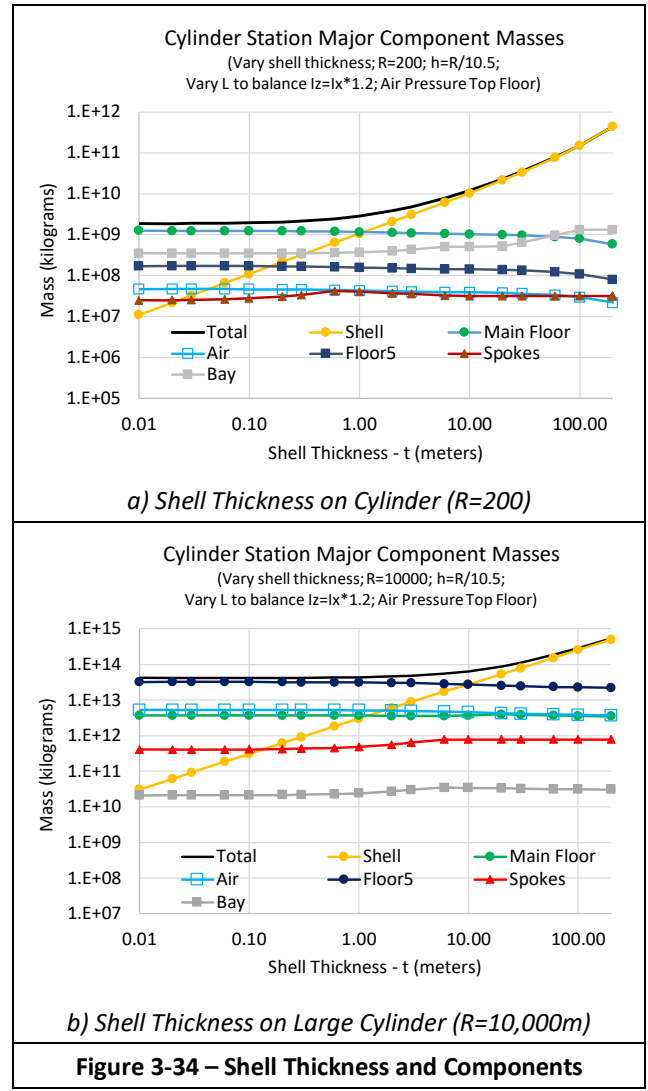

*a) Shell Thickness on Cylinder (R=200)*

*b) Shell Thickness on Large Cylinder (R=10,000m)*

**Figure 3-34 – Shell Thickness and Components**

Another example is the shuttle bay. The bay shell thickness increases with the station shell thickness until it reaches 20 meters. The bay is not a permanent living quarter and does not require additional thickness. The changing bay mass is more noticeable with the small cylinder size in Figure 3-34a than in the larger cylinder in Figure 3-34b.

The station radii are constants in this analysis, but the cylinder lengths vary to maintain the station stability. The cylinder length is reduced to maintain the rotational balance. The station mass is maintained as the mass changes from the reduced length and the larger thickness balance each other. With the 200-meter radius cylinder, the length (L) of the small station drops from 306 meters to 143 meters as the thickness increases to 200 meters. This decrease is necessary to maintain the rotational stability ($Iz/Ix=1.2$) with the changing L/R. The decreasing floor length offsets the mass gains on the main floor from the increasing thickness. The spokes, floors, and air mass are reduced with the station's thickest shell.

In these 200-meter and 10,000-meter examples, the mass of the multiple floors is almost 200,000 times greater in the larger station. The smaller cylinder station only has 4 floors, while the larger station has 187 floors. The increased radius and lengths also contribute to the larger multiple-floor mass.

### 3.9.6 Station Mass Summary

Stations with radii between 224 and 10,000 meters require 3e9 to 1e14 kilograms of material. There are many small near Earth asteroids to create the smaller stations. Only a few asteroid sources can provide the upper limit of material. Only those few large asteroids, moons, or Ceres are potential sources for such large stations.

The strength, stress, and mass of a space station's material must be carefully considered to ensure that the station can function safely and effectively in the challenging space environment. Adaptability will be essential when constructing the space station from an asteroid. The range of asteroid material strengths would change the size of the station components. The construction process must adapt to the real-time analysis of the material and component strengths. Even with this analysis and adaptation, there should only be modest changes to the station's size.

### 3.9.7 Constraints for Kalpana Station Example

Globus and his co-authors designed the Kalpana cylinder station to avoid a rotational stability risk [Globus et al. 2007]. They analyzed the stability of a thin shell cylinder station. They found that the cylinder length-to-radius ratio (L/R) should be 1.3 to create a stable station ($Iz/Ix=1.2$). Their Kalpana cylinder had a radius of 250 meters and a length of 325 meters. They expected the single-floor station to support a population of 3,000. Our analysis extends this work to include a thick shell, multiple floors, and other station components. The analysis of the multiple component station found it unstable using the L/R=1.3 because Iz/Ix was less than the 1.2 stability limit. Reducing the geometry ratio L/R to 1.24, the stability becomes $Iz/Ix=1.2$. The cylinder length would be reduced to 308 meters. The outer rim floor of this smaller cylinder can support 3357 people, about the same as the Kalpana cylinder. A multiple-floor design with five floors spaced 5 meters apart increases the population. The top floor can be 24 meters above the outer rim and all floors would be in a habitable gravity range from 0.95g to 1.05g. This design provides floor space to support over 12,160 people using 144.2 square meters per person.

## 3.10 Station Constraints and Population

Ultimately, the station population is a defining metric for these stations. This section presents population results using all the constraints including gravity, air pressure, and stability. All four geometries are analyzed and presented. The analysis considers the effects from varying the station radius and the mass.

### 3.10.1  Vary Radius for Population

This subsection varies the radius of the four station geometries and compares their mass and population. This analysis applies the previously covered constraints. These include using all the station components and design limits from gravity, air pressure, and rotational stability. The stations have 20-meter-thick shells and multiple floors spaced 5 meters apart. The floor space density is 144.2 square meters per person to provide comfortable living.. The stations are rotationally stable with their geometry ratio adjusted to obtain $Iz/Ix>=1.2$. The stations have a centripetal gravity of 1.05g at the outer rim. A gravity of 0.95g limits the position of the top floor on smaller stations. For larger stations, the air pressure limits the position of the top floor. Multiple floors are between the outer rim and the top floor and are separated by 5 meters. These stations use a 20-meter-thick shell to provide protection from radiation and space debris.

Figure 3-35a shows the populations supported with the various geometry stations at specific radii. The y-axis shows the population of the station on a logarithmic scale. The x-axis shows the radius of the stations on a logarithmic scale in meters. The five geometries again segregate into two groups. The dumbbells support smaller populations for a given radius than the other geometries. In the other group, the supported populations increase from the lowest with the torus geometry, then the ellipsoid, and finally the highest with the cylinder geometry.

With the logarithmic scales in Figure 3-35a it is difficult to discern some of the population trends. Figure 3-35b shows normalized population values from Figure 3-35a. The chart shows the same geometries, constraints, and x-axes. The y-axis shows the population normalized to the single dumbbell values. Figure 3-35b shows that the populations of the other geometries are at least 20 to 70 times the population of the single dumbbell. It also shows that the elliptical torus and ellipsoid support about the same population. The cylinder supports the largest population for a given radius. The other geometries have curved shell edges that tend to reduce the floor area and population. In decreasing supported

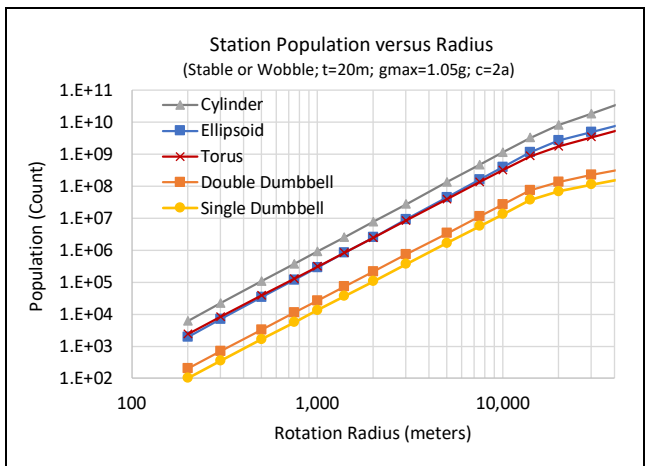

*a) Population of Geometries versus Rotation Radius*

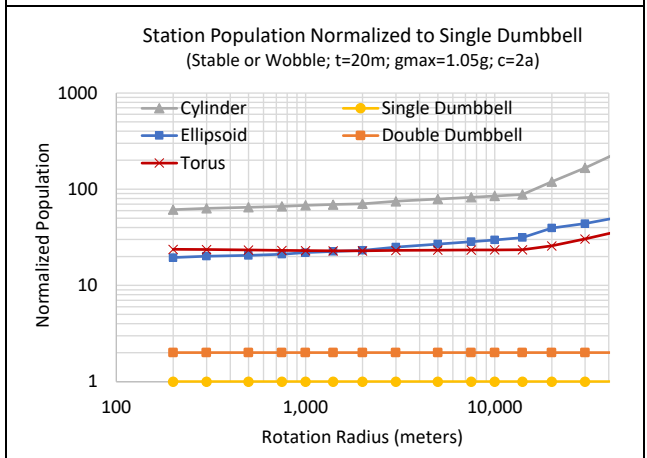

*b) Normalized Populations versus Rotation Radius*

**Figure 3-35 – Population vs Radius**

population by radius, the chart shows the cylinder, ellipsoid, torus, double dumbbell, and single dumbbell.

### 3.10.2 Vary Mass to Population

This subsection shows the potential station population as a function of the construction mass. The results vary the mass of the four station geometries and compare their populations. We advocate that comparing the station geometries using their mass instead of their radius is more valuable. Given an asteroid or source of material, there is typically a fixed mass available to build the station. Knowing the supported size and desired population helps select the station geometry.

Figure 3-36 is an example from [Jensen 2023] and compares the population supported by four geometries. Again, these show the mass of the construction material from five asteroids along the chart's horizontal axis. Many parameters, such as the radius, floor count, and support structures, vary with the available mass. The chart in Figure 3-36 combines all those parameters. It combines many assumptions, requirements, and design decisions. Their values are more for comparative than absolute results.

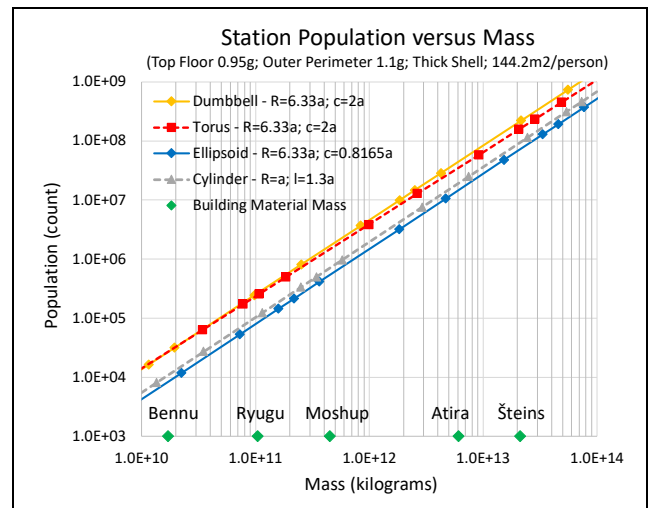

*Credit: Adapted from Figure 3-20 in [Jensen 2023] [CC BY-SA 4.0]*

**Figure 3-36 – Earlier Population vs Mass Results**

The charts in Figure 3-37 compare the population supported by the same four geometries and include the double dumbbell geometry. They show the station population as a function of the station mass. This analysis uses the same previously covered constraints for the station designs in Figure 3-35. Like with the previous analysis, the charts in Figure 3-37 combine many parameters and assumptions. There are quantitative differences between the charts of Figure 3-36 and Figure 3-37a. The data in Figure 3-37a are from a more detailed analysis but qualitatively follow the trends of the previous study. Scaling the dumbbells' dimensions causes trend differences with the small radius station.

The chart in Figure 3-37a shows a logarithmic horizontal x-axis measuring mass in kilograms. The masses of 5 asteroids are marked along the x-axis for reference. They use 1/3 of the bulk material from these asteroids to create the station [Jensen 2023]. The logarithmic vertical y-axis shows the supported population of the four station geometries. For the five asteroid masses, the maximum populations range from hundreds to over 100 million people.

Figure 3-37b shows the same data but normalizes the populations to the single dumbbell geometry to better compare the different geometries. The y-axis shows the station's population on a logarithmic scale ranging from 0.1 to 10. The x-axis again shows the mass of the building material available to construct the stations.

The five geometry populations organize into nearly the same two groups as with the radii analysis in Figure 3-35. We observe several trends over most of the mass range. At small masses and extremely large masses, the trends vary because of non-linear effects and material limitations. The ellipsoid geometry consistently provides the smallest population over most of this range of masses. The cylinder and torus provide about the same population for a given mass. The chart data shows that the dumbbell geometries support at least twice the populations of the other three geometries for smaller masses

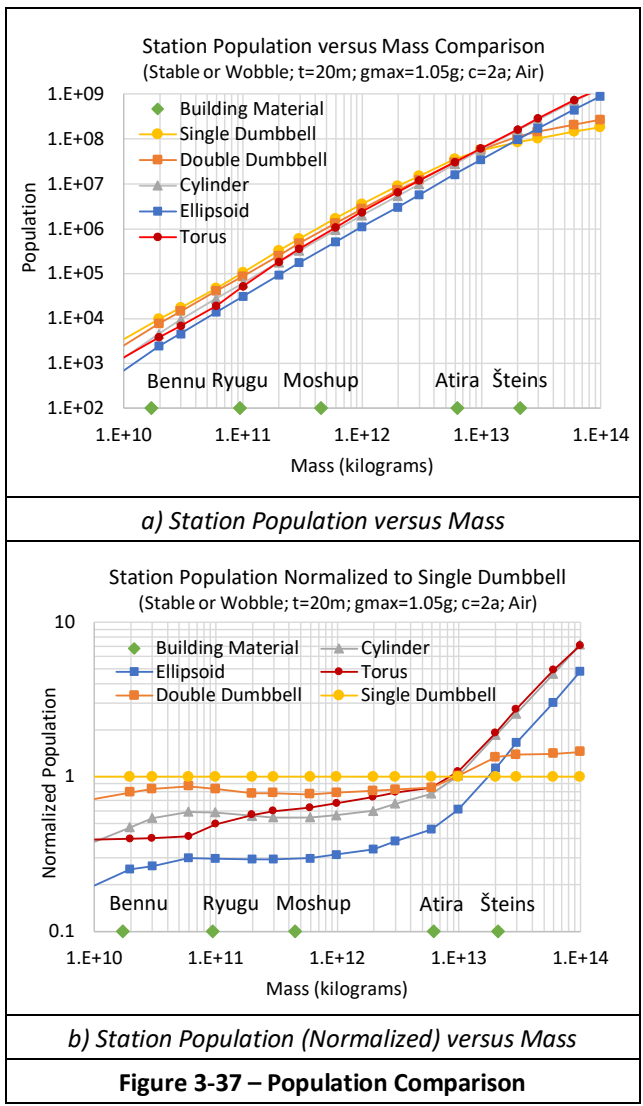

*a) Station Population versus Mass*

*b) Station Population (Normalized) versus Mass*

**Figure 3-37 – Population Comparison**

of building material. Except for large masses, the double dumbbell population is about 0.75 times the single dumbbell population. This dumbbell population ratio was shown and described with Figure 3-15.

With very large dumbbells, the double dumbbell population becomes larger than the single dumbbell in Figure 3-37. This data uses the top floor height being air pressure limited. With large stations, the number of floors in the multiple-floors component reaches a maximum because of this air pressure limit. With this maximum size limit, the double dumbbell has twice the number of nodes. Eventually, it has a greater population than the single dumbbell despite the smaller radius and node size. With the gravity limit and airtight floors, the number of floors continues to increase with the radius. The single dumbbell population continues to be greater than the double dumbbell population with the gravity limit.

The dumbbell spokes increase in size to provide sufficient strength to support the stresses from the rotating dumbbell nodes. The other geometries are designed to support the hoop stress of the rotating shell; as such, their spokes do not increase in size as much as the dumbbell spokes. This dumbbell spoke mass reduces the mass available for the habitable regions. As such, the other geometries support more population in large stations. Again, additional analysis of the material strength and the rotating structure stresses is recommended to further validate these dumbbell results.

### 3.11 Station Constraints Summary

This section covered multiple constraints limiting the size and geometry of a rotating space station. These included limits from gravity, population, materials, geometry, air pressure, and rotational stability. It provided details on the station mass and air pressure for those geometries and overviewed rotational stability concepts. Results from these constraints are now summarized.

Rotating the space station provides centripetal gravity. It would be best to provide one Earth gravity; the station rotation radius should be greater than 200 meters. To provide 1g, smaller stations would spin too fast and cause nausea for most residents [Globus and Hall 2017]. A rotation rate slower than 1 rpm is recommended to not exceed the material strength of an integrated station shell [Graem 2006].

We advocate using anhydrous glass produced from asteroid regolith to build the station. Asteroids typically have 10 times the amount of oxides (for glass) than metal for the station construction. The strength of this glass material could support a station of 10 to 20 kilometers in radius. It is possible that highly refined anhydrous glass could support a station with a radius of over 100 kilometers. For schedule and cost reasons, we recommend a smaller station radius closer to 3 or 4 kilometers. These larger stations support multiple floors and increase the available floor space. The extra floor space can support higher populations, lower population density, more agriculture, and increased industry production.

A refined set of space station geometries was reviewed. These refinements address rotational stability and gravity issues. Spherical stations become ellipsoidal stations; long cylindrical stations become short hatbox stations; circular cross-section torus stations become elliptical cross-section torus stations, and dumbbell structures are doubled for rotational stability. Determining the Moments of Inertia (MOIs) to compute the stability requires knowing the mass of the components. The masses of its major components and the station support the selection of an appropriate-size asteroid or solar system body. The size and mass of the station and its components also support the development of a construction schedule and equipment count; see [Jensen 2023].

Centripetal gravity pushes the station atmosphere towards the outer rotating rim. This produces air density and pressure that varies with height, similar to Earth. By design, the station's outer rim has sea-level air pressure. The minimum acceptable air density is Denver air pressure, at an Earth height of 1609 meters. Without using airtight layers, this creates a limit on the maximum number of floors in large stations.

This section reviewed passive stability for rotating stations. Perturbations can cause rotating systems in space to eventually rotate about the axis with the greatest angular moment of inertia [Globus et al. 2007]. There is the risk that an improperly designed station could catastrophically fail and tumble end-over-end. A key stability rule used in this analysis was that the desired axis of rotation should be 1.2 times greater than the other rotation axes [Brown 2002]. This stability design work was extended to all four geometries. The analysis evaluated thick shells, gravity and air pressure constraints, multiple floors, and more components in our stations. All the geometries (cylinder, ellipsoid, torus, and double dumbbell) can be passive rotationally stable.

## 4 Space Station Construction

The previous sections covered human frailty and station design constraints for space stations. This section considers the limitations imposed by construction space station approaches. It includes two recent space station construction approaches. This section also reviews the characteristics and limitations of very large space stations. These limitations are then applied to large historic stations. This section concludes with a summary of these construction concepts.

### 4.1 Construction Approaches

Constructing the space station introduces limits and constraints. For context, the first subsection reviews a historic construction approach. Authors have evolved their construction approach using manned, teleoperation, semi-autonomous, and autonomous systems. This subsection covers manual, teleoperation, and automated construction approaches.

*4.1.1 Construction – Historic Example*

A 1973 NASA study [Nishioka et al. 1973] evaluated the feasibility of mining lunar resources for earth use in the year 2000. Figure 4-1 is from that study and shows the exterior layout of a lunar mining facility. The facility includes crew quarters, a kitchen, dining, recreation, a hospital, and storage for the crew. It also includes manufacturing, equipment storage, environmental control, a spaceport, and a terminal. They also included a nuclear power plant located away from the primary base for safety reasons. They planned for a crew of 150 and provided approximately 35 cubic meters of space per person. They planned for 5 meters of lunar regolith to cover the inhabited regions of the facility for radiation protection. Regions not routinely inhabited would only be covered with 1 meter of regolith. Considerable detail went into the 162-page document [Nishioka et al. 1973].

Unfortunately, the 1973 NASA study [Nishioka et al. 1973] found that it would not be commercially feasible to mine, refine, and bring lunar minerals back to Earth. They found the mineral cost would be approximately two orders of magnitude higher than similar Earth mineral costs for the year 2000. One of their artworks is included to show a broad range of concepts including astronauts, rovers, power plans, linear accelerators, power plant, processing facility, and radiation protection. Although not cost-effective, their study provides a wealth of concepts for current studies.

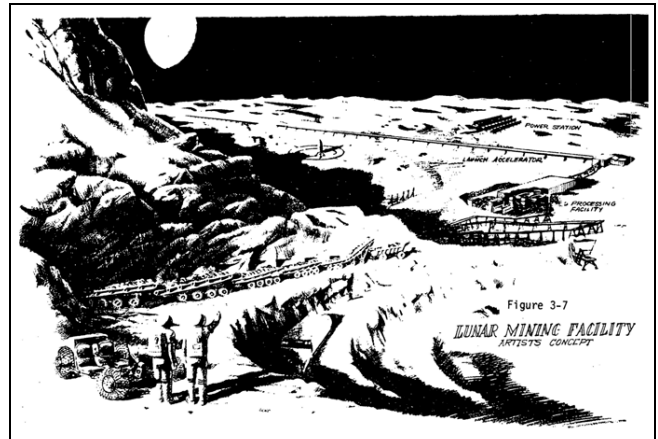

Credit: [Nishioka et al. 1973] Figure 3-7 [NASA Report Public Domain]

**Figure 4-1 – 1973 Vision of Lunar Mining Facility in the year 2000**

*4.1.2 Construction – Manual*

Manual construction has existed since prehistoric days. Metals, tools, and architecture techniques were being used even in 3000 BC. Manual labor with tools was exclusively used until the first Industrial Revolution. Large-scale production of iron and steam engines took place in the late 1700s. Mass production of building materials (bricks and timber) replaced labor-intensive construction in the 1800s. The second Industrial Revolution began in the 1880s and further developed steel and introduced electricity. Theory and innovations refined construction technology. The assembly line was introduced in the early 1900s. Millennials of groundwork poised the world for automation. The automobile industry first used the term automation around 1946 to describe the automatic devices and controls replacing manual tasks in their production lines.

Historically, authors first planned to build space stations manually. Even in the 1970s, O'Neill estimated space station construction costs using manual labor. He assumed 42 tons constructed per person-year using a metric comparable to large-scale construction on Earth [O'Neill 1974]. This resulted in cost estimates of $30 billion in 1972. This is comparable to the NASA Apollo program, which would be $200 billion in 2019 dollars. Even now, the NASA Artemis program is expected to cost $93 billion [Artemis 2022]. One obvious approach to reduce these costs is to use more automation.

Figure 4-2 provides another example where NASA produced an artist's concept of an astronaut manually tethering an asteroid [Lind 2011]. Manual labor in space is unhealthy for humans. There are risks from the vacuum environment with no oxygen or pressure. Health issues from radiation and microgravity were covered in *§2 Human Frailty Limitations*. The following subsections review how to eliminate human manual labor from space construction using teleoperation and automation.

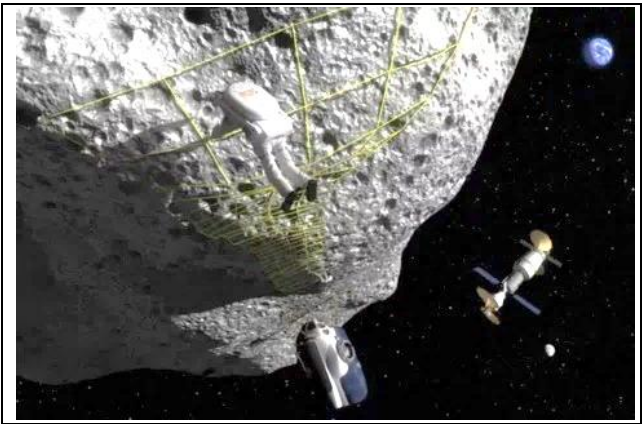

*Credit: NASA from [Lind 2011] [NASA Image Public Domain]*

**Figure 4-2 – Manual Work on Asteroid**

*4.1.3 Construction – Teleoperation*

Teleoperation is defined as the operation of a system or machine at a distance [Wiki Teleoperation 2023]. It is most commonly associated with robotics and mobile robots. Teleoperation allows equipment to be operated in space while the operator remains on Earth. Many authors consider this as a construction phase between manual and autonomous operations. The time lag from the Earth to the Near Earth Asteroid could be many seconds to several minutes. At these distances, remote operation may not be efficient. Some researchers are investigating approaches to overcome this time lag with software enhancements and predictive algorithms [Kajima et al. 2017].

In 2006, a researcher stated that until automated space mining technology improves sufficiently, the mining facilities on an asteroid would need to be accompanied by a sealed environment habitat for personnel [Vandenbos 2006]. This would put the tele-operators close to the equipment. In 2012, a team from the Kennedy Space Center outlined a self-replicating lunar factory system. Philip Metzger was the lead author of that paper [Metzger et al. 2012]. In their plans, equipment would initially be teleoperated and slowly transitioned to complete autonomy over time. Robotic advancements were essential to his vision. In 2016, Metzger argued that robotics was already more than adequate for the early stages of the space industry [Metzger 2016].

We agree that human interaction would be useful, but robotics and artificial intelligence have advanced enough to advocate a completely autonomous operation with minimal remote human oversight.

*4.1.4 Construction – Automation*

On Earth, automation benefits workers and companies [Lamb 2013] [Lee 2018]. For workers, such benefits include eliminating hard physical or monotonous work, eliminating assignments in dangerous environments, and reducing some occupational injuries. For companies, automation benefits include increased throughput or productivity, improved quality or increased predictability of quality, increased consistency of output, reduced direct human labor costs and expenses, and performing tasks beyond human capabilities.

The idea of using automation in space station construction is not new. In 1974, O'Neill stated: *"In the long run, space-colony construction is ideally suited to automation. A colony's structure consists mainly of cables, fittings and window panels of standard modular form in a pattern repeated thousands of times. The assembly takes place in a zero gravity environment free of the vagaries of weather. By the time that the colonies are evolving to low population density, therefore, I suspect that very few people will be involved in their construction."* [O'Neill 1974].

Multiple researchers have detailed how to automate space construction completely.

A 1980 summer study at the NASA Ames Research Center [Freitas and Gilbreath 1982] details advanced automation for space missions. The report includes a section on a self-replicating Lunar Manufacturing Facility (LMF) generated from an initial seed of 100 metric tons of technology and supplies. Their large monolithic factory can produce a duplicate of itself in a year.

Philip Metzger and his team outlined a self-replicating lunar factory system [Metzger et al. 2012]. Their approach was founded on bootstrapping. The first robotic machines on the Moon would fabricate a set of 1700s-era machines. With additional launches and supply deliveries, these machines would be steadily enhanced. Their ability would grow through the equivalent of the 1800s, 1900s, and finally into the 2000s. Their analysis suggests it has become feasible to bootstrap a self-replicating space-based manufacturing industry in just a few decades.

Justin Lewis-Weber also developed concepts for a lunar-based self-replicating solar factory [Lewis-Weber 2016]. He used simulators to analyze the complex self-replicating systems, He envisioned a system that could self-replicate, build solar space power components, and create a mass driver to deliver the components to a geostationary Earth orbit [Lewis-Weber 2016]. One small, lightweight seed factory would be sent to the moon to reduce launch costs. The goal was to exploit the exponential growth from self-replicating systems to produce energy at a dramatically reduced price.

A Boeing Technical Fellow, Brian Breshears, advocates that autonomous systems will be required for deep-space manufacturing [Frost 2011] [Barron 2017]. The distances involved will introduce minutes of communication latency, preclude teleoperation, and mandate advancement in autonomy.

*4.1.5 Automation – Recent Examples*

This subsection reviews two recent automation examples. One is our asteroid restructuring approach [Jensen 2023]. The other is spin construction [Miklavčič et al. 2022].

**Asteroid Restructuring:** In restructuring an asteroid, robotics is used to completely automate the building process. Figure 1-1 shows a rendering of an envisioned station created from an asteroid. Having crewed systems on the asteroid or in orbit around the asteroid would dramatically increase the cost and jeopardize the crew's health. We advocate that this system needs to be autonomous. It must operate according to a plan, adapt to its environment, and make decisions with minimal or no human intervention.

The restructuring mission and process goes through a series of activities to create an enclosed space station framework with a wealth of inventoried supplies [Jensen 2023]. The process first lands a probe on an asteroid. Using a modest seed package of materials and tools, robotic workers use the asteroid material to create copies of themselves, tools, vehicles, and automata. The initial probe and seed package are built with state-of-the-art 21st-century technology. The materials, tools, vehicles, and automata produced on the asteroid will resemble 18th and 19th-century technologies. An initial set of four general-purpose 21st-century robots use exchangeable probes, some supplies, and primarily in-situ material to construct a few thousand similar robots. They also build 10s of thousands of simple mechanical automata helpers from the asteroid material for additional productivity. These robots and helpers construct the station using asteroid material.

Analysis found that restructuring an asteroid like Atira will require thousands of worker robots and tens of thousands of pieces of support equipment. Operating these devices remotely with a crew of thousands of operators stationed on Earth would be cost-prohibitive. The time lag from Earth to the Near Earth Asteroid could be many seconds to several minutes, making teleoperation difficult.

One can estimate the schedule to restructure an asteroid into a rotating space station framework. The larger Atira station and the middle-size Ryugu station were considered in detail. Our analysis and simulations found that it will take five years for a Stanford-size station (Asteroid Ryugu) and twelve years for an Asteroid Atira-size station [Jensen 2023]. Figure 4-3 shows a chart with the population and build time of torus stations as a function of the volume of material used in their construction. As portions of the station are completed, crews on new missions will develop the critical infrastructure to support human habitation such as heating, cooling, atmosphere, and lighting for the station interior. The station would provide gravity and radiation protection for these first crews. Our Asteroid Restructuring paper [Jensen 2023] provides more details on this approach.

**Spin Construction:** A recent paper presented an approach to convert a 300-meter radius asteroid into a 3-kilometer-radius cylindrical station with 2-meter-thick outer walls [Miklavčič et al. 2022]. Figure 4-4 shows one of their rendering of this station. In this paper, the authors considered expanding a rubble pile asteroid to fill an exterior containment structure.

The rubble pile asteroid is converted into a hollow cylindrical shell by spinning the asteroid beyond its yield point. The asteroid's spin increases and the resulting centrifugal forces cause the rubble to break apart and spin outward [Miklavčič et al. 2022]. A containment structure surrounds the asteroid and captures the rubble. The structure expands to create the final cylindrical station. We assume nearly all this construction will be performed autonomously.

The Miklavčič researchers used finite element analysis to evaluate the spin construction approach. They considered multiple asteroid sizes and produced stations with radii ranging from 1400 to 4020 meters [Miklavčič et al. 2022]. Their simulations evaluated that the hoop stress of the containing structure would be approximately 130 MPa [Miklavčič et al. 2022]. They show that this can be built with existing construction materials. They found that the required strength increases with larger station radii; however, their finite element simulations suggested lower stresses than results from analytic stress equations. They found that the hoop stresses were lower with thicker exterior layers [Miklavčič et al. 2022].

The authors found that the spinup of a 300-meter asteroid would occur in only a few months. The external capturing structure expands with the outgoing rubble material. It expands to a cylinder with a length and radius of 3 kilometers. This paper did not discuss or analyze the cylinder endcaps. This cylinder rotates and has centripetal gravity and radiation shielding for the subsequent phases of enclosing and constructing the habitat's interior.

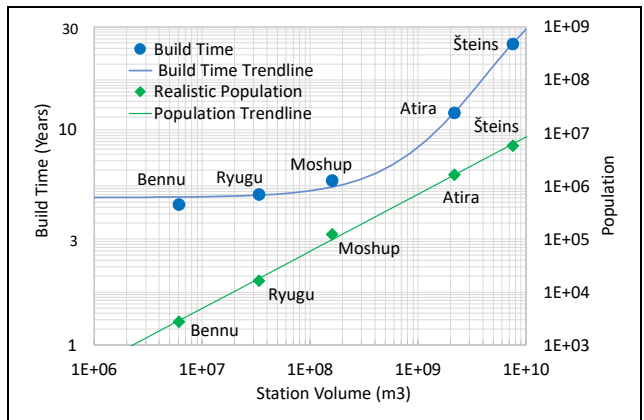

Credit: Figure 5-2 from [Jensen 2023] [CC BY-SA 4.0]

**Figure 4-3 – Station Material Volume and Resulting Population and Build Time**

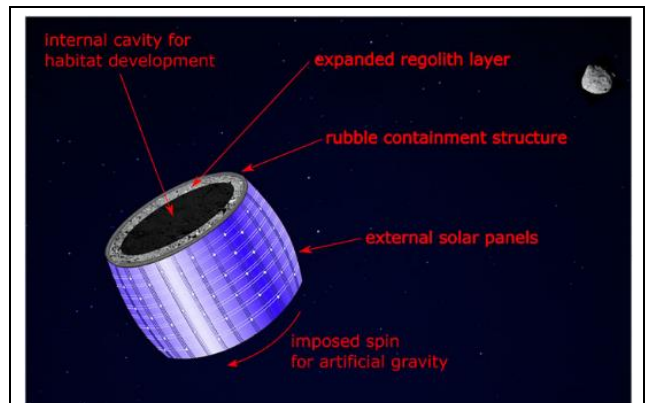

Credit: [Miklavčič et al. 2022] [CC BY-4.0]

**Figure 4-4 – Spin Construction of Habitat Bennu**

## 4.2 Very Large Space Stations

This section considers issues specific to very large space stations. Earlier analysis using the anhydrous material strength from [Bell and Hines 2012] suggests stations from regolith material could be 100 kilometers in radius; see *§3.4.3 Material Strength*. Such stations become feasible with autonomous construction approaches. The first subsection considers where to acquire the material for the very large stations. The construction assumes the stations are created with nearly unlimited material from an asteroid, moon, or dwarf planet. Additional subsections present air pressure and population limitations in these very large station. This section concludes with details on an example very large space station.

*4.2.1 Large Station Material Source*

These large stations require a large source of building material. The Atira torus station described earlier had a major radius of 2400 meters. This used an asteroid with a mean radius of about 2 kilometers and a mass of 4.1e13 kilograms. A 35,000-meter radius torus station would use an estimated 4.4e14 kilograms and likely require an asteroid of at least 1.3e15 kilograms. The 35,000-meter cylinder station would require a bigger asteroid of 4.8e15 kilograms. Multiple sources can be considered for these stations. Either station could be created from the asteroid Eros, with a mass of 6.7e15 kilograms. The Mars moon Phobos, with 10.8e15 kilograms, has sufficient mass to produce either of these stations. The dwarf planet Ceres has a mass of 9.4e20 kilograms and could support many sets of these huge stations.

Consider a large station with a more "realistic" radius of 10,000 meters. This radius is also at our conservative material strength limit; see Figure 3-33. The 10-kilometer radius cylinder station would weigh 1.0e14 kilograms. A 10-kilometer radius torus station would weigh 3.6e13 kilograms. These would take an asteroid that weighs about 3e14 kilograms. An asteroid with a diameter of about 7 or 8 kilometers would weigh close to that amount. No known near-earth asteroids are this size; however, many asteroids are greater than 7 kilometers [JPL SBD Search Engine 2023]. Those asteroids are often in the asteroid belt or are highly inclined; as such, they would be less desirable. Additional evaluation would be necessary to identify an asteroid appropriate for a 10-kilometer radius station.

A researcher is considering the dwarf planet Ceres to construct a large space station [Janhunen 2021]. He presents a megastation with 121 connected rotating cylinders. Each cylinder is 1 kilometer in radius and 10 kilometers in length. The megastation would support almost 7 million people. He also documents an even larger station with a radius of 25,000 kilometers that supports 800 billion people. He chose Ceres because of its vast resources, including nitrogen for the station's atmosphere [Janhunen 2021].

We are intrigued with the concept of using Ceres as our source for station material. The 10-kilometer radius station would use a tiny percentage of the material. A single-floor cylinder station of this size could support over 5 million people using 144.2 square meters per person. A multiple-floor cylinder station could support almost 1 billion people. Most space station construction will use solar power. It may be better to consider asteroids closer to the Earth's orbit. At Ceres, the sun would only be about 15 percent as bright as at the Earth orbit. Parabolic mirrors could be used to compensate for the dimmer light but might become too cumbersome for our planned restructuring approach [Jensen 2023].

*4.2.2 Large Stations and Air Pressure*

At a height of 1600 meters, air pressure is like Denver, Colorado. We use this height as our constraint for minimum air pressure. This study often uses a gravity constraint of 1.05g on the outer rim and 0.95g on the top floor. In large designs, there is an uninhabitable low air pressure zone between the air pressure limit and the gravity limit. This begins to occur beyond the cylinder radius of 16,900 meters; see Figure 3-10. As previously described in *§3.3.4 Airtight Layers of Floors*, we fill that region with floors in airtight layers in the station.

Figure 3-12 illustrated these cylinder station limits with a radius of 35,000 meters. The chart showed the gravity floor limit at 3333 meters and the air pressure floor limit at 1567 meters. This large cylinder included two airtight layers in its design at heights of 1567 and 3127. Above and below these airtight floors, the air density follows the same air density equations; however, the height resets to zero at each of the layer heights. No additional floors are constructed above the gravity limit of 0.95g, and the air density drops monotonically to the station center. Filling the low-pressure zone with floors increases the potential population of these large stations.

*4.2.3 Large Station and Population*

Figure 4-5 shows the population that could be supported with very large stations. These are extensions to Figure 3-37 with larger masses along the x-axis. Figure 4-5a shows the population as a function of the building material mass. The y-axis shows the population on a logarithmic scale and ranges from 100 to 100 trillion people. A solid black line shows 8 billion people on the chart. This represents the current population of the world in 2023. The x-axis shows the mass of construction material for the stations. This is on a logarithmic scale and ranges from 1e10 to 1e18 kilograms. A series of 7 astronomical bodies are across the x-axis. The green diamonds show the available building material from these asteroids (and moon). Ceres has a mass of 9.4e20 kilograms, and the charts only show that 1e18 kilograms of the outer surface would be used to build a single station that could support the population of 1000 Earths. Those charts should be taken skeptically without additional analysis of material strengths, construction times, and viable asteroids. They are useful for a very preliminary evaluation of very large stations. Such large stations are likely a little absurd, and we advocate building stations of a more modest size.

The normalized population chart in Figure 4-5b shows the same results as in Figure 4-5a. The populations are normalized to the torus population. Trends and differences are more easily seen in the normalized data. As shown and discussed earlier, for a given station mass of less than 1e13 kilograms, the dumbbells support a larger population than the other geometries. The ellipsoid supports the least population.

Figure 4-5a shows that for much larger radius stations, the population growth of the dumbbells slows past the 1e14 kilogram mass. The spoke dimensions and mass become very large to support the large rotating nodes and high rotational stress. The figures show that the ellipsoids and tori support similar populations. Both have curved elliptical shells limiting the surface area of the floors. This also indicates that the flat endcaps of the cylinder provide a population advantage over those curved shells. Figure 4-5b clearly shows that the constant floor width of the cylinder is an advantage for large stations and populations. There are data in Figure 4-5 that exceed other limitations and should be removed. Figure 3-33 clearly shows these limitations. With very large asteroids, the stresses in the spokes with large dumbbell radii would exceed known material strengths. Dumbbell population data points should be removed from the chart beyond the 1e14 kilogram mass where they exceed the 10K material limit. Conversely, non-dumbbell stations with masses less than 4e10 kilograms require radii smaller than 200 meters. These small stations would rotate too fast and cause nausea for most individuals. The non-dumbbell geometry type population data points should be removed from the chart below the 4e10 kilogram mass. These two limitations would also remove data points from the chart in Figure 3-33. There are approaches to address some of these limits. As an example, the radii could be increased by using more of the asteroid or by reducing the shield thickness from 20 meters to 8 meters.

These charts clearly show that large stations can support huge populations. Of course, one must ignore material strength and assume novel materials will exist. Previous sections have shown how gravity and air pressure can limit the number of floors and population. It is also important to consider human psychology and the aspect that people may not like living in the lower "bowels" of the station. Our restructuring asteroid study discussed and analyzed "reasonable" populations [Jensen 2023]. Figure 4-6 was derived from that work, and the reasonable population is shown as a function of the rotation radius. The x-axis shows the radius of the station on a logarithmic scale ranging from 100 to 1 million meters. The y-axis shows a percentage representing the reasonable population over the maximum (gravity-limited) population. All the floors can be used until constrained by air pressure limits. That effect begins when the top floor is at Denver's air pressure and would begin to occur at heights of about 1500 meters or a station radius of around 10,000 meters. The graph shows the reasonable population decrease begins earlier at about a radius of 3000 meters. Only a few top floors are used in the large stations. Although there could be over 300 floors in a station with a radius of 15,000 meters, psychological limits may reduce that number to 20 to 30 floors. We don't advocate putting a Morlock set of people in the bottom floors [Wells 1898].

Figure 4-7 compares the populations using several of the expected constraints on the station design. The x-axis of the chart shows the rotation radius on a logarithmic scale ranging from 10 to 1 million meters. The y-axis shows the population on a logarithmic scale ranging from 100 to 100 quadrillion people. Such sizes are likely absurd; the world's population is currently 8 billion people, and it could be supported by a rotating cylinder with a 20 to 60-kilometer radius. The largest populations shown in the chart are with the top floor limited only by the gravity range constraint. The next lower populations are with the top floor limited by the gravity and air pressure limits. Using the reasonable floor scaling shown in Figure 4-6 produces the next lower population. The chart also includes the population using only the outer rim of the station as a projected single floor.

The available floor space and supported population are significantly constrained for very large stations. A station with a radius of 35,000 meters could support a population of 85

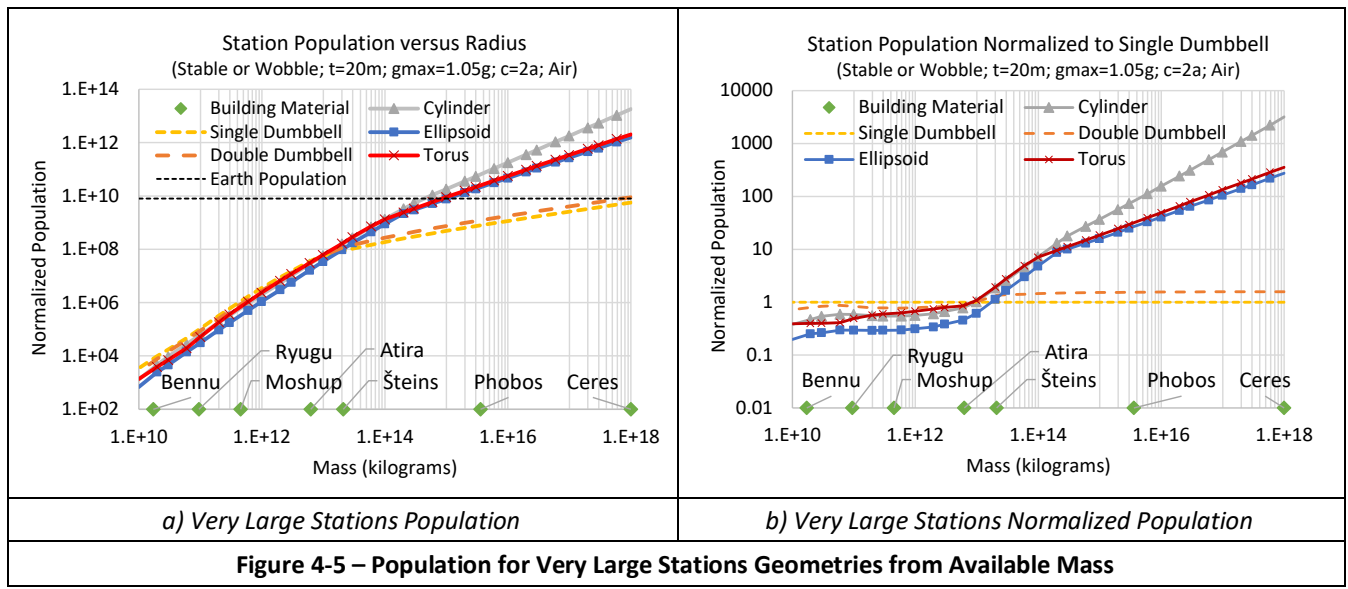

*a) Very Large Stations Population*     *b) Very Large Stations Normalized Population*

**Figure 4-5 – Population for Very Large Stations Geometries from Available Mass**

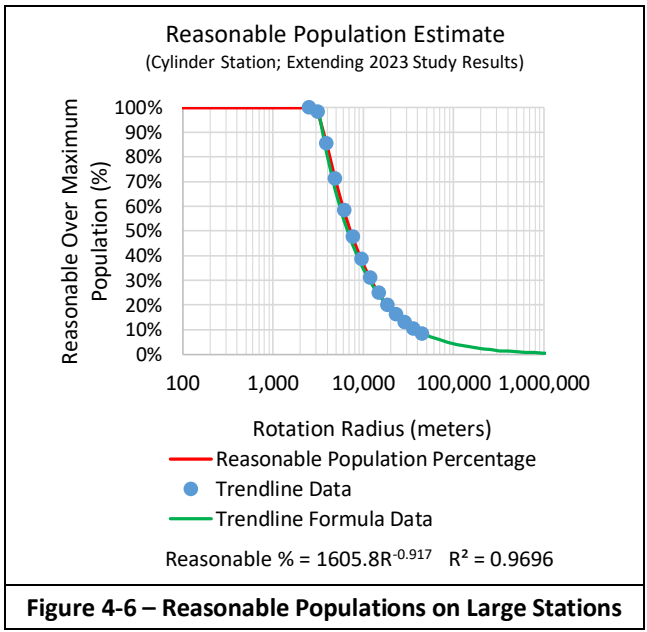

Figure 4-6 – Reasonable Populations on Large Stations

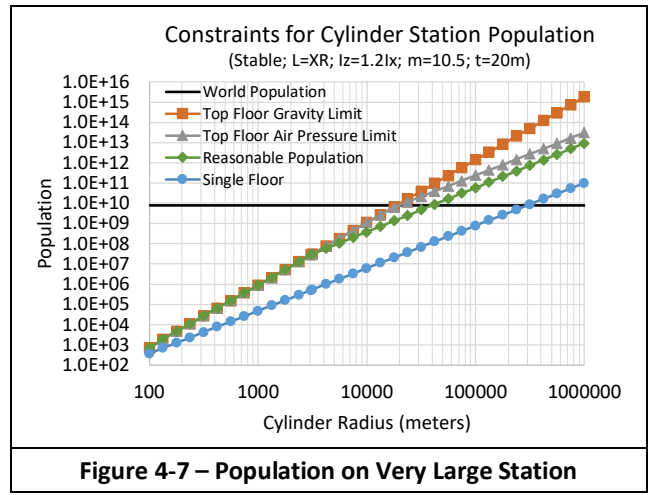

Figure 4-7 – Population on Very Large Station

million on the outer rim as a single floor. There could be over 600 floors in the gravity range floor limit using multiple floors. This could support over 54 billion people using all the floors. Realistically, Figure 4-7 shows that it would support more like 4.8 billion people. This assumes 144.2 square meters of surface area for each individual. Lower levels would be available and habitable and could be used for agriculture or future industries. These stations could be very spacious and still support an absurd number of people.

The purpose of this subsection is to investigate very large stations. There is considerable volume that is not used in these large stations. In the 35,000-meter example station, the centripetal gravity range limits the top floor at a height of 3333 meters. Habitable air pressure limits the top floor to 1568 meters. To increase the surface area and population, one could increase the number of floors by increasing the pressure (and density) in the lower floors. This would require more atmosphere mass (cost), and the lowest floors would become uncomfortable for human occupation. In addition, the psychological effects could limit the station to less than 50 usable floors. Even with these constraints, these large stations support a huge population.

*4.2.4 Airtight Layers and Population*

Air pressure is one of the more limiting constraints on the number of floors and the population in large stations. Multiple airtight layers in these very large stations could further increase the population. Earlier sections introduced the concept of multiple airtight layers. This subsection provides more details on this concept for specific large-radius stations. If the layers are large enough, each layer could have sufficient open space on each top floor to address the psychological constraint. Figure 4-8a shows a small cross-section of the tube in a very large torus space station. This torus has an outer radius of 35,000 meters. The drawing shows detail at the base of one of the torus spokes. It labels many station parts such as the shell, spoke, and floors. It includes values for the height above the outer shell, centripetal gravities, air pressures, and distances from the center. The station would have over 750 floors. Each of the lower layers has a 280-meter-high ceiling and 240 floors separated by 5 meters. The top layer has 50 floors and a 4772-meter-high ceiling.

The outer rim of the torus tube is at 35,000 meters, and the inner rim is at a radius of 25,456 meters or a height of 9544 meters. The top floor is limited by gravity and at a height of 4772 meters. The gravity on that floor is at 0.95g and is 1.1g at the outer rim. Typically, air pressure would limit the top floor to a height of 1600 meters. The diagram shows 4 airtight layers in the torus tube. The top floor of each layer is at least at the air pressure (Denver limit).

This design includes at least a 280-meter open space above the top floor of the 4 layers. The vista on the lowest airtight layer would be 8,987 meters. The top layer floor has a vertical open space of 4772 meters. On the top airtight layer, the vista would be 32,602 meters. In contrast, on an Earth ocean beach, an average height person can see about 4,800 meters to the horizon. The vistas on this station would be spectacular and would likely be limited by atmospheric haze or clouds instead of being limited by the curvature of the Earth.

Figure 4-8b shows details on the same 35,000-radius torus design. This provides a similar but magnified view like Figure 3-12. Recall the top floor (constrained by gravity) would not have habitable air pressure in these large stations without using layers. The top floor is constrained to a height of 4772 meters to provide habitable gravity. The outer cylinder radius is at 35,0000. The outer rim is at 1.1g and this height provides a gravity of 0.95g. This is the scaling factor of 7.33 between the floor height and the outer radius; see Table 3-1.

The air pressure is set to 1 atmospheric pressure at the outer radius (101,350 Pa). Figure 4-8b shows the station air pressure (without layers) as a solid red line. The graph shows that at a height of about 1500 meters, the air pressure has dropped to the Denver air pressure. Without airtight layers the top floor would be at 60,000 Pa air pressure. This chart shows 3 airtight layers at 1400, 2800, and 4200 meters. Each airtight layer can be pressurized separately.

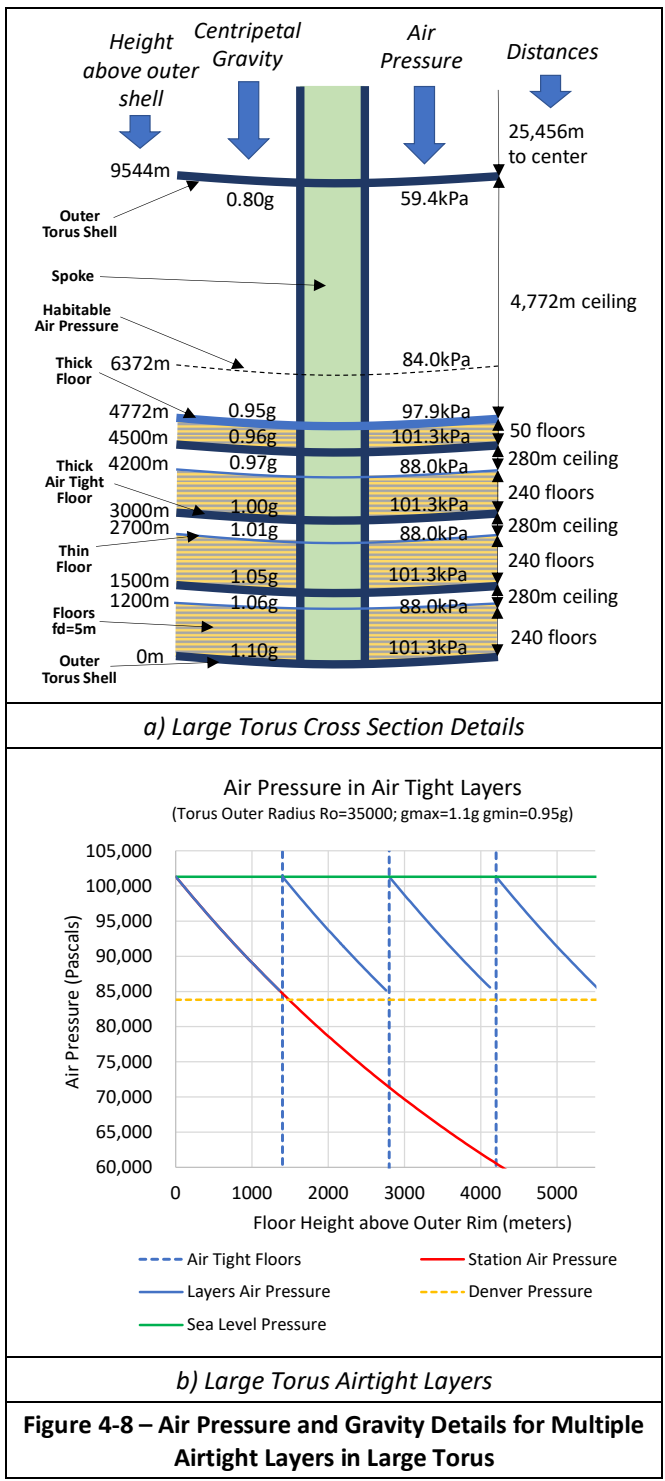

*a) Large Torus Cross Section Details*

*b) Large Torus Airtight Layers*

**Figure 4-8 – Air Pressure and Gravity Details for Multiple Airtight Layers in Large Torus**

*4.2.5 Example Large Stations*

This subsection focuses on torus and cylinder geometries. Figure 4-9 shows the cross-sections of two very large stations. The station in Figure 4-9a is the cross-section of a large torus and is similar to Figure 4-8a. Again, the radius is 35000 meters. This exceeds our conservative 10000-meter radius; however, this size could be possible with refined materials and perhaps even strengthened anhydrous glass material.

Figure 4-8b shows this station provides 0.95g on the top floor and 1.1g on the outer rim floor. This defines the top floor at 4772 meters above the outer rim. This size station rotates once about every 6 minutes.

The station in Figure 4-9b is a cross-section of a large cylinder. Again, the radius to the outer rim is 35,000 meters. This station also provides 0.95g on the top floor and 1.1g on the outer rim floor. The top floor would be 4772 meters above the outer rim.

The stations in Figure 4-9 show many floors between the top floor and the outer rim. Figure 4-8b showed without using airtight layers, the top floor at 4772 meters would have less than 57,000 Pascals of air pressure. This pressure would be equivalent to 4603 meters above sea level. This is higher than the tallest mountain in the Rocky Mountains and would cause altitude sickness for most people without sufficient training and acclimation. At 4772 meters there is habitable gravity but not habitable air pressure.

The layers provide habitable air pressure on all the floors. Without the airtight layers there would only be floors to a height of about 1500 meters. There would be only 300 floors to the Denver air pressure limit. Instead, in four layers the example stations would have 770 floors shown in Figure 4-8.

Figure 4-9 includes other station metrics. All the stations have a maximum gravity of 1.1g on their outer rim. The left column has common metrics for the torus and cylinder. The right column has metrics for the large cylinder station. The two middle columns have two versions of the torus station. The left middle column shows the torus version that uses the standard elliptical torus minor axes ratio of c=3a. This produces a torus station that is significantly smaller in mass and population than the cylinder station. As an example, the population drops from 7.0e10 (cylinder) to 2.7e10 people (torus). The right middle column torus version uses an elliptical torus minor axes ratio of c=7.15a. This ratio increases the mass of the torus to match the cylinder station mass of 2.78e15 kilograms. The torus population is 6.30e10 and is much closer to the cylinder population of 7.0e10. With this geometry design change, the torus station stability Iz/Ix drops from 1.78 to 1.23. Both are larger than our 1.2 minimum stability metric.

Cumulus, stratus, and nimbostratus can form below 4772 meters on Earth. The gravity in the space station at the higher elevations is less than on Earth. With this lower gravity, the station would not hold onto its gaseous envelope as tightly as on Earth. Clouds might form at higher station altitudes. Research is still necessary to determine the clouds and weather in these large space stations.

These 35,000-meter radius stations hold huge populations. The population on Earth is 8 billion people, so these stations hold almost 8 Earths of population. This is with a density of 144.2 square meters per person. The amount of open space triples with the layers and could improve the density. Although it is fun to dream about such large stations, more realistic sizes would be more appropriate in the near future.

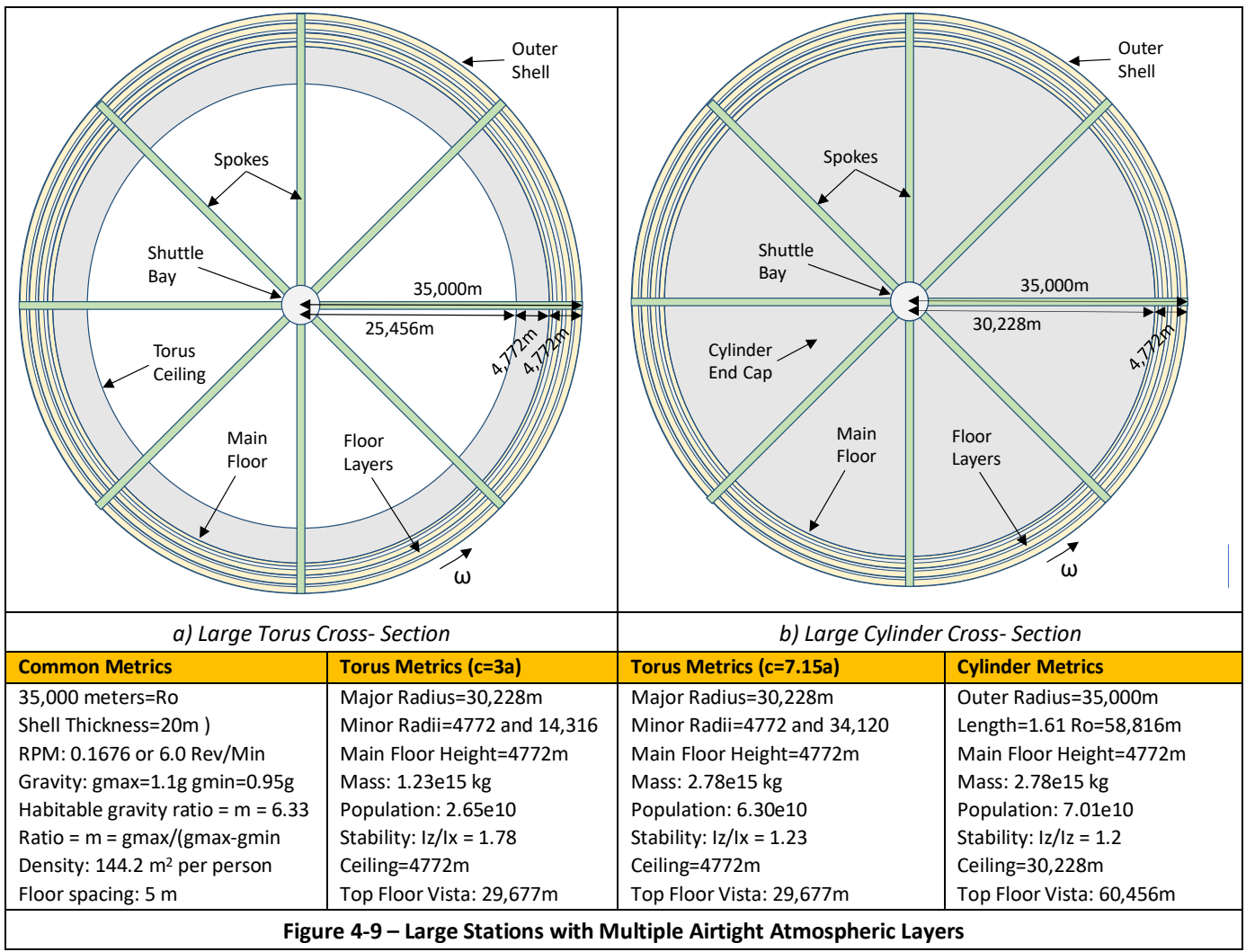

| Common Metrics | Torus Metrics (c=3a) | Torus Metrics (c=7.15a) | Cylinder Metrics |
|---|---|---|---|
| 35,000 meters=Ro | Major Radius=30,228m | Major Radius=30,228m | Outer Radius=35,000m |
| Shell Thickness=20m ) | Minor Radii=4772 and 14,316 | Minor Radii=4772 and 34,120 | Length=1.61 Ro=58,816m |
| RPM: 0.1676 or 6.0 Rev/Min | Main Floor Height=4772m | Main Floor Height=4772m | Main Floor Height=4772m |
| Gravity: gmax=1.1g gmin=0.95g | Mass: 1.23e15 kg | Mass: 2.78e15 kg | Mass: 2.78e15 kg |
| Habitable gravity ratio = m = 6.33 | Population: 2.65e10 | Population: 6.30e10 | Population: 7.01e10 |
| Ratio = m = gmax/(gmax-gmin | Stability: Iz/Ix = 1.78 | Stability: Iz/Ix = 1.23 | Stability: Iz/Iz = 1.2 |
| Density: 144.2 m² per person | Ceiling=4772m | Ceiling=4772m | Ceiling=30,228m |
| Floor spacing: 5 m | Top Floor Vista: 29,677m | Top Floor Vista: 29,677m | Top Floor Vista: 60,456m |

**Figure 4-9 – Large Stations with Multiple Airtight Atmospheric Layers**

## 4.3 Large Historic Stations

Most historic studies use only the outer shell of their geometry for living space. This section considers some large historic stations or increases the size of moderate-sized stations. These stations use multiple floors, provide a reasonable gravity range and air pressure, and ensure rotational stability. The following subsections focus on one large station example from each of the four geometries.

### 4.3.1 Cylinder Station Example

A recent paper investigated the rotational instability of long rotating cylinder space stations [Globus et al. 2007]. That study imposed a limit on the length to passively control the imbalance of the rotating structure [Globus et al. 2007]. To meet this imbalance metric, they found the cylinder length should be less than 1.3 times the radius.

Arthur C. Clarke wrote Rendezvous with Rama and described a cylinder station with a radius of 8 kilometers and a length of 50 kilometers [Clarke 1973]. Figure 4-10 includes artwork of the Rama interior. Using a single floor on the station radius, the station would have 2.5 billion square meters of floor space. Allocating 144.2 square meters per person, the Rama station could house 17 million people.

This analysis applies the rotational stability limits to the Rama cylinder station with multiple components. For the station to be rotationally balanced, the length should be 1.48 times the radius. This reduces the length from 50,000 meters to 11,843 meters. This length is longer than the length found using the Globus thin shell balance ratio of 1.3 [Globus et al. 2007]. The station would rotate once every 2.92 minutes, the outer hull would have 1.05g, and the top floor would have 0.95g. Using this gravity range, the station has a multiple-floor height of 762 meters. The station would have about 150 floors with a spacing of 5 meters between the floors. Air pressure would be fine on all floors. These floors would have 87 billion square meters of floor space. Using all the floors with 144.2 square meters per person, the station could house 600 million people. Multiple floors more than compensate for the population lost from the reduced cylinder length.

### 4.3.2 Sphere Station Example

Researchers and writers have considered spherical space stations for over 100 years. Gerard O'Neill proposed a spherical

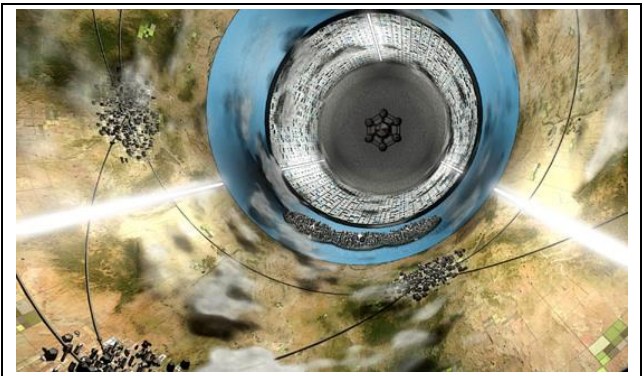
*Credit: Artist: James A. Ciomperlik; [Ciomperlik 2007] [Public domain]*

**Figure 4-10 – Large Cylinder Station Interior**

station with a diameter of 1800 meters to support 75,000 people [O'Neill 1976]. NASA and artist Rick Guidice created artwork illustrating the interior of such a spherical station; see Figure 4-11.

Thin shell analysis shows that spherical space stations are rotationally unstable [Globus et al. 2007]. Analysis of ellipsoid space stations with multiple components shows they can be rotationally stable. In fact, specific radius spherical stations can be rotationally stable. This analysis uses multiple floors, a 20-meter-thick shell, air pressure top floor limit, and gmax=1.05g. With those constraints and using all the components in the ellipsoid station, the analysis shows a spherical station that is 12,194 meters in radius would be rotationally stable with Iz/Ix=1.2.

John Bernal proposed an 8000-meter radius sphere as a space habitat [Bernal 1929]. He suggested a population of 20,000 to 30,000 people. Limiting gravities between 0.95g and 1.05g, the 762 meters between the outer radial rim towards the center could be habitable. This station's habitable single-floor area would be 243 million square meters using the projected area equation from [Johnson and Holbrow 1977]. This would support about 1.7 million people using the 144.2 square meters per person.

Using the 762 meters of habitable space as multiple floors, the station would have up to 153 floors of space. Our analysis shows this Bernal station would have radial axes, a and b, of 8000 meters in length, and the polar axis, c, would be 7829 meters. This is almost spherical with a ratio of a=1.02c and it is rotationally stable with Iz/Ix=1.2. The multiple floors are the dominant mass. They are around the outer equator and improve the stability inertias.

This oblate ellipsoid rotates about the polar axis. The outer hull would have a radius of 8000 meters, and the top floor would be 762 meters. The top floor would have 295 million square meters of floor space. Air pressure would be fine from the lowest floor to the top floor. The gravity would range from 0.95g to 1.05g with the station rotating once every 2.92 minutes. There would be 152 floors with a total surface area of 31 billion square meters. This station could support 219 million people at 144.2 square meters per person.

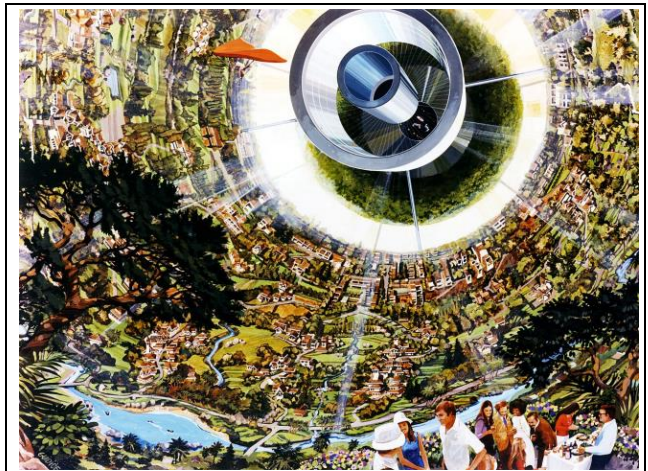
*Credit: NASA Ames Research Center; Artist: Rick Guidice; [Guidice 1970s] [NASA Image Public Domain]*

**Figure 4-11 – Bernal Sphere Interior**

*4.3.3 Torus Station Example*

The space station in the 2013 movie Elysium was a large rotating torus with a major radius of 30 kilometers and a minor radius of 1.5 kilometers. It rotated about once every 6 minutes to create an Earth-like gravity on the surface [Brody 2013]. This station would have at least 565 square kilometers of surface area with one floor. Elysium supports 500,000 people; as such, it has a population density of 1131 square meters per resident. The station from the movie did not have an enclosed ceiling. Figure 4-12 illustrates an enclosed torus space station that is the size of Elysium.

For a multiple-component design, the torus station typically sets the main floor at the major radius. The design would have multiple floors from the main floor to the outer rim. With the large major radius, the gravity range would be from 0.98g on the main floor to 1.03g at the outer rim. The gravity range scaling factor, m, is 20 and much bigger than our typical 6.33 or 9.5 used in this study. The m=20 results in a much more narrow gravity range.

With the minor radius of 1500 meters, 300 floors could be in the station using the 5-meter spacing between floors. This is a radius where the air pressure is acceptable on all floors. This would result in 13.6 billion square meters of floor space and support a population of 945 million. The perceived height of the ceiling would be 1500 meters. The vista in the rotation direction would be 18,735 meters.

Living deep in such a station could cause psychological issues. This analysis considers an alternative that uses only the top 20 of the multiple floors. Elysium would increase its floor area from 565 square kilometers to 11,300 square kilometers. With 144.2 square meters per person, the 20 floors of Elysium would support 82 million people. Using the movie's generous 1131 square meters per person, it would still support over 10 million people. Using only the top 20 floors of the available 300-floor height, aesthetics could be improved for the ceiling height and the vista. The gravity

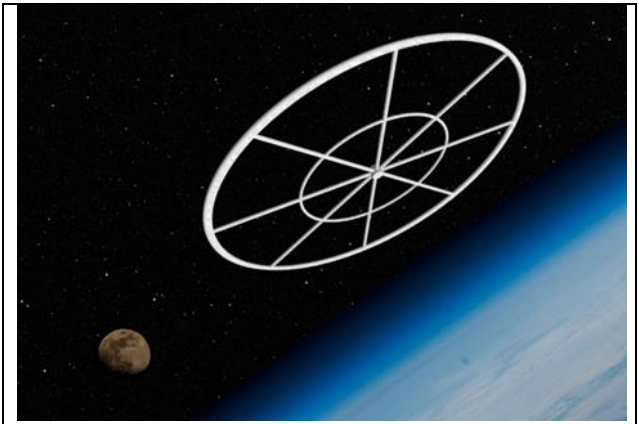

*Credit: Self-produced using Blender and Gimp. Background: Milky Way from NASA SVS Deep Star Maps, Earth and Moon from NASA Explore Image Article; [NASA Image Public Domain]*

**Figure 4-12 – Elysium Size Torus Space Station**

range could be reduced, and air pressure would be near sea level across all the floors.

### 4.3.4 Dumbbell Station Example

The 1977 NASA study [Johnson and Holbrow 1977] included a dumbbell using two spherical nodes with radii of 316 meters to support a population of 10,000 people. They allocated 67 square meters for each individual. In that study, they used a projected surface area equal to the center cross-section of the dumbbell sphere. A much larger dumbbell design with multiple floors is described in [Jensen 2023].

Figure 4-13 provides a rendering of that dumbbell. It is roughly to scale for a station created from the Atira asteroid, which has 4.1e13 kilograms of mass. The station was designed to use 3.9e12 kilograms of material. A simple stress analysis finds that the dumbbell's tether or trusses need an equivalent cross-section of at least 12,800 square meters using a tensile strength of 1500 MPa (filled structure metric). Figure 4-13 shows multiple truss tethers to provide this cross-section.

The dumbbell would have over 160 floors. Its inhabitants would have a dome over them with a ceiling 820 meters above. This station could support a realistic population of almost 2 million people. The dumbbell would provide a vista of 1.6 kilometers by 0.8 kilometers from the center of the main floor. The design is quite habitable and offers open space for psychological well-being.

### 4.3.5 Large Station Adaptations

Previous subsections and examples characterized large space stations using multiple components. It is possible to maintain acceptable gravity levels from the top floor to the outer rim (0.95g to 1.05g). A station designed to be rotationally stable typically has a smaller dimension on its rotation axis. Multiple floors increase the population and compensate for that smaller dimension. With the increased floor space, the station design can be simplified by bringing the agriculture into the station (instead of an external "crystal palace") [O'Neill 1976]. Even with the simpler design and the internal agriculture, the stations can support greater populations than the historic single-floor design. The large stations begin to have an excessive number of floors that could cause psychological well-being issues for individuals living deep in the station. Using fewer floors and higher ceilings could address those issues and yet continue to support large populations.

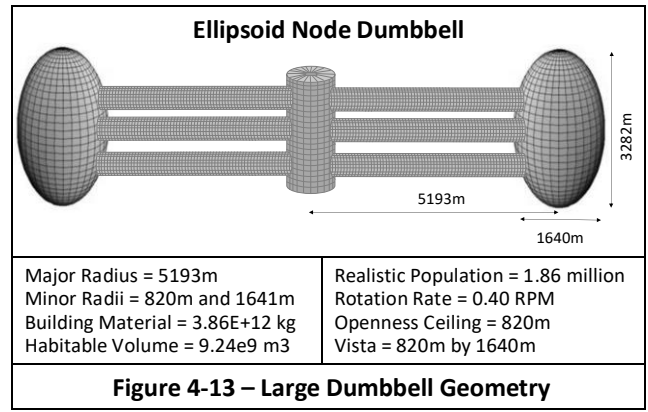

| | |
|---|---|
| Major Radius = 5193m | Realistic Population = 1.86 million |
| Minor Radii = 820m and 1641m | Rotation Rate = 0.40 RPM |
| Building Material = 3.86E+12 kg | Openness Ceiling = 820m |
| Habitable Volume = 9.24e9 m3 | Vista = 820m by 1640m |

**Figure 4-13 – Large Dumbbell Geometry**

## 4.4 Construction Summary

This section reviewed the evolution of space station construction. It reviewed how construction approaches have evolved from using manned, teleoperation, semi-autonomous, and autonomous systems. It included two recent autonomous construction approaches.

This section reviewed the characteristics of very large space stations. Previous chapters have shown the four typical geometries. Thick shields provide protection from radiation and debris impacts. Three key constraints limit the size of large space stations. First, the source of material (available asteroids) limits the size. A 10,000-meter radius station would require an asteroid with a diameter of about 8,000 meters. There are no known near-Earth asteroids of this size. Adapting our autonomous system to operate in the asteroid belt region may be possible. Second, the air pressure in large stations limits the number of floors. Even this limit has a solution, which is to create multiple airtight layers of floors. Third, psychological issues could limit the habitable region to these large stations' top 20 to 50 floors. Another subsection presented these design limitations and extensions applied to historic stations. This presented the size and population changes with these enhanced historic stations.

Despite the challenges, multiple authors believe space station construction is feasible. We include several quotes in historical order from some of these authors:

In the 1970s, a NASA study concluded *"On the basis of this 10-week study of the colonization of space there seems to be no insurmountable problems to prevent humans from living in space….The people of Earth have both the knowledge and resources to colonize space."* [Johnson and Holbrow 1977].

Metzger believes that hardware adaptations are mature enough for early use in the space industry [Metzger 2016]. He wrote, *"There is nothing to suggest we should wait for* *robotic autonomy to improve before we start the project.*" [Metzger 2016].

The authors of the spin construction approach state *"our results indicate that the basic physics of transforming small asteroids into human habitats is feasible."* [Miklavčič et al. 2022].

We stated in our restructuring asteroids paper *"We are at the stage where it appears that the restructuring process is viable. … The restructuring process offers humanity the opportunity to truly become a space faring society."* [Jensen 2023].

## 5  Limits Summary

This section summarizes the constraints identified in this paper. It includes constraint charts illustrating these limits as a function of station radius. It concludes with a subsection covering our conclusions from this analysis.

Many constraints limit the size of space stations. Human frailty creates limits from constraints such as gravity range, air pressure, and radiation. Structure design creates limits from constraints such as material, geometry, and rotational stability. Human psychology creates limits such as surface area, surface usage, and vistas. The following subsections cover these limits.

### 5.1  Human Frailty Limits

This subsection covers human frailty limits from gravity, air pressure, radiation, population, and psychology.

*5.1.1 Gravity Limits*

Centripetal forces from rotation create artificial gravity in the rotating space station. Limiting the station rotation speed to 2 rpm or less prevents motion sickness [Hall and Globus 2017]. Stations need a 223-meter or greater radius to provide this rotation speed and produce Earth-equivalent centripetal gravity. Another study defined a maximum rotation velocity of 1 rpm or less when using an attached shield [Graem 2006]. This implies a minimum rotation radius of almost 900 meters.

When possible, our study aims for a gravity range of 0.95g to 1.05g in the most occupied areas of the station. This gravity range should minimize the health risk to the residents. Regions with gravity closest to Earth-like (1g) are used for living, recreational, and working quarters. The centripetal gravity increases with depth into the station. Higher-gravity regions could provide high-strength training, filtration systems, and higher-gravity research. A recent space station design used an entire deck as a large ventilation duct [Soilleux and Gunn 2018]. One could place this ventilation deck (or two) in the higher-gravity region in these large stations. This saves space on the 1g regions to support a larger population. Other functions could be placed in the lower, high-gravity deck including machinery, storage, and agriculture. With today's understanding, working in the higher and lower gravities should be kept to a minimum and avoided or perhaps done in shifts. Our hope is the chosen gravity range and limited exposure to lower and higher gravities will have no impact on the inhabitants' health.

*5.1.2 Air Pressure Limits*

Air is obviously required to support our station's population. Early space missions used low-pressure pure oxygen to reduce weight and still provide a breathable atmosphere. Ideally, the station will have an earthlike mixture of nitrogen, oxygen, and carbon dioxide.

Human bodies have adapted to a range of air pressures on Earth's surface. This ranges from slightly less than sea level to altitudes where severe altitude sickness can begin to occur (above 3,500 meters). The station design will use an acceptable air pressure range for the colonists' health. Sea level is the high air pressure limit. A low air pressure limit is 83,728 Pascals. This is the air pressure of Denver, which is 1,609 meters above sea level. These limits are conservative, and nearly all individuals would breathe comfortably. Lower air pressure and density can reduce mass, costs, and stresses.

Text for Figure 3-10 described the top floor gravity limit of 0.95g in a station with a radius of 16,900 meters. The station would have a top floor at 1609 meters above the outer rim. The air pressure at this station's height would be like that of Denver, Colorado. With large stations, air pressure limits the habitable floors more than the gravity limits. One choice to increase the number of floors and population is to lower the air pressure on the top floor or increase the air pressure on the outer rim. In very large stations, another alternative is to create multiple airtight layers of floors within the gravity-limited regions.

*5.1.3 Radiation Limits*

There is significant radiation in space that can damage the DNA in cells [Blanchett and Abadie 2018]. Long-term effects from this radiation include cancer, sterility, cataracts, cardiovascular damage, cognitive impairment, and memory deficits. Various studies suggest using a layer of regolith to provide protection from cosmic rays. The depth of the regolith ranges from 2.5 meters [O'Neill 2008] to more than 8 meters [Turner and Kunkel 2017] to reduce the radiation to an Earth background level. Researchers have found 6 meters of lunar regolith would eliminate the radiation effects [De Angelis et al. 2002]. Our designs use greater thickness on the outer walls to provide additional integrity for potential debris, small meteoroids, and ship collisions. Our designs and analysis typically use a shell thickness of 20 meters.

*5.1.4 Population Limits*

Station population limits were introduced in [Jensen 2023]. For all four geometries, using multiple floors significantly improved the supported population. For a single-floor geometry, the population is proportional to the radius squared (area function). For multiple-floor geometries, the population is nearly proportional to the radius cubed (volume function). Multiple floors take advantage of that internal volume in the station geometry and yet leave open aesthetic views.

The population density can be improved with multiple floors and increased surface area. Instead of 67 square meters per resident, as used in earlier NASA studies [Johnson and Holbrow 1977], our studies offer 144 square meters per person. A larger design can offer a population density of many hundreds of square meters per resident. This is not unexpected. As stated in the NASA SP-413 study [Johnson and Holbrow 1977], habitats with large living space *"would be settled with much lower population densities, so as to permit additional "wild" areas and parkland."*

These population increases from multiple floors are common on the Earth. As summarized in [Jensen 2023]: *"Adding floors to a structure greatly increases the available floor space. Examples of multiple floor structures include underground cities, submarines, cruise ships, and skyscrapers. Historic and modern underground cities exist. Entrepreneurs have begun to convert abandoned military missile silos into multiple floor homes and underground cities. Limited space and costs in urban environments promote high rise living."*

### 5.1.5 Psychological Limits

Living in a space station could cause psychological issues for the inhabitants. NASA studies have considered the oppressive closed-quarters ambience of a space station to be a risk to the colonists' psychological well-being [Johnson and Holbrow 1977][Keeter 2020]. These stations are smaller than our envisioned large stations. There are still other issues with creating and using many floors.

Researchers have studied the psychological issues associated with multiple-story buildings and underground cities. Researchers and developers believe the biggest problems for underground cities are not technical but social [Garrett 2019]. Studies have found that people living in high-rises suffer from greater mental health problems, higher fear of crime, fewer positive social interactions, and more difficulty with child-rearing [Barr 2018]. These researchers offer approaches to address these risks. With proper planning and space allocation, using many floors is acceptable for space station habitation.

Open vista aesthetics should address the psychological limits of a rotating space station with multiple floors. Our designs reserve at least the space station's upper half, low-gravity regions to serve as open space. This open space provides good vistas and beneficial aesthetics for the residents and visitors. Sphere, cylinder, and dumbbell geometries all have similar aesthetics. The interiors have open space overhead. For the same population (or construction mass), the torus typically has lower ceilings than the other geometries. The torus has a longer vista when looking down the torus tube in the rotation direction.

## 5.2 Station Design Limits

This section covers station design limits from geometry, rotational stability, materials, and mass.

### 5.2.1 Geometry Limits

The typical space station geometries were refined to address multiple floors, rotational stability, and gravity issues. Spherical stations became ellipsoidal stations; long cylindrical stations became short hatbox stations; circular cross-section torus stations became elliptical cross-section torus stations; and dumbbell structures were doubled. Air pressure becomes a limitation with very large stations. These large stations provide open views and vistas, providing the inhabitants with psychological well-being. The large stations can produce an excessive number of floors, and living deep in the station could cause psychological issues for the inhabitants. Creating open spaces in the lower decks might provide pleasing aesthetics for those inhabitants. Given the excessive number of floors, reducing their number or creating lower-level open spaces are viable approaches to address those psychological issues.

### 5.2.2 Rotational Stability Limits

There is a risk for rotation instabilities with asymmetric geometries. This instability can cause abrupt changes in orientation between two rotational states. This abrupt change would not be desirable and would likely be catastrophic for a space station. Our stations are symmetric; however, one must still design the geometry of the space station to avoid this behavior. Experience with spin-stabilized spacecraft suggests that the desired axis of rotation should have an angular moment of inertia at least 1.2 times greater than any other axis [Brown 2002]. Refining the set of space station geometries helps to address rotational stability.

### 5.2.3 Material Limits

Previous subsections reviewed multiple material types, availability, strength, and stresses. That study found that all the materials considered supported their working stress below the rotation radius of 10,000 meters. Our study often uses this radius as a maximum value because of the material strength review.

The exterior station shell could be over 20 kilometers in radius using melted asteroid material (basalt rods). Instead of basalt building material, using anhydrous glass might support even larger stations. This paper reviewed and included anhydrous glass data from multiple researchers [Blacic 1985] [Carsley, Blacic, and Pletka 1992] [Bell and Hines 2012]. The tensile strength of refined anhydrous glass suggests stations up to 125,000 meters could be feasible. Our designs assume a truss-like structure of anhydrous glass filled with crushed regolith. The tensile strength of this structure is defined as 1500 MPa, and the density is set to 1720 kilograms per cubic meter. Table 3-6 provides an estimate that this structure could support a station with a 21.1-kilometer radius.

Perfect anhydrous glass tiles will not be produced using our somewhat primitive manufacturing techniques. Before building large stations, it would be wise to focus conservatively on smaller station radii (3 to 4 kilometers).

Some of these stations could have other structural issues. More research and analysis are clearly required in this area. These early results provide confidence to recommend additional studies and that such stations can be constructed.

*5.2.4 Mass Limit*

Stations with radii between 224 and 10,000 meters require 2e10 to 3e13 kilograms of material. Many small near-Earth-asteroids would provide the 2e10 kilogram lower limit. There are few asteroids to provide the 3e13 kilogram limit. A moon or Ceres are potential sources for the material in stations of this size.

## 5.3 Constraints Overview Charts

This paper has presented many constraints on space station size. This subsection presents graphical overviews of those constraints.

*5.3.1 Limits and Population*

Figure 5-1 contains a column chart showing the population of two different size cylinder stations. The chart shows the effect from geometry size, single and multiple floors, air pressure limit, and gravity limit. The y-axis shows the population on a logarithmic scale ranging from 100 to 1 trillion people. It includes a dotted line at 8 billion people to reference today's world population. The x-axis shows four sets of station data. The data includes results from cylinders with a rotation radius of 2400 meters and 35,000 meters. Previous chapters and sections have covered these size stations.

The chart includes the population of a single-floor cylinder station. This is the population on the outer rim surface area of the cylinder. The leftmost columns show the population for a long cylinder with a length of 10 times the radius. This is the "traditional" geometry used in early station designs [O'Neill 1976]. The single floor 2400 meter radius cylinder with L=10R can support 2.5 million people with 144.2 square meters per person. The 35,000-meter radius with L=10R station can support 534 million people on the single-floor design. These stations are not rotationally stable.

Figure 5-1 also includes a cylinder station with a length of 1.3 times the radius. This station is described as a hatbox shape. This single-floor design is rotationally stable. As expected, the population is less than the long cylinder geometry. The 2400-meter radius cylinder can support 326 thousand people on the outer rim of a single floor. The 35,000-meter radius station can support 69 million people.

The chart includes two designs using multiple components. The multiple-component cylinder designs have different length-to-radius ratios to produce rotationally stable designs. The L/R ratios are larger than the single-floor design ratios, and they range from 1.36 to 1.66. One of the components is the multiple floors. The data shows a significant increase in the supported population with the multiple floors.

The chart shows two limits on the top floors of the stations. One limit is set by the Denver air pressure constraint of about 1600 meters. The other limit shows the top floor constrained only by gravity. Both these cases use a gravity range from 1.05g on the outer (lowest) floor and 0.95g on the top floor. This floor would be located at the height of the rotation radius divided by 9.5.

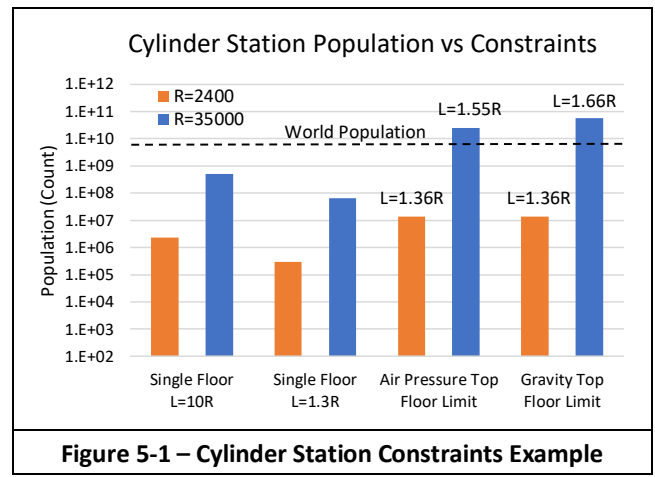

**Figure 5-1 – Cylinder Station Constraints Example**

The chart shows that the 2400-meter radius cylinder can support 13.5 million people. The gravity range limits the population to 42 floors. The gravity would be 0.95g on the top floor, and the air pressure is higher (better) than the Denver limit. The air pressure does not limit the top floor of the 2400-meter radius station.

When limited to the Denver air pressure at a station height of 1567 meters, the 35,000-meter radius station could have 310 floors. That design can support 25 billion people. The top floor gravity limit is at a height of 3333 meters. The air pressure at this height would be uncomfortable for most people. Creating multiple layers of floors would allow the floors to extend to the gravity limit. With 4 airtight layers, the 35,000-meter radius cylinder station can support over 50 billion people on 663 floors. This is becoming a bit absurd; recall that Earth's population is 8 billion.

*5.3.2 Habitable Constraints*

Only two cylinder station sizes were reviewed in the previous subsection. This subsection shows many more constraints and their effect on a range of cylinder and torus station designs. These charts are in Figure 5-2. This subsection first describes common elements from these charts. Table 5-1 summarizes these common elements.

Rotating the space station provides centripetal gravity. Small stations would spin too fast to provide 1g and cause inner ear disturbances for most residents. This causes motion sickness and disorientation. A minimum station rotation radius should be greater than 200 meters. Figure 5-2 shows this limit as a solid yellow line at 200 meters on the rotation radius x-axis. This limits the left edge of the green habitable area.

Our designs impose an acceptable gravity range limit on habitable areas. The top floor of the multiple floors is 0.95g, and the bottom floor is 1.05g. This gravity range produces a limit on the top floor at the rotation radius divided by 10.5 for cylinders and 9.5 for tori. This is shown as the solid, thick orange line limiting the top edge of the green habitable area.

| Table 5-1 – Common Station Limits and Constraints ||
|---|---|
| Constraint | Limit |
| Rotation | Motion Sickness If R<200m |
| Gravity Range | 0.95g to 1.05g |
| Top Floor Gravity | h = R/10.5 or R/9.5 |
| Top Floor Air pressure | h = 1609 m (Denver Altitude) |
| Multiple Layers of Floors | Layers of h=1600 m for air pressure |
| Materials Strength | Radius = 10K (Conservative) |
| Psychology | Good aesthetics if vista>130m |
| Cylinder Vista | Top floor to center |
| Torus Vista | Long view down torus tube |

Station designs also impose a habitable minimum air pressure limit matching the city of Denver's air pressure. This limit is shown as a green dashed line at 1608 meters on the y-axis. This typically limits the top floor height and the top edge of the green habitable area. The charts include the concept of using multiple airtight layers of floors. The charts show this concept as green stair steps beyond the 1600-meter air pressure limit. This extends the green habitable area into its upper right corner. This is only valid for very large stations with radii greater than 16,000 meters.

The charts include two material strength limits. One is a conservative limit of a 10,000-meter rotation radius. The other is a less conservative 40,000-meter rotation radius using values for anhydrous glass. This 40,000-meter radius is less than the 125,000-meter radius using perfect anhydrous glass. These material strengths limit the right side of the green habitable area.

The habitable region for cylinders is shown in Figure 5-2a. It shows cylinder radii from 200 meters to 40,000 meters are habitable. The top floor ranges from a height of 19 meters for the 200-meter station to 1600 meters for the 16,800-meter station. Without airtight layers, the top floor height remains at 1600 meters up to the 40,000-meter station.

The habitable region for tori is shown in Figure 5-2b. It shows torus radii from 200 meters to 40,000 meters are habitable. Like with the cylinder, the top floor ranges from a height of 19 meters for the 200-meter station to 1600 meters for the 16,800-meter station. Without airtight layers, the top floor height remains at 1600 meters up to the 40,000-meter station radius.

One of the few differences shown between Figure 5-2a and Figure 5-2b is the vista. The charts assume the psychology limit is an open vista minimum of 130 meters. This would be a 9 or 10-story building, and it is the height used in the Stanford Torus [Johnson and Holbrow 1977]. A solid dark blue line shows the best vista on each chart. A dotted blue line shows the radius where the psychology limit is met.

The enclosed cylinder station vista extends from one side of the top floor, through the cylinder center, and to the same floor on the opposite side. The vista is always greater than 130 meters, even with the smallest cylinder station. The psychology radius limit is shown at the minimum x-axis value. A cylinder would be smaller than 65 meters in radius and spin rapidly to provide the 1g on the outer rim.

The torus has an enclosed tube that provides a much lower ceiling. The gravity range determines the size of the tube. In this example, the tube's minor radial axis length is set to the major radius length divided by 9.5. A better vista in the torus is available by looking down the tube in the rotation direction and discussed in *§2.6.2 Open Vistas Aesthetics*. The chart shows that the psychological vista limit has a negligible impact on the size of the torus's habitable area. It shows that the torus radius must be greater than 145.5 meters to provide a sufficient vista. In this case, the motion sickness minimum radius of 200 meters is more limiting.

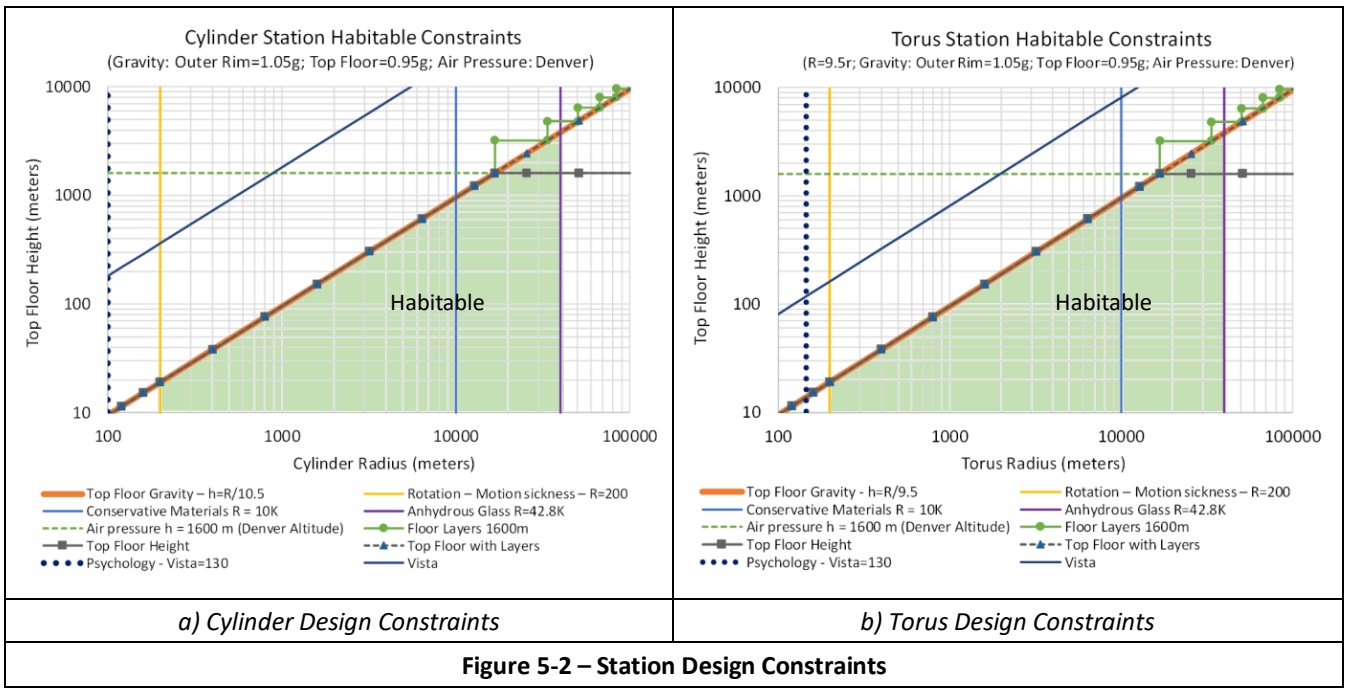

| a) Cylinder Design Constraints | b) Torus Design Constraints |

**Figure 5-2 – Station Design Constraints**

## 5.4 Conclusions from Limits

The limits presented in this paper provide guidance on the size and geometry of the space station. Our analysis and details do not cover all issues associated with a large project like this restructuring effort. Exploring large project issues such as design, construction, management, and perception would be valuable. This list of issues is not presented to provide ammunition to doubters; instead, this is a list of solvable research challenges for a realistic project.

Even with these issues, researchers and entrepreneurs are still considering how to implement these Space Station designs. Jeff Bezos is the founder of Amazon. He later founded Blue Origin to "*build a road to space so our children can build the future* [Blue Origin 2021]." He has proposed space stations like the O'Neill colonies [Blitz and Orf 2019]. Those authors wrote that Bezos stated *"How are we going to build O'Neill colonies? I don't know and no one in this room knows"* [Blitz and Orf 2019]. Blue Origin recognizes the many issues with space exploration and advocates *gradatim ferociter* or *Step by Step, Ferociously*. Recent research construction approaches offer new means to build these space stations. We acknowledge there are many issues but believe that the solutions are within humanity's grasp.

[Diamandis 1987] Reconsidering Artificial Gravity for Twenty-First Century Space Habitats. Peter H. Diamandis, Space Manufacturing 6: Nonterrestrial Resources, Biosciences, and Space Engineering, B. Faughnan, G. Maryniak (Eds.), 6:55-68. AIAA, 1987, https://ssi.org/programs/ssi-g-lab-project/artificial-gravity-peter-diamandis-1987/

[Dunbar 2019] Why Space Radiation Matters, NASA Official: Brian Dunbar, 8 October 2019, National Aeronautics and Space Administration, Analog Missions, https://www.nasa.gov/analogs/nsrl/why-space-radiation-matters

[Dyson 1960] Search for Artificial Stellar Sources of Infra-Red Radiation, Freemann J. Dyson, 3 June 1960, Science. 131 (3414): pp. 1667-1668, Bib-code:1960Sci...131.1667D. doi:10.1126/science.131.3414, https://fermatslibrary.com/s/search-for-artificial-stellar-sources-of-infra-red-radiation

[Fitzpatrick 2011] Newtonian Dynamics, Richard Fitzpatrick, 2011 March 31, Rigid Body Rotation, Rotational Stability, The University of Texas at Austin, https://farside.ph.utexas.edu/teaching/336k/Newton/node71.html

[Fitzpatrick 2023] Dumbbell Space Stability, Richard Fitzpatrick, 2023 June 28, Personal Correspondence with Professor University of Texas

[Freitas and Gilbreath 1982] Advanced Automation for Space Missions: Proceedings of the 1980 NASA/ASEE Summer Study held at the University of Santa Clara, Robert A. Freitas, Jr., and William P. Gilbreath, June 23-August 29, 1980; NASA Conference Publication 2255, 15 May 1981, NASA Ames Research Center, Moffett Field, Publication Date: 1 November 1982, https://ntrs.nasa.gov/archive/nasa/casi.ntrs.nasa.gov/19830007077.pdf

[Frost 2011] Chapter: Challenges and Opportunities for Autonomous Systems in Space, Chad R. Frost, 2011, NASA Ames Research Center, Frontiers of Engineering: Reports on Leading-Edge Engineering from the 2010 Symposium (2011), https://www.nap.edu/read/13043/chapter/17

[Garrett 2019] A cinema, a pool, a bar: inside the post-apocalyptic underground future, Bradley L Garrett, 16 December 2019, The Guardian, https://www.theguardian.com/cities/2019/dec/16/a-cinema-a-pool-a-bar-inside-the-post-apocalyptic-underground-future

[Globus 2006] Contest-Driven Development of Orbital Tourist Vehicles, Al Globus, American Institute of Aeronautics and Astronautics Space 2006, San Jose, California, 19-21 September 2006. https://pdfs.semanticscholar.org/4ca2/10f824b5c00c02b6a3f7ee8cef572f5bf649.pdf

[Globus and Hall 2017] Space Settlement Population Rotation Tolerance, Al Globus and Theodore Hall, June 2017, NSS Space Settlement Journal, https://space.nss.org/wp-content/uploads/NSS-JOURNAL-Space-Settlement-Population-Rotation-Tolerance.pdf

[Globus et al. 2007] The Kalpana One Orbital Space Settlement Revised, Al Globus, Nitin Arora, Ankur Bajoria, and Joe Strout, 2007, American Institute of Aeronautics and Astronautics, http://alglobus.net/NASAwork/papers/2007KalpanaOne.pdf

[Graem 2006] Visions 2200 - A Perspective on the Future Space Habitat, H Graem, 2006, http://visions2200.com/SpaceHabitat.html, Wayback Machine Access: https://web.archive.org/web/20080209111914/http://visions2200.com/SpaceHabitat.html

[Guidice 1970s] Space Colony Art from the 1970s, NASA Ames Research Center, Artist: Rick Guidice, 1970s, NASA ID Number: AC76-0628, https://web.archive.org/web/20200617054429/https://settlement.arc.nasa.gov/70sArtHiRes/70sArt/art.html;

[Hale 1869] The Brick Moon, Edward Everett Hale, October 1869, The Atlantic Monthly, No. 144, https://www.gutenberg.org/ebooks/1633

[Hall 1991] The Architecture of Artificial Gravity: Mathematical Musings on Designing for Life and Motion in a Centripetally Accelerated Environment, Theodore W. Hall, November 1991, Space Manufacturing 8 Energy and Materials from Space, Proceedings of the Tenth Princeton/AIAA/SSI Conference, 15-18 May 1991, https://citeseerx.ist.psu.edu/viewdoc/download?doi=10.1.1.551.3694&rep=rep1&type=pdf

[Hall 1993] The Architecture of Artificial Gravity: Archetypes and Transformations of Terrestrial Design, Theodore W. Hall, 12-15 May 1993, Space Manufacturing 9, The High Frontier Accession, Development and Utilization, Proceedings of the Eleventh SSI-Princeton Conference, p. 198-209, http://www.artificial-gravity.com/SSI-1993-Hall.pdf

[Hall 1997] Artificial Gravity and the Architecture of Orbital Habitats, Theodore W. Hall, 20 March 1997, Proceedings of 1st International Symposium on Space Tourism, Daimler-Chrysler Aerospace GmbH., http://www.spacefuture.com/archive/artificial_gravity_and_the_architecture_of_orbital_habitats.shtml

[Hall 2006] Artificial Gravity Visualization, Empathy, and Design, Theodore Hall, 19-21 September 2006, AIAA 2006-7321, Session: SAS-3: Space Architecture Symposium: Gravity Regime Architecture and Construction, https://doi.org/10.2514/6.2006-7321, http://www.artificial-gravity.com/AIAA-2006-7321.pdf

[Hand and Finch 1998] Analytical Mechanics, Louis N. Hand and Janet D. Finch, 1998, Cambridge University Press. p. 267. ISBN 978-0-521-57572-0, https://www.iaa.csic.es/~dani/ebooks/Mechanics/Analytical%20mechanics%20-%20Hand,%20Finch.pdf

[Hirt et al. 2013] New ultrahigh-resolution picture of Earth's gravity field; Christian Hirt, Sten Claessens, Thomas Fecher, Michael Kuhn, Roland Pail, and Moritz Rexer; 28 August 2013; Geophysical Research Letters. 40 (16): 4279–4283, https://doi.org/10.1002%2Fgrl.50838

[Janhunen 2021] Terraforming the dwarf planet: Interconnected and growable Ceres megasatellite world, Pekka Janhunen, 10 May 2021, https://arxiv.org/pdf/2011.07487

[Jensen 2023] Autonomous Restructuring of Asteroids into Rotating Space Stations, David W. Jensen, 23 February 2023, https://arxiv.org/abs/2302.12353

[Jensen 2024s] Space Station Rotational Stability, David W. Jensen, 22 July 2024, Submitted to arXiv.org

[Johnson and Holbrow 1977] NASA SP-413: Space Settlements: A Design Study, Eds: Richard D. Johnson and Charles Holbrow, 1977, Scientific and Technical Information Office, https://ntrs.nasa.gov/citations/19770014162

[JPL SBD Search Engine 2023] JPL Small-Body Database Search Engine, NASA Jet Propulsion Laboratory, Solar System Dynamics, Accessed 2023, https://ssd.jpl.nasa.gov/sbdb_query.cgi

[Kajima et al. 2017] Development and evaluation of the innovative remote construction system by cooperation of remote control and automatic control, Kajima Corporation, Shibarua Institute of Technology, Kyoto University, and The University of Electro-Communications, JAXA, Automatic and autonomous exploration technology / Solution Creating Research, Research on the unmanned construction system of a manned lunar outpost, http://www.ihub-tansa.jaxa.jp/english/files/report/business-overview_4.pdf

[Kawano et al. 2022] Rotations: A tumbling T-handle in space: the Dzhanibekov effect, Daniel T. Kawano, Alyssa Novelia, and Oliver M. O'Reilly, 10 March 2022, https://rotations.berkeley.edu/a-tumbling-t-handle-in-space/

[Keeter 2020] Long-Term Challenges to Human Space Exploration, Bill Keeter, 4 September 2020, National Aeronautics and Space Administration, https://www.nasa.gov/centers/hq/library/find/bibliographies/Long-Term_Challenges_to_Human_Space_Exploration

[Koelle and Williams 1959] Project Horizon: Volume II, Technical Considerations & Plans, H. H. Koelle and F. L. Williams, 9 June 1959, A U. S. Army Study For The Establishment of a Lunar Outpost, United States Army, https://history.army.mil/faq/horizon/Horizon_V2.pdf

[Lamb 2013] Industrial Automation: Hands On, Frank Lamb, 2013, McGraw Hill Professional, ISBN 007181647X, 9780071816472, https://books.google.com/books/about/Industrial_Automation_Hands_On.html?id=H977GaxHwaAC

[Lee 2018] Total Automation: The Possibility of Lights-Out Manufacturing in the Near Future, Noah K. Lee, May 2018, Missouri University of Science and Technology, https://scholarsmine.mst.edu/cgi/viewcontent.cgi?article=1025&context=peer2peer

[Lente and Ősz 2020] Barometric formulas: various derivations and comparisons to environmentally relevant observations, Gábor Lente and Katalin Ősz, 4 April 2020, ChemTexts 6, 13 (2020). https://doi.org/10.1007/s40828-020-0111-6, https://link.springer.com/article/10.1007/s40828-020-0111-6

[Lewis-Weber 2016] Lunar-Based Self-Replicating Solar Factory, Justin Lewis-Weber, 2016, New Space, vol. 4, issue 1, pp. 53-62, https://www.researchgate.net/publication/297746685_Lunar-Based_Self-Replicating_Solar_Factory
July 2024     Design Limits on Large Space Stations     61

# 7 License Types

This document is Copyright © 2024 to David W. Jensen. All self-produced material is licensed under a Creative Common License CC BY-SA 4.0 [CC BY-SA 4.0]. Licensing for material from other sources is referenced from the text and are described in this section.

[CC BY-4.0] Attribution 4.0 International (CC BY 4.0), Creative Commons, You are free to share and adapt for any purpose, even commercially. You can copy and redistribute the material in any medium or format. You can remix, transform, and build upon the material. You must give appropriate credit, provide a link to the license, and indicate if changes were made; https://creativecommons.org/licenses/by/4.0

[CC BY-NC 4.0] Attribution-NonCommercial 4.0 International (CC BY-NC 4.0), Creative Commons, You are free to copy and redistribute the material. You may remix, transform, and build on the material. You must give appropriate credit, provide a link to the license, and indicate if changes were made. You may not use for commercial purposes; https://creativecommons.org/licenses/by-nc-sa/4.0/

[CC BY-SA 4.0] Attribution-ShareAlike 4.0 International (CC BY-SA 4.0), Creative Commons, Free to Share and Adapt; You must give appropriate credit, provide a link to the license, and indicate if changes were made; You may adapt the material for any purpose, even commercially; ShareAlike: You must distribute your contributions under the same license as the original, https://creativecommons.org/licenses/by-sa/4.0/

[Facts] Facts and data are generally not eligible for copyright. "Charts, graphs, and tables are not subject to copyright protection because they do not meet the first requirement for copyright protection, that is, they are not "original works of authorship," under the definitions of 17 U.S.C. § 102(a). Essentially, that means that a graph, chart, or table that expresses data is treated the same as the underlying data. Facts, data, and the representations of those facts and data are excellent examples of things that require much "sweat of the brow" to create, but yet still do not receive copyright protection." https://deepblue.lib.umich.edu/bitstream/handle/2027.42/83329/copyrightability_of_tables_charts_and_graphs.pdf

[NASA Image Public Domain] National Aeronautics and Space Administration, Accessed 21 November 2021, NASA content is generally not subject to copyright in the United States. You may use this material for educational or informational purposes, including photo collections, textbooks, public exhibits, computer graphical simulations and Internet Web pages, https://images.nasa.gov/, https://www.nasa.gov/multimedia/guidelines/index.html

[NASA Report Public Domain] NASA Scientific and Technical Information Program, Accessed 21 November 2021, Generally, United States government works (works prepared by officers and employees of the U.S. Government as part of their official duties) are not protected by copyright in the U.S. (17 U.S.C. §105) and may be used without obtaining permission from NASA, https://ntrs.nasa.gov/, https://sti.nasa.gov/disclaimers/

[Public Domain] Public domain consists of all the creative work to which no exclusive intellectual property rights apply. Those rights may have expired, been forfeited, expressly waived, or may be inapplicable.

# 8 Contents – Design Limits on Large Space Stations